\documentstyle[newamssymbols]{birk-ot}
\Year{2002}
\begin{document}
%%%%%%%%%%%%%%%%%%%%%%%%%%%%%%%%%%%%%%%%%%%%%%%%%%%%%%%%%%%%%%%%%%%%%%%%%%%%%%%
\Title{Integrable Systems and Factorization Problems}
\Shorttitle{Integrable Systems and Factorization Problems}
\By{{\sc M. ~A. ~Semenov-Tian-Shansky}}
\Names{Semenov-Tian-Shansky}
\Email{semenov@u-bourgogne.fr}
\maketitle
%%%%%%%%%%%%%%%%%%%%%%%%%%%%%%%%%%%%%%%%%%%%%%%%%%%%%%%%%%%%%%%%%%%%%%%%%%%%%%%
%                                                                             %
%                   Please insert now the article body.                       %
%                                                                             %
%%%%%%%%%%%%%%%%%%%%%%%%%%%%%%%%%%%%%%%%%%%%%%%%%%%%%%%%%%%%%%%%%%%%%%%%%%%%%%%
\begin{abstract}
The present lectures were prepared for the Faro International
Summer School on Factorization and Integrable
Systems in September 2000. They were intended for participants with
the background in Analysis and Operator Theory but without special knowledge
of Geometry and Lie Groups. The text below represents a sort of compromise:
it is certainly impossible not to speak about Lie algebras and Lie groups at
all; however, in order to make the main ideas reasonably clear, I tried to
use only matrix algebras such as $\frak{gl}(n)$ and its natural subalgebras;
Lie groups used are either $GL(n)$ and its subgroups, or \emph{loop groups }%
consisting of matrix-valued functions on the circle (possibly admitting an
extension to parts of the Riemann sphere). I hope this makes the environment
sufficiently easy to live in for an analyst. The main goal is to explain how
the factorization problems (typically, the matrix Riemann problem)
generate the entire small world of Integrable Systems along with the
geometry of the phase space, Hamiltonian structure, Lax representations,
integrals of motion and explicit solutions. The key tool will be the \emph{%
classical r-matrix} (an object whose other guise is the well-known Hilbert
transform). I do not give technical details, unless they may be exposed in a
few lines; on the other hand, all motivations are given in full scale
whenever possible. I hope that this choice agrees with the spirit of the
Faro School and will help to bridge the gap between
different branches of Mathematical Analysis.
\end{abstract}

%\tableofcontents

\newsection{Introduction}

The story of the discovery of the modern theory of Integrable Systems is
certainly too long (and too well-known), and I can hardly add anything new;
so let me tell just a few words before addressing the bulk of the subject.
The study of completely integrable systems goes back to the classical papers
of Euler,  Lagrange, Jacobi, Liouville and others on
analytical mechanics. By the end of the XIX-th century all interesting examples
seemed to have been exhausted, and the interest has shifted to the qualitative
study of chaotic behaviour. The new age in the study of integrable systems has
begun with the famous paper \cite{GGKM} on the KdV equation, which was the first
example of an \emph{infinite dimensional} dynamical system with nontrivial but highly
regular behaviour and with a rich excitation spectrum. Amplifying the
earlier results of Gardner, Greene, Kruskal and Miura and
of Peter Lax \cite{PL} Faddeev and Zakharov \cite{ZF} have shown
in 1971, exactly 30 years ago, that the KdV equation is in fact a
\emph{completely integrable Hamiltonian system} in a technical sense.
Within a short time, many more examples have been discovered, notably,
the sine--Gordon equation, the first ever example of a \emph{relativistic} completely
integrable system \cite{ZTF}.
These discoveries were particularly exciting in view of the possible physical
applications: while it was of course clear that ``generic'' nonlinear
equations are non-integrable, it has been argued that Fundamental Physics
always deals with highly \emph{non-generic} equations which might be
integrable in some sense or other. These initial hopes have been fulfilled
only partially; one major obstacle is that the new technique does not apply
to (natural) non-linear equations in realistic space-time dimension. On the
other hand, the mathematics of complete integrability has proved to be
extraordinarily rich, bringing together Functional Analysis, Algebraic
Geometry, Lie Groups, Representation Theory, Symplectic Geometry and much
more. The analytic machinery used in the initial papers was based on the
Inverse Scattering Problem; the subsequent developments allowed to single
out the basic geometric ideas of the theory and to provide a unified basis
for different examples. One of the key ingredients of this geometric
approach are Infinite Dimensional Lie Groups; in some loose sense, one can
say that Integrable Systems always possess some rich \emph{hidden symmetry}.
One may recall that one of the original motivations of S.Lie has been the
use of continuous transformation groups for the study of differential equations. With
the modern methods at hand, we came closer to that goal; it has now become
clear, however, that almost all classical examples of integrable mechanical
systems from the XIX  century textbooks, as well as the infinite
dimensional systems associated with integrable PDE's are related not to
finite-dimensional Lie groups, but rather to their \emph{%
infinite-dimensional analogs}.

The general geometric construction that we shall discuss below allows to
unify the following characteristic features that are typical for all
examples known so far:

\begin{enumerate}
\renewcommand{\theenumi}{\roman{enumi}} \renewcommand{\labelenumi}{(%
\theenumi)}
\item  The equations of motion are compatibility conditions for a certain
auxiliary system of linear equations.

\item  They are Hamiltonian with respect to a natural Poisson bracket.

\item  Integrals of motion are spectral invariants of the auxiliary linear
operator. They are in involution with respect to the Poisson bracket
referred to above.

\item  The solution of the equations of motion reduces to some version of
the Riemann-Hilbert problem.
\end{enumerate}

Depending on the nature of the auxiliary linear problem, the associated
nonlinear equations may be divided into the following three groups:

a) Finite-dimensional systems,

b) Infinite-dimensional systems with one or two spatial variables,

c) Integrable systems on one-dimensional lattices.

\noindent In case (a) the auxiliary linear problem is the eigenvalue problem
for a finite-dimensional matrix (possibly depending on an additional
parameter). In case (b) the associated linear operator is differential. In
case (c) it is a difference operator.

As it happens, the key properties 1--4 referred to above are corollaries of
a single general theorem which may be adapted to numerous concrete
applications. The original idea of this theorem is due to B.Kostant and
M.Adler; its important amplification brings in the notion of \emph{classical
r-matrix} which provides a link between abstract Riemann-Hilbert problems
and the ideas of Quantum Group Theory. The statement and proof of this
theorem are particularly simple for systems of types (a) and (b). (Lattice
systems require a special treatment, since the associated Poisson brackets
belong to a different and more sophisticated class. We shall discuss this
case later on, but it is natural to begin with the simpler cases (a) and
(b)).

One more word of caution: while it is aesthetically very attractive to
deduce a large variety of examples together with their explicit solutions
from a simple general construction, there is one important disadvantage: for
a given dynamical system (even if it is known to be completely integrable!)
it is very difficult to tell \emph{a priori,} what is the underlying Lie
group, or Lie algebra. The practical way around this difficulty is to look
at various examples associated with different Lie algebras; with some skill,
one manages to recognize among these examples both classical and new
integrable systems which admit physical interpretation. The list of
interesting Lie algebras includes:

\begin{enumerate}
\item  Finite dimensional semisimple Lie algebras. Associated integrable
systems include open Toda lattice and other finite dimensional Hamiltonian
systems which may be integrated in \emph{elementary functions} (rational
functions of $\exp t$,    or $t$,    where $t$ is the time variable).

\item  Loop algebras, or affine Lie algebras. Associated integrable systems
are finite dimensional Hamiltonian systems which may be integrated in \emph{%
Abelian functions} of time. Integrable tops, as well as almost all classical
examples from the XIXth century Analytical Mechanics find their place here.

\item  Double loop algebras and their central extensions. This class of
algebras accounts for integrable PDE's admitting the zero curvature
representation (such as the Nonlinear Schroedinger equation, the Sine-Gordon
equation and many others).

\item  Algebras of pseudodifferential operators. The KdV equation and its
higher analogs come from this example, although it is more practical to
derive them from double loop algebras.\footnote{A further generalization
is possible: we can add one more spatial variable and consider the loop
algebra based on the algebra of pseudodifferential operators; this yields
the so called KP equation for functions of \emph{three} variables.}

\item  Algebra of vector fields on the line. This algebra, or rather its
central extension \emph{(the Virasoro algebra)} and the associated loop algebra
again are related to the KdV equation.
\end{enumerate}

As we see from this list, the choice of the Lie algebra determines not only the
possible kinematics (i.e., the structure of the phase space) of the
dynamical systems which admit a realization based on this algebra, but also
the functional class of the possible solutions. In all cases, it is very
important to examine \emph{central extensions} of the algebras in question
(if any), as well as their non-trivial automorphisms: they usually lead to
new examples. Non-trivial central extensions exist in cases 2, 3, 4, 5, each
one leading to a non-trivial theory. As for outer automorphisms of these
algebras, they serve, for instance, to define \emph{twisted loop algebras }%
with important range of applications (see Section 7.1) and are also
used in the study of nonlinear finite-difference equations.

It is virtually impossible to provide these lectures with a comprehensive
list of references. The point of view adopted here is certainly rather
subjective. I have included a few references to the old original papers as
well as to several reviews.

\newsection{A few Preliminaries: Poisson Brackets,
Coadjoint Orbits, etc.}

The first question that is prior to the study of the dynamics of a
mechanical system is that of its \emph{kinematics}, i.e. of the structure of
its phase space. Typically, the phase space of an individual dynamical
system should be a symplectic manifold. However, an important conclusion
which may be drawn from the practical study of numerous examples is that
integrable systems associated with auxiliary linear problems always arise in
\emph{families.} The appropriate geometrical setting for the study of such
families is provided by the theory of \emph{Poisson manifolds}.
%\footnote{A good reference book for the geometric notions discussed is \cite{arn}}.
Let me
recall that a Poisson bracket on a smooth manifold $M$ is the structure of a
Lie algebra on the space $C^{\infty }(M)$;  moreover, the Poisson bracket
satisfies the Leibniz rule, i.e., it is a derivation with resect to each
argument. In local coordinates, a Poisson bracket is written as
\[
\left\{ \varphi ,\psi \right\} (x)=\sum_{i,j}\pi _{ij}(x)\frac{\partial
\varphi }{\partial x_{i}}\frac{\partial \psi }{\partial x_{j}},
\]
where $\pi _{ij}$ is an antisymmetric tensor \emph{(Poisson tensor)}
satisfying a quadratic differential constraint which assures the Jacobi
identity. When the manifold $M$ is symplectic, i.e., admits a nondegenerate
closed 2-form $\omega =\sum \omega ^{ij}dx_{i}\wedge dx_{j}$,    the associated
Poisson tensor $\pi _{ij}=\left( \omega ^{ij}\right) ^{-1}$.   Reciprocally,
whenever the Poisson tensor is nondegenerate, its inverse is a symplectic
form. In general, a Poisson manifold is not symplectic; the fundamental
theorem which goes back to Lie asserts that it always admits a
stratification whose strata are already symplectic manifolds (\emph{the
symplectic leaves} of our Poisson manifold). The geometrical meaning of this
decomposition is very simple. Any $H\in C^{\infty }(M)$ defines on $M$ a
Hamiltonian vector field which acts on $\varphi \in C^{\infty }(M)$ via
%\begin{equation*}
\[X_{H}\varphi =\left\{ H,\varphi \right\} ; \]
%\end{equation*}
for a given point $x\in M$ the tangent vectors $X_{H}\left( x\right) $ span
a linear subspace in the tangent space $T_{x}M$; this is precisely the
tangent space to the symplectic leave passing through $x$.\footnote{%
The integrability condition which assures the local existence of the
submanifold with the given tangent distribution immediately follows from the
Jacobi identity; the subtle part of the proof consists in checking that the
possible jumps of the rank of the Poisson tensor $\pi $ do not lead to
singularities of the leaves; one can check that the Lie derivative
of $\pi $ along any Hamiltonian vector field is zero, and hence its rank is
constant along the leaves (but may jump in the transversal direction).} By
construction, Hamiltonian vector fields are tangent to symplectic leaves,
and hence the Hamiltonian flows are preserving each leaf separately. A
closely related property of general Poisson manifolds is the existence of
\emph{Casimir functions.} By definition, a function $H\in C^{\infty }(M)$ is
called a Casimir function if it lies in the center of the Poisson bracket;
equivalently, Casimir functions define trivial Hamiltonian equations on $M$.
Restrictions of Casimir functions to symplectic leaves in $M$ are constants;
reciprocally, the common level surfaces of Casimir functions define a
stratification of $M$; typically, this stratification is more coarse than
the stratification into symplectic leaves (i.e., symplectic leaves are not
completely separated by the values of the Casimirs), but in many
applications the knowledge of Casimir functions yields a sufficiently
accurate description.

A very typical example of a Poisson manifold is the \emph{dual space of a
Lie algebra}. I shall briefly recall the corresponding construction, since
it proved to be very important for the study of integrable systems. Let $%
\frak{g}$ be a Lie algebra, $\frak{g}^{*}$ its dual space, $P\left( \frak{g}%
^{*}\right) $ the space of polynomial functions on $\frak{g}^{*}$.   By the
Leibniz rule, a Poisson bracket on $P\left( \frak{g}^{*}\right) $ is
completely determined by its value on the subspace of linear functions $%
\frak{g}\subset P\left( \frak{g}^{*}\right) $;  for $X,Y\in \frak{g}$ let us
set simply
\begin{equation}
\left\{ X,Y\right\} \left( L\right) =\left\langle L,\left[ X,Y\right]
\right\rangle ,\;L\in \frak{g}^{*}.  \label{lp}
\end{equation}
The Jacobi identity for the bracket (\ref{lp}) follows from that for the Lie
bracket in $\frak{g}$;  since $P\left( \frak{g}^{*}\right) $ is dense in $%
C^{\infty }\left( \frak{g}^{*}\right) $,    it canonically extends to all
smooth functions. Explicitly, we have
\begin{equation}
\{\varphi _{1},\varphi _{2}\}\left( L\right) =\left\langle L,\left[ d\varphi
_{1}\left( L\right) ,d\varphi _{2}\left( L\right) \right] \right\rangle .
\label{lpg}
\end{equation}
(Note that $d\varphi _{i}\left( L\right) \in ({\frak{g}}^{*})
^{*}\simeq \frak{g}$,    and hence the Lie bracket is well defined). The
bracket (\ref{lpg}) is usually called the \emph{Lie--Poisson bracket}. Its
properties are closely related to the distinguished representation of the
associated Lie group, the \emph{coadjoint representation.} Let $G$ be a Lie
group with Lie algebra $\frak{g};\;\exp :\frak{g} \rightarrow G$ the
exponential map. The adjoint and coadjoint representations of $G$ acting in $%
\frak{g}$ and $\frak{g}^{*}$,    respectively, are defined by
\begin{eqnarray*}
Ad\,
\,g\cdot X &=&\left( \frac{d}{dt}\right) _{t=0}g\cdot \exp tX\cdot
g^{-1},\;X\in \frak{g}, \\
\left\langle Ad\,^{*}g\cdot L,X\right\rangle &=&\left\langle L,%
Ad\,
\,g^{-1}\cdot X\right\rangle ,\;X\in \frak{g},\;L\in \frak{g}^{*}.
\end{eqnarray*}
Set
\[
ad\,X\cdot Y=\left( \frac{d}{dt}\right) _{t=0}%
Ad\,
\exp tX\cdot Y,\;ad^{*}\,X\cdot L=\left( \frac{d}{dt}\right)
_{t=0}Ad^{*}\exp tX\cdot L.
\]
Clearly, one has $ad\,X\cdot Y=\left[ X,Y\right] , ad^{*}\,X=-\left(
ad\,X\right) ^{*}$.   The following fundamental theorem again goes back to
Lie; it was rediscovered by Kirillov and Kostant in 1960's:

\begin{theorem}
(i) Symplectic leaves of the Lie-Poisson bracket coincide with $G$-orbits in
$\frak{g}^{*}$ \emph{(coadjoint orbits).} (ii) Casimir functions of the
Lie-Poisson bracket are precisely the coadjoint invariant functions on $%
\frak{g}^{*}$.
\end{theorem}
It is very easy to verify a somewhat weaker property.
\begin{proposition}
\label{h}Let $\varphi \in C^{\infty }(\frak{g}^{*})$ be an arbitrary
function; the Hamiltonian equation of motion defined by $\varphi $ with
respect to the Lie-Poisson bracket may be written in the following form:
\begin{equation}
\frac{dL}{dt}=-ad^{*}\,d\varphi \left( L\right) \cdot L,\;L\in \frak{g}^{*};
\label{H}
\end{equation}
in other words, the velocity vector, associated with any Hamiltonian
equation on $\frak{g}^{*}$ is automatically tangent to the coadjoint orbit
passing through $L$.
\end{proposition}

In the context of integrable systems coadjoint orbits are of particular
importance: in many applications, the phase spaces of integrable systems
\emph{are} coadjoint orbits for some appropriate Lie group\footnote{%
It's probably worth making some precisions: when we deal with concrete
equations, coadjoint orbits are almost allways a good starting point, but it
may be quite useful to enrich our tools. In some cases, an orbit is too
\emph{big} for our purpose and it is possible to cancel out some degrees of
freedom by passing to the quotient space over some manifest symmetry group.
On the other hand, in some cases, an orbit is too \emph{small} and it's
more practical to use a bigger phase space which is mapped onto the orbit in
a way which is compatible with its Poisson structure. Finally, there are
classes of examples when the context of Lie algebras appears to be too
restrictive and we have to deal with \emph{nonlinear} Poisson brackets from
the very beginning. We shall comment on these examples later on
(see Section 9).}. Classification of coadjoint orbits for particular Lie
groups is a good exercise (which may be quite involved depending on the
nature of the Lie group); let us just quote a few examples which will be
useful in the sequel.

\begin{example}
Let $\frak{g}=\frak{gl}(n)$ be the full matrix algebra; its dual space $%
\frak{g}^{*}$ may be canonically identified with $\frak{g}$ by means of the
invariant inner product\footnote{%
Throughout these lectures \emph{inner product} means a nondegenerate
symmetric bilinear form ($\Bbb{C}$-bilinear in the case of complex
algebras); over the reals we do not impose any positivity condition.}
\begin{equation}
\left\langle X,Y\right\rangle =%
{\rm tr}\,XY.  \label{tr}
\end{equation}
Thus the adjoint and the coadjoint representations of the corresponding Lie
group $G=GL(n)$ are identical; we have $%
Ad\,
^{*}g\cdot L=gLg^{-1}$;  the coadjoint orbits consist of conjugate
(isospectral) matrices; their classification is given by the Jordan normal
form. Casimir functions are spectral invariants of matrices; their level
surfaces consist of a finite number of coadjoint orbits.
\end{example}

\begin{example}
Let $\frak{b}_{+}\subset \frak{g}$ be the subalgebra of upper triangular
matrices; the pairing (\ref{tr}) allows to identify its dual with the space $%
\frak{b}_{-}$ of lower triangular matrices. The coadjoint representation of
the corresponding Lie group $B_{+}$ of upper triangular invertible matrices
is given by
\[ Ad\,
^{*}b\cdot F=P_{-}\left( b\cdot F\cdot b^{-1}\right) ,\;b\in B_{+},\;F\in
\frak{b}_{-}, \]
where $P_{-}:\frak{g}\rightarrow \frak{b}_{-}$ is the projection operator
which replaces by zeros all matrix coefficients above the principal diagonal.
\end{example}

In this example the adjoint and the coadjoint representations are \emph{%
inequivalent.}

After this brief discussion of Poisson geometry and coadjoint orbits let us
return to the study of integrable systems. The use of linear Poisson
brackets and of coadjoint orbits seems a good guess to get a proper
kinematical description of our future examples; this suggestion is further
supported by the following simple observation which specializes proposition
\ref{h} above.

\begin{proposition}
\label{one}Assume that $\frak{g}=\frak{gl}(n)$ is identified with its dual
space and equipped with the Lie--Poisson bracket. For any $\varphi \in
C^{\infty }(\frak{g})$ the Hamiltonian equation of motion is written in the
form
\begin{equation}\label{ham}
\frac{dL}{dt}=-\left[ d\varphi \left( L\right) ,L\right] ;
\end{equation}
hence all Hamiltonian flows on $\frak{g}$ preserve spectral invariants of
matrices.
\end{proposition}

Equation \reff{ham} looks exciting: one is tempted to compare it with the
famous Lax equations. A closer look on the picture reveals, however, that
proposition \rref{one} leads to a deception. Indeed, spectral invariants of
matrices are \emph{Casimir functions} for the Lie-Poisson bracket; their
conservation is a trivial fact which has nothing to do with integrability of
equation \reff{ham}. There is very little chance that this equation with
arbitrary Hamiltonian $\varphi $ will be completely integrable; on the other
hand, the spectral invariants themselves which seem to be natural candidates
to provide integrable systems, generate \emph{trivial} flows, in view of the
following simple fact:

\begin{proposition}
\label{casimirs}For any Lie algebra $\frak{g}$ a function $\varphi \in
C^{\infty }(\frak{g}^{*})$ is a Casimir function for the Lie--Poisson
bracket on $\frak{g}^{*}$ if and only if
\[ ad^{*}\,d\varphi \left( L\right) \cdot L=0 \]
for any $L\in \frak{g}^{*}$.   (Note that $d\varphi \left( L\right) \in (\frak{%
g}^{*})^{*}\simeq \frak{g}$;  when $\frak{g}$ and $\frak{g}^{*}$ are
identified, this relation is reduced to $\left[ d\varphi \left( L\right)
,L\right] =0.)$
\end{proposition}

Despite this initial setback, the original idea to use Lie--Poisson brackets
and coadjoint orbits can be saved. However, instead of the initial
Lie-Poisson bracket which provides the set of spectral invariants but does
not yield any nontrivial dynamics associated with them, we must find a \emph{%
different} one. It's at this point that the classical r-matrix is brought
into play.

%%%%%%%%%%%%%%%%%%%%%%%%%%%%%%%%%%%%%%%%%%%%%%%%%%%%%%%%%%%%%%%%%%%%%%%%%%%

\newsection{Classical r-matrices and Lax Equations\label{rmat}}

Let $\frak{g}$ be a Lie algebra We shall say that $r\in {\mathrm {End}}\,
\frak{g}$ is a \emph{classical r-matrix} if the bracket

\begin{equation}
\left[ X,Y\right] _{r}=\frac{1}{2}\left( \left[ rX,Y\right] +\left[
X,rY\right] \right)  \label{rbr}
\end{equation}
is a Lie bracket, i.e. if it satisfies the Jacobi identity. The skew
symmetry of (\ref{rbr}) is obvious for any $r$.   We denote the Lie algebra
with the bracket (\ref{rbr}) by $\frak{g}_{r}$ and say that $\left( \frak{g},%
\frak{g}_{r}\right) $ is a \emph{double Lie algebra.}

If $\frak{g}$ is a double Lie algebra, there are \emph{two} different
Poisson brackets in the space $\frak{g}^{*}$,    namely, the Lie-Poisson
brackets of $\frak{g}$ and $\frak{g}_{r}$.   The latter bracket will be
referred to as the\emph{\ r-bracket}, for short.

A class of double Lie algebras which is of great importance for applications
is constructed as follows. Assume that there is a vector space decomposition
of $\frak{g}$ into a direct sum of two Lie subalgebras, $\frak{g}=\frak{g}%
_{+}\dotplus \frak{g}_{-}$.   Let $P_{\pm }$ be projection operators onto $%
\frak{g}_{\pm }$ parallel to the complementary subalgebra;  set
\begin{equation}\label{r}
r=P_{+}-P_{-}.
\end{equation}
In this case, bracket (\ref{rbr}) is given by

\begin{equation}
\lbrack X,Y]_{r}=[X_{+},Y_{+}]-[X_{-,}Y_{-}],  \label{stand}
\end{equation}
where $X_{\pm }=P_{\pm }X,Y_{\pm }=P_{\pm }Y$.   In other words, the bracket (%
\ref{rbr}) is the difference of Lie brackets in $\frak{g}_{+}$ and $\frak{g}%
_{-}$.   The Jacobi identity for $\frak{g}_{r}$ is obvious from (\ref{stand}).

As discussed in Section 6, in typical applications the Lie algebra $\frak{g}$
is a \emph{loop algebra}, i.e., an algebra of matrix-valued functions on the circle,
and  $\frak{g}_{\pm}$ its subalgebra consisting of functions which are analytic
inside (resp., outside) the circle. In that case, the classical r-matrix (\ref{rbr})
is precisely the Hilbert transform. Of course, general classical r-matrices need not have
this simple form (although (\ref{rbr}) is by far the most important example of all).
We shall discuss the general theory of r-matrices a little later; let us first state
the key theorem which motivates the definition.

%%%%%%%%%%%%%%%%%%%%%%%%%%%%%%%%%%%%%%%%%%%%%%%%%%%%%%%%%%%%%%%%%%%%%%%%%%%

\newsubsection{Involutivity Theorem}

Let $I(\frak{g}^{*})$ be the ring of Casimir functions on $\frak{g}^{*}$
(with respect to the original Lie-Poisson bracket); equivalently, $I(\frak{g}%
^{*})\subset C^{\infty }(\frak{g}^{*})$ is the set of coadjoint invariants.

\begin{theorem}
\label{AKS}(i) Functions in $I(\frak{g}^{*})$ are in involution with respect
to the r-bracket on $\frak{g}^{*}$.   (ii) The equations of motion induced by $%
h\in I(\frak{g}^{*})$ with respect to the r-bracket have the form

\begin{equation}
\frac{dL}{dt}=-ad_{\frak{g}}^{*}M\cdot L,\ M=r(dh(L))\   \label{lax}
\end{equation}

\noindent If $\frak{g}$ admits a nondegenerate invariant bilinear form, so
that $ad_{\frak{g}}^{*}\simeq  ad_{\frak{g}}$,    equations (2.4) have the
\emph{Lax form}
\[ \frac{dL}{dt}=\left[ L,M\right] . \]
\end{theorem}

\begin{proof} (i) Let $h_{1},h_{2}\in  I(\frak{g}^{*})$;  set $dh_{i}\left(
L\right) =X_{i}$ for brevity. By definition,
\[\arr{2.0}{l} {
\left\{ h_{1},h_{2}\right\} _{r}\left( L\right) \ds =\left\langle L,\left[
X_{1},X_{2}\right] _{r}\right\rangle \\
\ds =\frac{1}{2}\left\langle L,\left[ rX_{1},X_{2}\right] +\left[
X_{1},rX_{2}\right] \right\rangle \\
\ds =\frac{1}{2}\left\langle ad_{\frak{g}}^{*}X_{2}\cdot L,rX_{1}\right\rangle
-\frac{1}{2}\left\langle ad_{\frak{g}}^{*}X_{1}\cdot L,rX_{2}\right\rangle
=0,}
\]
since, by proposition \ref{casimirs}, $ad_{\frak{g}}^{*}X_{2}\cdot L=ad_{%
\frak{g}}^{*}X_{1}\cdot L=0$.   (ii) We have
\[ \frac{dL}{dt}=-ad_{\frak{g}_{r}}^{*}dh\left( L\right) \cdot L; \]
(\ref{rbr}) implies that
\begin{equation}
ad_{\frak{g}_{r}}^{*}X\cdot L=\frac{1}{2}\left( ad_{\frak{g}}^{*}rX\cdot
L+r^{*}\left( ad_{\frak{g}}^{*}X\cdot L\right) \right) ;  \label{co}
\end{equation}
since $h\in  I(\frak{g}^{*})$,    the second term in (\ref{co}) vanishes.
\end{proof}

\begin{remark}
The matrix $M$\ in \reff{lax} is not uniquely defined: one can always add
to it something which commutes with $L$.   Here is a useful option: set
\begin{equation}
M_{\pm }=\pm \frac{1}{2}\left( r\pm Id\right) \left( dh\left( L\right)
\right) ;  \label{m}
\end{equation}
equation \reff{lax} holds with \emph{any}\ of these two operators. Below,
we shall see that this form of M-operator appears naturally from the global
formula for the solutions.
\end{remark}

Theorem \rref{AKS} has a transparent geometrical meaning: it shows that the
trajectories of the dynamical systems with Hamiltonians $h\in  I(\frak{g}%
^{*})$ lie in the intersection of \emph{two families of orbits} in $\frak{g}%
^{*}$,    the coadjoint orbits of $\frak{g}$ and $\frak{g}_{r}$.   Indeed, the
coadjoint orbits of $\frak{g}_{r}$ are preserved by all Hamiltonian flows in
$\frak{g}_{r}$. On the other hand, because of (2.4), the flow is always
tangent to the $\frak{g}$-orbits in $\frak{g}^{*}$.

In many applications the intersections of orbits are precisely the
``Liouville tori'' for our dynamical systems.

%%%%%%%%%%%%%%%%%%%%%%%%%%%%%%%%%%%%%%%%%%%%%%%%%%%%%%%%%%%%%%%%%%%%%%%%%

\newsubsection{Factorization Theorem}

The scheme outlined so far incorporates only two of the three main features
of the inverse scattering method: the Poisson brackets and the Lax form of
the equations of motion. As it happens, the most important feature of this
method, the reduction of the equations of motion to the Riemann problem, is
already implicit in our scheme. An abstract version of the Riemann problem
is provided by the \emph{factorization problem} in Lie groups.

We shall state a factorization theorem, which is the global version of
Theorem \ref{AKS}, for the simplest r-matrices of the form (\ref{r}). Let $G$
be a connected Lie group with Lie algebra $\frak{g}$,    and let $G_{\pm }$ be
its subgroups corresponding to $\frak{g}_{\pm }$.

\begin{theorem}
\label{fact}Let $h\in I\left( \frak{g}^{*}\right) ,X=dh\left( L\right) $.
Let $g_{\pm }(t)$ be the smooth curves in $G_{\pm }$ which solve the
factorization problem
\begin{equation}
\exp tX=g_{+}(t)\cdot g_{-}(t)^{-1},\;g_{+}(0)=e.  \label{exp}
\end{equation}
Then the integral curve $L(t)$ of equation (2.4) with $L(0)=L$,    is given by
any of the two formulae,
\begin{equation}
L(t)=%
Ad\,
_{G}^{*}\,g_{+}(t)^{-1}\cdot L=%
Ad\,
_{G}^{*}\,g_{-}(t)^{-1}\cdot L.  \label{two}
\end{equation}
\end{theorem}

\begin{proof} Differentiating \reff{two} with respect to $t$ we get
\[ \frac{dL}{dt}=-ad^{*}\left(g_{+}^{-1}\dot{g}_{+}\right) \cdot
L=-ad^{*}\left( g_{-}^{-1}\dot{g}_{-}\right) \cdot L. \]
We shall check that $g_{\pm }^{-1}\dot{g}_{\pm }=M_{\pm }$,    where $M_{\pm }$
are the M-operators from (\ref{m}). Due to our special choice of $r$ we have
\[
M_{\pm }\left( t\right) =\pm P_{\pm }X\left( t\right) ,
\]
where $X\left( t\right) =dh\left( L\left( t\right) \right) $.   The $Ad^{*}G$%
-invariance of $h$ implies that
\[
X(t)=%
Ad\,
_{G}g_{\pm }(t)^{-1}\cdot X.
\]

\begin{exercise}     %{exercise}
Check this formula in matrix case (``gradients of invariant functions are
covariant'').
\end{exercise}       %{exercise}

Writing \reff{exp} in the form $g_{+}(t)\exp tX=g_{-}(t)$ and
differentiating with respect to $t$,    we get
\[ g_{+}^{-1}\dot{g}_{+}+%
Ad\,
_{G}g_{-}(t)^{-1}\cdot X=\dot{g}_{-}g_{-}^{-1} \]
Since $g_{\pm }^{-1}\dot{g}_{\pm }\in \frak{g}_{\pm }$,    this implies $g_{\pm
}^{-1}\dot{g}_{\pm }=\pm P_{\pm }X\left( t\right) $,    as desired.
\end{proof}

Note that the two possible choices of sign in \reff{two} are equivalent
precisely because $X(t)$ belongs to the centralizer of $L(t)$,    i.e., $%
ad^{*}X(t)\cdot L(t)=0$ (in fact, this is the characteristic property of
Casimir functions).

By the implicit function theorem, the factorization \reff{exp} exists for $%
t $ sufficiently small; note that in our proof we need \emph{not} assume
that this factorization exists globally for all $t$.   Geometrically, this
means that the solution of the Lax equation exists as long as the curve $%
\exp tX$ remains in the ``big cell'' $G_{+}\times G_{-}\subset G$;  in other
words, the flow associated with the Lax equation is not necessarily
complete. One can show in typical examples that the curve intersects
``complementary'' cells of positive codimension transversally and returns
back to the big cell; for the exceptional values of $t$ the solution
``escapes to infinity'', i.e., displays a pole in $t$.

%%%%%%%%%%%%%%%%%%%%%%%%%%%%%%%%%%%%%%%%%%%%%%%%%%%%%%%%%%%%%%%%%%%%%%%%%%%

\newsubsection{Factorization Theorem and Hamiltonian Reduction\label{red}}

A more geometric proof of Theorem \rref{fact} is based on \emph{Hamiltonian
reduction}. Recall that the Hamiltonian reduction applies to Hamiltonian
dynamical systems with high degree of symmetry; it allows to exclude certain
redundant degrees of freedom. Classically, the use of reduction is to
simplify multidimensional systems getting quotient, or reduced, systems of
lower dimension. However, as pointed out in \cite{KKS}, one can reverse this
reasoning and use Hamiltonian reduction in the opposite direction, starting
with a simple multidimensional system with high symmetry (``free dynamics'')
and getting a complicated reduced system as an output. In order to apply
this idea, one has to answer the following questions:

\begin{enumerate}
\item  Find a ``big'' phase space and suggest the ``free dynamics'' which
will yield Lax equations as the quotient system.

\item  Make sure there is an expected high degree of symmetry for the free
system.

\item  Perform the reduction.
\end{enumerate}

Although the proof based on this approach is much longer than the elementary
computation presented above, it is more transparent and explains the origin
of the result. A simple candidate for the big phase space is the \emph{%
cotangent bundle} $T^{*}G$ equipped with the canonical symplectic structure.
Let us first of all describe the ``free dynamics'' on $T^{*}G$.   The group $G$
acts on itself by left and right translations; these actions naturally lift
to $T^{*}G$; both actions are Hamiltonian with respect to the canonical
symplectic structure. Let us identify $T^{*}G$ with $G\times \frak{g}^{*}$
by means of \emph{left translations}.

\begin{exercise}  %{exercise}
In left trivialization the action of $G$ on $T^{*}G\simeq G\times \frak{g}%
^{*}$ by left (right) translations is given by
\begin{eqnarray}
\lambda \left( g\right) &:&\left( x,L\right) \longmapsto \left( gx,L\right) ,
\label{act} \\
\rho \left( g\right) &:&\left( x,L\right) \longmapsto \left(
xg^{-1},Ad^{*}g\cdot L\right) .  \nonumber
\end{eqnarray}
\end{exercise}    %{exercise}

(In \emph{left} trivialization the action of $G$ by \emph{right }%
translations induces a nontrivial action in the fiber $\frak{g}^{*}$;  it is
easy to check that it is precisely the coadjoint action.) Left-invariant
functions on $T^{*}G$ are identified with functions on $\frak{g}^{*}$.   Since
the canonical Poisson bracket of left-invariant functions is also
left-invariant, this induces a Poisson structure on $\frak{g}^{*}$;  it is
easy to check that it coincides with the \emph{Lie-Poisson bracket}. Casimir
functions on $\frak{g}^{*}$ canonically lift to smooth functions on $T^{*}G$
which are $G$-\emph{biinvariant}. For $h\in I(\frak{g}^{*})$ let us denote
by $\hat{h}\in C^{\infty }\left( T^{*}G\right) $ the corresponding
biinvariant Hamiltonian on $T^{*}G$.

\begin{lemma}
The Hamiltonian flow on $T^{*}G$ defined by $\hat{h}$ is given (in left
trivialization) by
\begin{equation}
F_{t}:\left( x,L\right) \longmapsto \left( x\cdot \exp tdh(L),L\right)
,\;x\in G,\;L\in \frak{g}^{*}.  \label{F}
\end{equation}
In other words, integral curves of $\hat{h}$ project to left translates of
one-parameter subgroups in $G$;  the choice of $h$ determines the (constant)
velocity vector $dh(L)$ which depends only on the initial data.
\end{lemma}

Since the ``free Hamiltonian'' $\hat{h}$ is $G$-biinvariant, the flow $%
F_{t}$ admits reduction with respect to \emph{any} subgroup $U\subset
G\times G$.   There is, at this stage, a very large freedom in the choice of
such a subgroup which all lead to different but meaningful quotient systems.
The particular choice which is imposed by the choice the r-matrix (\ref{r})
is $U=G_{+}\times G_{-}$.   By (\ref{act}), with our choice of trivialization,
the action of $G_{+}\times G_{-}$ on $T^{*}G\simeq G\times \frak{g}^{*}$ is
given by
\begin{equation}
\left( g_{+},g_{-}\right) :\left( x,L\right) \longmapsto \left(
g_{+}xg_{-}^{-1},%
Ad\,
_{G}^{*}g_{-}\cdot L\right) .  \label{a}
\end{equation}

We now turn to the reduction procedure. In textbooks the reduction is
usually described in a rather complicated way (which involves the moment map
and a good deal of symplectic geometry) (see, e.g., \cite{arn}). Here is
an elementary substitute. Let us ask what is bad about the naive suggestion:
consider a group action $G\times M\rightarrow M$ on a symplectic manifold
and take the projection $\pi :M\rightarrow M/G$ onto the quotient space? The
answer is: the quotient space is no longer symplectic, since  symplectic
forms transform by pullback, and there is no natural symplectic form on $M/G$
(it in not even in general even dimensional!) But on the other hand, the
quotient space \emph{does} carry a Poisson bracket (which transforms by
push-forward!). The difficult part of reduction consists in the description
of the particular \emph{symplectic leaves} of this quotient Poisson bracket;
it's at this stage that one needs the moment map and all other machinery. If
we do not want a too detailed description, or if we manage to guess the
symplectic leaves in some other way, everything becomes simple! In the
present case, we can display a map $\pi $ which is constant on the orbits of
$G_{+}\times G_{-}$ in $M$ and hence its image yields a \emph{model} of the
quotient space; the Poisson structure on this quotient is easy to recognize.

For $x\in G$ we denote by $x_{\pm }$ the solution of the factorization
problem
\begin{equation}
x=x_{+}\cdot x_{-}^{-1},\;x_{+}\in G_{+},\;x_{-}\in G_{-}.  \label{f}
\end{equation}

\begin{lemma}
(i) The map $\pi :T^{*}G\longrightarrow \frak{g}^{*}:\left( x,L\right)
\longmapsto %
Ad\,
_{G}^{*}x_{-}^{-1}\cdot L$ is constant on the orbits of $G_{+}\times G_{-}$
in $T^{*}G$.   (ii) If $G$ is globally diffeomorphic to $G_{+}\times G_{-}$,
i.e., the factorization problem (\ref{f}) is always solvable, $\pi $ is a
global cross-section of this action. (iii) The induced Poisson structure on $%
T^{*}G/G_{+}\times G_{-}\simeq \frak{g}^{*}$ coincides with the Lie-Poisson
bracket for $\frak{g}_{r}\simeq \frak{g}_{+}\oplus \frak{g}_{-}$.
\end{lemma}

The check of (i) and (ii) is immediate; the proof of (iii) requires a little
knowledge of symplectic geometry (or else a three-line computation); we
shall not present it here (see \cite{RS}).

\begin{lemma}
\label{q}The flow (\ref{F}) \emph{factorizes over} ${\frak{g}}\,_{r}^{*}$;
in other words, there exists a natural flow $\bar{F}_{t}:\frak{g}\,_{r}^{*}{%
\longrightarrow \frak{g}}\,_{r}^{*}$ (called the \emph{quotient} flow) which
makes the following diagram
\begin{center}
\setlength{\unitlength}{3947sp}%
\begin{picture}(4818,1904)(1501,-3635)
\put(4600,-1900){\makebox(0,0)[lb]{$T^{*}G$}}
\put(4600,-3000){\makebox(0,0)[lb]{${\frak{g}}\,_{r}^{*}$}}
\put(2200,-1900){\makebox(0,0)[lb]{$T^{*}G$}}
\put(2200,-3000){\makebox(0,0)[lb]{${\frak{g}}\,_{r}^{*}$}}
\put(3400,-2900){\makebox(0,0)[lb]{${\bar{F_t}}$}}
\put(2400,-2400){\makebox(0,0)[lb]{${\pi}$}}
\put(4800,-2400){\makebox(0,0)[lb]{${\pi}$}}
\put(3400,-2050){\makebox(0,0)[lb]{${F_t}$}}
\put(2600,-1850){\vector(1,0){1900}}
\put(2600,-3000){\vector(1,0){1900}}
\put(4700,-2100){\vector(0,-1){700}}
\put(2300,-2100){\vector(0,-1){700}}
\end{picture}
\end{center}
commutative; the quotient flow $\bar{F}_{t}:\frak{g}\,_{r}^{*}{%
\longrightarrow \frak{g}}\,_{r}^{*}$ is given by
\begin{equation}
\bar{F}_{t}:L\longmapsto
Ad\,
\,_{G}^{*}\,g_{-}\left( t\right) ^{-1}\cdot L,  \label{minus}
\end{equation}
where, as in (\ref{exp}), $g_{+}\left( t\right) g_{-}\left( t\right)
^{-1}=\exp t\,dh(L)$.
\end{lemma}

The flow $\bar{F}_{t}$ is precisely the result of the reduction procedure.
Note that (\ref{minus}) involves only $g_{-}$;  this is due to our choice of
trivialization of $T^{*}G$;  trivialization by \emph{right} translations yields the
equivalent formula for the quotient flow $\bar{F}_{t}:L\longmapsto
Ad_{G}^{*}g_{+}\left( t\right) ^{-1}\cdot L$.

In general, of course, $G$ need not be diffeomorphic to $G_{+}\times G_{-}$;
still, $\frak{g}^{*}$ provides a model for a ``big cell'' in the quotient
space $T^{*}G/G_{+}\times G_{-}$;  one can show that under very mild
restrictions the action (\ref{a}) is proper and hence the quotient space is
a well-defined manifold; the quotient flow induced on this manifold is also
well defined and may be regarded as the natural completion of the incomplete
flow associated with Lax equations.

The choice of the subgroup $G_{+}\times G_{-}\subset G\times G$ as the
symmetry group for the ``free system'' may seem arbitrary; indeed, there are
many other possible choices leading to meaningful examples. (Among the
dynamical systems which may be obtained in this way, there are, e.g., the \emph{%
Calogero-Moser systems}, cf. \cite{KKS}.) The key property which
characterizes our special situation is the simple description of the
quotient space: as we see from Lemma \ref{q}, in this case the quotient
space is simply the dual of a Lie algebra with its Lie-Poisson bracket, and
its symplectic leaves (which are the\emph{\ symplectic quotients} of $%
T^{*}G) $ are the coadjoint orbits of $G_{r}$;  for other choices of the
symmetry group the description of the quotient space will be more
complicated and it may no longer be a homogeneous space.

%%%%%%%%%%%%%%%%%%%%%%%%%%%%%%%%%%%%%%%%%%%%%%%%%%%%%%%%%%%%%%%%%%%%%%%%%%%

\newsection{Classical Yang-Baxter Identity\label{YB}}

We have already mentioned that the most interesting r-matrices are
associated with decompositions of the Lie algebra into complementary Lie
subalgebras. However, it is worth examining the general conditions on $r$
which follow from the Jacobi identity for the r-bracket. These conditions
are known as the \emph{classical Yang-Baxter equations}; they were first
derived as a semi-classical approximation to the \emph{quantum Yang-Baxter
equations} which arise in the study of quantum completely integrable systems.

The restrictions on $r$ which follow from the Jacobi identity are quite easy
to establish.
For $r\in
 End\,%
\frak{g}$ set
\begin{equation}
B_{r}\left( X,Y\right) =[rX,rY]-r([rX,Y]+[X,rY]).  \label{Obstr}
\end{equation}

\begin{proposition}
The r-bracket \reff{rbr} satisfies the Jacobi identity if and only if, for
any $X,Y,Z\in \frak{g}$,
\begin{equation}
\lbrack B_{r}(X,Y),Z]+[B_{r}(Y,Z),X]+[B_{r}(Z,X),Y]=0.  \label{B}
\end{equation}
\end{proposition}

The proof is straightforward: just substitute \reff{rbr} into the Jacobi identity and
regroup the terms. The necessary and sufficient condition \reff{B} is usually replaced by
sufficient conditions which are \emph{bilinear} rather than trilinear. The
simplest sufficient condition is the so-called \emph{classical Yang-Baxter
equation} (CYBE)
\begin{equation}
B_{r}(X,Y)=0.  \label{cybe}
\end{equation}
Another important sufficient condition is the \emph{modified classical
Yang-Baxter equation} (mCYBE)
\begin{equation}
B_{r}\left( X,Y\right) =-c\,[X,Y],\;c=const.  \label{mcybe}
\end{equation}
By rescaling, we may always assume that $c=1$.   Note that the r-matrices \reff
{r} satisfy mCYBE with $c=1$.

The reason for the study of classical r-matrices satisfying the modified
classical Yang-Baxter identity is that although they do not in general have
the simple form \reff{r}, one can still associate with them a factorization
problem. By contrast, the ordinary classical Yang-Baxter identity \reff{cybe}
represents a degenerate case and does not lead to a factorization problem.

Let us briefly describe the corresponding construction. Given an r-matrix
which satisfies mCYBE, set
\begin{equation}
r_{\pm }=\frac{1}{2}\left( r\pm Id\right) .  \label{pm}
\end{equation}

\begin{proposition}
For each $X,Y\in \frak{g}$
\[ \lbrack r_{\pm }X,R_{\pm }Y] = r_{\pm }([X,Y]_{r}), \]
i.e., $r_{\pm }:\frak{g}_{R}\rightarrow \frak{g}$ are Lie algebra
homomorphisms.
\end{proposition}

Set $\frak{g}_{\pm }=%
\mathrm{Im}\,%
r_{\pm }$.   Clearly, $\frak{g}_{\pm }$ is a Lie subalgebra of $\frak{g}$.   If $%
r$ has the form (\ref{r}), then $r_{\pm }=\pm P_{\pm }$ and the subalgebras $%
\frak{g}_{\pm }=P_{\pm }(\frak{g})$ are complementary. In general case this
is no longer true. However, this difficulty may be resolved by passing to
the \emph{double} of $\frak{g}$.   By definition,
\[ \frak{d}=\frak{g}\oplus \frak{g} \]
is the direct sum of two copies of $\frak{g}$.

\begin{proposition}
The mapping
$i_{r}:\frak{g}_{r}\rightarrow \frak{g}\oplus \frak{g}:X\longmapsto
(r_{+}X,r_{-}X)$ is a Lie algebra embedding, and each $Y\in \frak{g}$ has a
unique decomposition, $Y=Y_{+}-Y_{-}$,    where $(Y_+,Y_{-})\in
\mathrm{Im}\, i_{r}$.
\end{proposition}

\noindent The last assertion follows immediately from the obvious identity $%
r_{+}-r_{-}=Id$.

Now, let $G,G_{r}$ be (local) Lie groups which correspond to $\frak{g}$,    $%
\frak{g}_{r}$.   The homomorphisms $r_{\pm }$ give rise to the Lie group
homomorphisms which we denote by the same letters. Put $G_{\pm }=r_{\pm}(G_{R})$.
The composition of maps
\[ i_{r}:G_{r}\longrightarrow G\times G:x\longmapsto (r_{+}x,r_{-}x) \]
is a Lie group embedding. Consider the map
\[ m:G\times G\longrightarrow G:(u,v)\longmapsto uv^{-1}. \]
Then $f=m\circ i_{r}  : G_{r}\longrightarrow G$ is a local homeomorphism, and
therefore an arbitrary element $y\in G$ which is sufficiently close to unity
admits a unique representation
$y=y_{+}y_{-}^{-1}
$
with $\left( y_{+},y_{-}\right) \in \mathrm{Im}\,i_{r}$.

The proof of Theorem \ref{fact} extends to the present setting with only
minor changes.

It is probably worth giving examples of r-matrices satisfying \reff{mcybe}
which are \emph{not} of the form \reff{r}. Let $\frak{g}=\frak{gl}(n)$ be
the matrix algebra; let us consider its decomposition $\frak{g}=\frak{n}_{+}%
\dot{+}\frak{h}\dot{+}\frak{n}_{-}\ $ into direct sum of upper triangular,
diagonal, and lower triangular matrices. Let $P_{\pm },P_{0}$ be the
corresponding projection operators. Let $r_{0}$ be the \emph{partially
defined} linear operator on $\frak{g}$ with domain $\mathrm{Dom}(r_{0})=%
\frak{n}_{+}\dot{+}\frak{n}_{-}$ given by
\[
r_{0} X = \left\{ \arr{1.0}{l}{
{\phantom -} X, \, \hbox{if } X \in \frak{n}_{+}, \\
-X ,  \, \hbox{if } X \in \frak{n}_{-}. }
\right.
\]
We want to extend $r_{0}$ to the entire linear space $\frak{g} $ in such a
way that it satisfies the modified classical Yang-Baxter identity. Let us
first drop the latter condition and consider \emph{all} linear operators $r%
\supset r_{0},\;\mathrm{Dom}(r)=\frak{g}$.   Let us set again $r_{\pm }=%
\frac{1}{2}\left( r\pm Id\right) ,i_{r}=r_{+}\oplus r_{-}$ and consider the
subspace $\frak{g}_{r}=\mathrm{Im}\,%
i_{r}$.   It is easy to see that this subspace is transversal to the diagonal
subalgebra $\frak{g}_{d}=\{\left( X,X\right) \in \frak{g}\oplus \frak{g}\}$;
conversely, all extensions $r \supset  r_{0}$ are in bijective
correspondence with linear subspaces $\frak{g}_{r}\subset \frak{g}\oplus
\frak{g}$ which contain $\frak{n}_{+}\oplus \frak{n}_{-}$ and are
transversal to the diagonal. All such subspaces may be parametrized in the
following way:
\[ \frak{g}_{r}=\left\{ \left( X_{+}+X_{0},X_{-}+\theta X_{0}\right) ;X\in
\frak{g}\right\} , \]
where $X_{\pm }=\pm P_{\pm }X,X_{0}=P_{0}X$ and $\theta \in \mathrm{End}\,\frak{h}$ is
a linear operator; equivalently, the extensions $r\supset r_{0}$ are
described by the von Neumann formulae,
\[\ \left\{ \arr{1.1}{l}{
{\phantom z} _{\phantom z} X = X_{+}-X_{-}+\left( Id-\theta \right) X_{0}, \\
r_{\theta }X  =X_{+}+X_{-}+\left( Id+\theta \right) X_{0}.}
\right. \]
The transversality condition is equivalent to the non-degeneracy of $%
Id-\theta $.   Since $\frak{h}\subset \frak{g}$ is an abelian subalgebra which
normalizes both $\frak{n}_{+}$ and $\frak{n}_{+}$,    it is easy to see that $%
\frak{g}_{r}\subset \frak{g}\oplus \frak{g}$ is a Lie subalgebra (and not
merely a linear subspace) for any $\theta $,    and hence all $r_{\theta }%
\supset r_{0}$ satisfy mCYBE. In more general examples, when the relevant
subalgebra is not abelian, there are additional algebraic constraints on $%
\theta $ which make the construction partially rigid.

\emph{The moral of the story:} One can construct new classical r-matrices
from the standard ones of the form \reff{r} by assuming that the simple
formula \reff{r} applies not globally, but only on some subspace of
positive codimension and then using the extension theory of linear
operators. The same trick works for loop algebras; in that case, the
relevant codimensions are finite. Let us set, for example, $\frak{G}=L\frak{g%
},\;\frak{g}=\frak{gl}\left( n\right) $; by definition, $\frak{G}$ is the
algebra of Laurent polynomials with matrix coefficients. Set $\frak{N}%
_{+}=\{X(z)=X_{0}+X_{1}z+X_{2}z^{2}+...;X_{0}\in \frak{n}_{+}\},\;\frak{N}%
_{-}=\{X(\lambda )=X_{0}+X_{1}z^{-1}+X_{2}z^{-2}+...;X_{0}\in \frak{n}_{-}\}$%
. Then $\frak{G}=\frak{N}_{+}\dot{+}\frak{h}\dot{+}\frak{N}_{-}$,    where $%
\frak{h}$ is the subalgebra of constant diagonal matrices. Hence we may
apply the previous construction without any modification. A classification
theorem, due to Belavin and Drinfeld, assures that all r-matrices on
semisimple Lie algebras and their loop algebras satisfying mCYBE and certain
additional conditions arise in a similar way; of course, the most difficult
part of the construction is the case when $\frak{h}$ is replaced by a
non-abelian subalgebra. (Cf. \cite{BD}, \cite{What}.)

%%%%%%%%%%%%%%%%%%%%%%%%%%%%%%%%%%%%%%%%%%%%%%%%%%%%%%%%%%%%%%%%%%%%%%%%%%%

\newsection{A Finite-dimensional Example}

Let again $\frak{g}=\frak{gl}(n)$ be the full matrix algebra. There are
several natural decompositions of $\frak{g}$ into direct sum of
complementary subalgebras, e.g.,
\begin{equation}
\frak{g}=\frak{b}_{+}+\frak{n}_{-},  \label{g}
\end{equation}
where $\frak{n}_{-}$ is the subalgebra of lower triangular nilpotent
matrices and $\frak{b}_{+}=\frak{h}+\frak{n}_{+}$ the complementary
subalgebra of upper triangular matrices. Another decomposition is
\begin{equation}
\frak{g}=\frak{b}_{+}+\frak{k},  \label{i}
\end{equation}
where $\frak{k}=so(n)$ is the Lie algebra of skew symmetric matrices. We can
associate two standard classical r-matrices with these decompositions:
\begin{eqnarray*}
r_{\mathrm{Gauss}} &=&P_{\frak{b}_{+}}-P_{\frak{n}_{-}}, \\
r_{\mathrm{Iwasawa}} &=&\tilde{P}_{\frak{b}_{+}}-P_{\frak{k}},
\end{eqnarray*}
where $P_{\frak{b}_{+}},P_{\frak{n}_{-}},\tilde{P}_{\frak{b}_{+}},P_{\frak{k}%
}$ are the respective projection operators (mind that of course $P_{\frak{b}%
_{+}}\neq \tilde{P}_{\frak{b}_{+}})$.   The Lie groups $G_{r_{\mathrm{Gauss}}}$
and $G_{r_{\mathrm{Iwasawa}}}$ are isomorphic to $B_{+}\times N_{-}$ and $%
B_{+}\times K$,    respectively. The associated factorization problems in the
general linear group $G=GL(n)$ are the Gauss decomposition of matrices,
\begin{equation}
g=bn^{-1},b\in B_{+},n\in N_{-},  \label{G}
\end{equation}
in the former case, and the Iwasawa decomposition
\begin{equation}
g=bk^{-1},b\in B_{+},k\in K,  \label{I}
\end{equation}
in the latter one. (Here $N_{-}\subset GL(n)$ denotes the subgroup of lower
triangular unipotent matrices, $B_{+}\subset GL(n)$ the subgroup of upper
triangular matrices, and $K=SO(n)\subset GL(n)$ the subgroup of orthogonal
matrices.) Note that in the Iwasawa case the product map $B_{+}\times
K\rightarrow G:\left( b,k\right) \mapsto bk^{-1}$ is a bijection onto $G$
and hence the factorization problem is always solvable. For the Gauss
decomposition the image of $B_{+}\times N_{-}\rightarrow G:(b,n)\longmapsto
bn^{-1}$ is an open dense subset in $G$,    and hence the factorization problem is
solvable for almost all (though not for all) initial data.

The dual space $\frak{g}^{*}$ is canonically identified with $\frak{g}$ by
means of the invariant inner product
\[ \left\langle X,Y\right\rangle =%
{\rm tr}\, XY; \]
the decompositions \reff{g}, \reff{i} give rise to the biorthogonal
decompositions
\[ \frak{g}^{*}=\frak{b}_{+}^{\bot }+\frak{n}_{-}^{\bot }=\frak{b}_{+}^{\bot }+%
\frak{k}^{\bot },
\]
which provide the models for dual spaces of the subalgebras,
\[ \frak{b}_{+}^{*}\simeq \frak{n}_{-}^{\bot }=\frak{b}_{-},\frak{n}%
_{-}^{*}\simeq \frak{b}_{+}^{\bot }=\frak{n}_{+} \]
in the case of the Gauss decomposition and
\[
\frak{b}_{+}^{*}\simeq \frak{k}^{\bot }=\frak{p},\frak{k}^{*}\simeq \frak{b}%
_{+}^{\bot }=\frak{n}_{+}
\]
in the case of the Iwasawa decomposition (here $\frak{p}\subset Mat(n)$
denotes the subspace of symmetric matrices).\footnote{%
Mind that the model for the dual of a subalgebra depends not only on the
subalgebra itself but also on the choice of its complement in the big Lie
algebra (and eventually also on the choice of the inner product whenever it
is not unique).} Let us focus on coadjoint orbits of $B_{+}$.   Denote by $%
\frak{d}_{p}\subset Mat(n),p\in \Bbb{Z}$,    the set of all matrices supported
on p-th diagonal,
\[
\frak{d}_{p}=\left\{ X\in Mat(n);X_{ij}=0\;\hbox{for }j-i\neq p\right\}
\]

\begin{exercise}%{exercise}
 (i) Let us model the dual space $\frak{b}_{+}^{*}$ on $\frak{b}_{-}$;  for
all $q\geq 0$ the subspaces $\frak{b}_{-}^{q}=\oplus _{p=0}^{q}\frak{d}%
_{-p}\subset \frak{b}_{-}$ are invariant with respect to the coadjoint
action of $B_{+}$.   (ii) Similarly, if $\frak{b}_{+}^{*}$ is modelled on $%
\frak{p}$,    the subspaces $\frak{p}^{q}=\oplus _{p=0}^{q}\left( \frak{d}_{p}+%
\frak{d}_{-p}\right) \cap \frak{p}$ are also invariant.
\end{exercise}%{exercise}

Orbits in the subspace $\frak{b}_{-}^{0}$ are all trivial (i.e., each point
in this subspace is stable with respect to the coadjoint action and hence
is a separate orbit); the subspace $\frak{b}_{-}^{1}$ contains orbits of
maximal dimension $2n-2$ (over $\C$ there is just one such orbit, over
the reals there is a finite number of them); the typical example is the
orbit ${\mathcal{O}}_{f}$ which contains the matrix
$f\in \frak{d}_{-1}\subset \frak{b}_{-}$,   %
\[
f=\left(
\begin{array}{llll}
0 & 0 & \cdots & 0 \\
1 & \ddots & \ddots & \vdots \\
\vdots & \ddots & 0 & 0 \\
0 & \cdots & 1 & 0
\end{array}
\right) ;
\]

\begin{exercise}%{exercise}
 (i) ${\mathcal{O}}_{f}$ consists of all matrices of the form
\[
l=\left(
\begin{array}{llll}
p_{1} & 0 & \cdots & 0 \\
b_{1} & \ddots & \ddots & \vdots \\
\vdots & \ddots & p_{n-1} & 0 \\
0 & \ddots & b_{n-1} & p_{n}
\end{array}
\right) ,b_{i}\neq 0,\sum p_{i}=0.
\]
(Over the reals, in addition, $\mathrm{sign}\,b_{i}=+1.\footnote{%
It is easy to see that in the real case the signs of all matrix coefficients
below the principal diagonal are preserved by the coadjoint action; hence there are
exactly $2^{n-1}$ open orbits in our subspace.})$ (ii) Lie-Poisson
brackets of coordinate functions $p_{i},b_{j}$ on ${\mathcal O}_{f}$ are
given by
\[
\left\{ p_{i},p_{j}\right\} =\left\{ b_{i},b_{j}\right\} =0,\left\{
p_{i},b_{i}\right\} =-\left\{ p_{i+1},b_{i}\right\} =b_{i}.
\]
(iii) If we set $b_{i}=\exp \left( q_{i}-q_{i+1}\right) $,    the coordinates $%
q_{i}$ have canonical Poisson brackets with momenta, $\left\{
p_{i},q_{j}\right\} =\delta _{ij}$.
\end{exercise}%{exercise}

Thus, as a symplectic manifold, ${\mathcal{O}}_{f}$ is isomorphic to the
standard phase space $\R ^{2n-2}$.   In the Iwasawa model, the same orbit
is realized by symmetric matrices,
\begin{equation}
L=\left(
\begin{array}{llll}
p_{1} & b_{1} & \cdots & 0 \\
b_{1} & \ddots & \ddots & \vdots \\
\vdots & \ddots & p_{n-1} & b_{n-1} \\
0 & \ddots & b_{n-1} & p_{n}
\end{array}
\right) .  \label{toda}
\end{equation}
Recall that our main theorem associates dynamical systems to coadjoint
orbits in $\frak{g}_{r}^{*}$;  since $\frak{g}_{r_{\mathrm{Gauss}}}$ and $%
\frak{g}_{r_{\mathrm{Iwasawa}}}\ $are direct sums,
\[
\frak{g}_{r_{\mathrm{Gauss}}}\simeq \frak{b}_{+.}\oplus \frak{n}_{-},\;\frak{%
g}_{r_{\mathrm{Iwasawa}}}\simeq \frak{b}_{+.}\oplus \frak{k},
\]
respectively. Coadjoint orbits of this bigger algebra are Cartesian products
of the coadjoint orbits of the factors. In the Iwasawa case the simplest
meaningful choice is to take the \emph{zero orbit} of $K$ in $\frak{k}%
^{*}\simeq \frak{n}_{+}$;  with this choice, \reff{toda} becomes the Lax
matrix for the associated dynamical systems. The Hamiltonians are spectral
invariants of $L$; taking, for instance,
\[
H=\frac{1}{2}%
{\rm tr}\,
L^{2}
\]
and expressing it in terms of the canonical variables $p_{i},q_{i}$,    we get
\[
H=\frac{1}{2}\sum_{i}p_{i}^{2}+\sum_{i}\exp 2\left( q_{i}-q_{i+1}\right) ,
\]
i.e., the Hamiltonian of the open Toda lattice. In the Gauss case, it is
also possible to take the zero orbit of the complementary subalgebra $\frak{n%
}_{-}$;  with this choice the Lax matrix $l$ will be lower triangular and
hence its spectral invariants will depend only on momenta $p_{i}$ and the
corresponding Hamiltonians will be trivial. Luckily, in this case there is
another option: we may take any one-point orbit $\{e\}$ of $N_{-}$
in $\frak{n}_{-}^{*}\simeq \frak{n}_{+}$;  the constant matrix $e$ is simply
added to the Lax matrix. This procedure does not add any new
degrees of freedom to our system, but it modifies the embedding of the
``little orbit'' ${\mathcal{O}}_{f}$ into the big algebra and hence the
spectral invariants of the Lax matrix. Specifically, set
\[
e=\left(
\begin{array}{llll}
0 & 1 & \cdots & 0 \\
0 & \ddots & \ddots & \vdots \\
\vdots & \ddots & 0 & 1 \\
0 & \cdots & 0 & 0
\end{array}
\right) ,\; L_{\mathrm{Gauss}}=l+e.
\]

\begin{remark}
The reader familiar with semisimple Lie algebras will notice that the Jordan
matrices $e,f$\ are \emph{principal nilpotent elements\ }of the general
linear algebra, which suggests the way to generalize the above construction
to other algebras.
\end{remark}

\begin{exercise}%{exercise}
 (i) $\left\{ e\right\} \subset \frak{d}_{1}\subset \frak{n}_{+}$ is a
one-point coadjoint orbit of $N_{-}$.   \newline
(ii) In canonical coordinates, the Hamiltonian $h=\frac{1}{2}%
{\rm tr}\,
L_{\mathrm{Gauss}}^{2}$ is given by
\[
h=\frac{1}{2}\sum_{i}p_{i}^{2}+\sum_{i}\exp \left( q_{i}-q_{i+1}\right) .
\]
\end{exercise}%{exercise}

Note that the Hamiltonians $H$ and $h$ (which are defined on the same
manifold ${\mathcal{O}}_{f}$) are \emph{different} (although, in this particular case,
 they happen to be related
by a simple canonical change of variables); so are the associated
factorization problems which solve the Toda equations.

Other spectral invariants, e.g., $h_{p}={\rm tr}\,%
L_{\mathrm{Gauss}}^{p},\,p=2,3,...$,    form a system of integrals of motion in
involution. Obviously, the number of independent integrals does not exceed $%
n-1$ (mind that $h_{1}$ reduces to a constant on ${\mathcal{O}}_{f}$).

\begin{exercise}%{exercise}
(i) Write down explicitly the two factorization problems. (ii) Find the
explicit relation between them. (iii) Show that over the reals the group
element $\exp tL_{\mathrm{Gauss}}(p,q)$ lies in the ``big cell'' for all $%
\left( p,q\right) \in \Bbb{R}^{2n-2}$ and for all $t\in \Bbb{R}$,    and hence
the Gauss factorization is always possible. (iv) Check that the integrals $%
h_{2},...,h_{n}$ remain functionally independent after restriction to $%
{\mathcal{O}}_{f}$.
\end{exercise}%{exercise}

In both cases, the entries of the factors are \emph{rational functions} of $%
\exp t$ (with coefficients depending on the initial data, i.e., on $p,q$);
thus the functional dependence of the solutions on the time variable is
fairly simple. It is easy to see that this simple behaviour is
characteristic for all Lax equations associated with factorization problems
in finite-dimensional Lie groups.

\begin{remark}
The subspaces $\frak{d}_{p}\subset \frak{g}$\ satisfy $\left[ \frak{d}_{p},%
\frak{d}_{q}\right] \subset \frak{d}_{p+q}$,   \ i.e., they form a \emph{grading\ }%
of the matrix algebra. As a consequence, the subspaces $\oplus _{p\geq n}%
\frak{d}_{p},\;n=0,1,2,...$,   \ are Lie subalgebras and form a decreasing
filtration of $\frak{b}_{+}$;  by duality, the subspaces $\frak{b}%
_{-}^{q}=\oplus _{p\leq -n}\frak{d}_{-p}\subset \frak{b}_{-}\simeq \frak{b}%
_{+}^{*},\;n=0,-1,-2,...$,   \ form an increasing filtration of $\frak{b}%
_{+}^{*}$\ by $Ad\, _{B_{+}}^{*}$-invariant subspaces. The orbit ${\mathcal{O}}_{f}$\ is the
``biggest'' orbit in the subspace $\frak{v}_{1}$; \ its choice is quite
natural, since $\dim {\mathcal{O}}_{f}=2n-2$\ is twice the number of
independent Hamiltonians, which is precisely the amount needed for complete
integrability. Dynamical systems associated with coadjoint orbits of higher
dimension (which are abundant) have not got enough ``obvious'' integrals of
motion to assure their \emph{Liouville integrability;}\ on the other hand,
all these systems are explicitly solvable by means of the factorization
problem. This queer situation is due to the resonance behaviour\ of these
systems: their trajectories span a submanifold in the phase space with
codimension higher than in the generic case.
\end{remark}

An important general conclusion to be held from this example is the role of
\emph{grading}: it provides a natural decomposition into subalgebras, as
well as plenty of invariant subspaces for the coadjoint action.

%%%%%%%%%%%%%%%%%%%%%%%%%%%%%%%%%%%%%%%%%%%%%%%%%%%%%%%%%%%%%%%%%%%%%%%%%%%

\newsection{\label{loop}Loop Algebras and the Riemann Problem}

As already mentioned, loop algebras provide a natural environment for the
study of numerous finite-dimensional systems. In this Section we shall
briefly outline the corresponding constructions.

Let $\frak{g}$ be a semisimple Lie algebra with invariant inner product ($%
\frak{g}=\frak{sl}(n)$ with the inner product $\left( X,Y\right) =%
{\rm tr}\,
XY$ is a good example; in the sequel we shall mainly deal with this standard
matrix case). Its loop algebra, $L\frak{g}$ is the Lie algebra of Laurent
polynomials with coefficients in $\frak{g}$,    $L\frak{g=g}\left[
z,z^{-1}\right] $ with pointwise commutator, $\left[ Xz%
^{n},Yz^{m}\right] =\left[ X,Y\right] z^{n+m}$,    or $%
\left[ x,y\right] \left(z\right)=
\left[ x\left( z\right) ,y\left( z\right) \right] $.   We may regard an element $x\in \frak{G}$ as a polynomial mapping
from the unit circle $\Bbb{T}$ into $\frak{g}$.   An invariant inner product
on $\frak{G}$ is given by
\begin{equation}
\left\langle x,y\right\rangle =\int_{\Bbb{T}}{\mathrm tr}\,x\left( z\right)
y\left( z\right) \frac{dz}{2\pi iz}.  \label{inn}
\end{equation}
Let $\frak{g}_{n}=\frak{g}\cdot z^{n}$;  clearly, $\left[ \frak{g}_{n},\frak{g%
}_{m}\right] \subset \frak{g}_{n+m}$,    and hence the decomposition\break\hfil
$L\frak{g=\oplus }_{n\in {\Z}}\,\frak{g}_{n}$ defines a grading (the so called
\emph{standard grading}) of the loop\break\hfil algebra. Set
\begin{equation}
L\frak{g}_{+}=\frak{\oplus }_{n\geq 0}\frak{g}_{n}=\frak{g}\left[ z\right] ,L%
\frak{g}_{-}=\frak{\oplus }_{n<0}\frak{\frak{g}_{n}}=z^{-1}\frak{g}\left[
z^{-1}\right] .  \label{split}
\end{equation}

\begin{proposition}
\label{loops}(i) $L\frak{g}_{+}$ and $L\frak{g}_{-}$ are graded subalgebras
of $L\frak{g}$,    and
\[ L\frak{g} = L\frak{g}_{+}{\dot{+}}L\frak{g}_{-}. \]
(ii) The inner product (\ref{inn}) sets $L\frak{g}$ into duality with
itself; in particular, $L\frak{g}_{\pm }$ is set into duality with $L\frak{g}%
_{\mp }$.   (iii) The ring of Casimir functions on $L\frak{g}^{*}\simeq L\frak{%
g}^{*}$ is generated by the functionals
\[ \Phi _{n,m}\left[ L\right] =\frac{1}{n}\mathrm{%
 Res}\,%
_{z=0}\,{\mathrm{tr}}\,L\left( z\right) ^{n}z^{-m-1}. \]
\end{proposition}

Note that these functionals are smooth in the sense of ordinary calculus of
variations and ${\mathrm{grad}}\,\Phi _{n,m}=\frac{1}{z^{m}}L^{n-1}$.

Let $P_{\pm }$ be the projection operators onto $L\frak{g}_{\pm }$ parallel
to the complementary subalgebra, $r=P_{+}-P_{-}$.   In analytic terms, $L\frak{%
g}$ consists of trigonometric polynomials on the circle, and $r$ is the
standard Hilbert transform. The Lie algebra $\left( L\frak{g}\right) _{r}$
is isomorphic to the direct sum $L\frak{g}_{+}\oplus L\frak{g}_{-}$.   For $%
X\in L\frak{g}$ we set $X_{\pm }=\pm P_{\pm }X$.

\begin{proposition}
The coadjoint action of $\left( L\frak{g}\right) _{r}$ on its dual is given
by
\[ ad^{*}X\cdot L=P_{+}\left[ X_{-},L\right] -P_{-}\left[ X_{+},L\right] ; \]
this action leaves  invariant the subspace of polynomial loops $L\frak{g}\subset \left( L%
\frak{g}\right) _{r}^{*}$.
\end{proposition}

Notice that both linear operators $L\mapsto P_{+}\left[ X_{-},L\right] $ and
$L\mapsto P_{-}\left[ X_{+},L\right] $ are Toeplitz.

Since $L\frak{g}$ is infinite-dimensional, the choice of the associated Lie
group becomes non-obvious. A reasonable option is to take the group $%
\mathcal{G}_{W}$ of all Wiener maps $g:\Bbb{T}\rightarrow G$.   The Lie
algebra $L\frak{g}$ may also be replaced by its appropriate completion, e.g.
the Wiener algebra $L\frak{g}_{W}$.   Of course, with this choice the full
dual space $L\frak{g}_{W}^{*}$ becomes a rather complicated object; the
point is that the set of polynomial loops $L\frak{g}\subset L\frak{g}%
_{W}^{*} $ is invariant with respect to the coadjoint action of $\left( L%
\frak{g}\right) _{r}$ (though of course not with respect to the coadjoint
action of $L\frak{g}_{W}$!). Let ${\mathcal{G}}_{W}^{+}\subset {\mathcal{G}}_{W}$
be the subgroup of Wiener maps $g:\Bbb{T}\rightarrow G$ which are
holomorphic in the unit circle, and ${\mathcal{G}}_{W}^{\_}\subset {\mathcal{G}}%
_{W}$ the subgroup of maps which are holomorphic outside the unit circle and
satisfy the normalization condition $\lim_{z\rightarrow \infty }g\left(
z\right) =id. $ The Lie group which corresponds to $\left( L\frak{g}\right)
_{r}$ may be identified with ${\mathcal{G}}_{W}^{r}={\mathcal{G}}_{W}^{+}%
\times {\mathcal{G}_{W}}^{-}$;  its coadjoint action is given by
\begin{eqnarray*}
\left( Ad^{*}h\cdot L\right) &=&P_{+}\left( h_{-}\left( z\right) L\left(
z\right) h_{-}^{-1}\left( z\right) \right) -P_{-}\left( h_{+}\left( z\right)
L\left( z\right) h_{+}^{-1}\left( z\right) \right) ;\; \\
h &=&\left( h_{+},h_{-}\right) \in {\mathcal{G}}_{W}^{+}\times {\mathcal{G}}%
_{W}^{-}.
\end{eqnarray*}

\begin{exercise} %{exercise}
For $\frak{g}=\frak{sl}(n)$ describe all coadjoint orbits of ${\mathcal{G}}%
_{W}^{r}{\mathcal{\ }}$in the subspace of Laurent polynomials of the form
\[ L\left( z\right) =l_{-1}z^{-1}+l_{0}+l_{1}z. \]
\end{exercise} %{exercise}

\begin{remark}
An analyst may feel disappointed with our choice of the `restricted dual' $L%
\frak{g}\subset L\frak{g}_{W}^{*}$\ of the Wiener algebra; indeed, the
coadjoint orbits which are contained in this space are modelled on (matrix)
polynomial functions; even general rational functions are not allowed, to
say nothing of more interesting classes of analytic functions. The reason
for this deliberate restriction is very simple: we are willing to get
dynamical systems which admit a simple parametrization and (possibly) some
physical interpretation; practical experience shows that examples which are
physically interesting are usually associated with coadjoint orbits of the
lowest possible dimensions. (By contrast, most of the orbits which lie in
the `exotic' part of the full dual space are infinite-dimensional.) This
does not mean of course that dynamical systems which are modelled on
analytic functions of a more complicated nature are totally uninteresting;
but once again, `good' examples are associated not with generic coadjoint
orbits in the `very big dual', but rather with well-embedded
finite-dimensional ones. (Below we shall see how to construct such examples
using a different choice of the basic Lie algebra.) One more reason to
single out the finite-dimensional orbits is the possibility to bring into
play the highly powerful machinery of Algebraic Geometry: we shall see below
that polynomial (or, more generally, rational) Lax matrices give rise to
algebraic curves of \emph{finite genus} and Lax equations are linearized on
their Jacobians. Lax matrices associated with infinite-dimensional orbits
will lead to curves of \emph{infinite genus}.
\end{remark}

The specialization of our main theorem to the present situation may be
stated as follows:

\begin{theorem} \label{Riemann}
(i) Invariant functionals $\Phi _{n,m}$ give rise to
Hamiltonian equations of motion on $L\frak{g}\subset \left( L\frak{g}\right)
_{r}^{*}$ with respect to the Lie-Poisson bracket of $L\frak{g}_{r}$;  these
equations may be written in the Lax form,
\begin{equation}
\frac{dL}{dt}=\left[ L,M_{\pm }\right] ,M_{\pm }=\pm P_{\pm }\left( \mathrm{%
grad}\,\Phi _{n,m}\right) .  \label{laxeq}
\end{equation}
(ii) The integral curve of \reff{laxeq} with origin $L_{0}$ is given by
\begin{equation}
L\left( t,z\right) =g_{\pm }\left( t,z\right) ^{-1}L_{0}g_{\pm }\left(
t,z\right) ,  \label{curve}
\end{equation}
where $g_{+}\left( t\right) ,g_{-}\left( t\right) $ solve the matrix Riemann
problem
\begin{equation}\label{riem}
\arr{2.0}{c}{\ds
\exp t\,{\mathrm {grad}}\,\Phi _{n,m}\left[ L_{0}\right] \left( z\right)
= g_{+}\left( t,z\right) g_{-}\left( t,z\right) ^{-1},\\
\ds g_{+}\left( t\right)
\in {\mathcal {G}}_{W}^{+}, \;g_{-}\left( t\right) \in {\mathcal {G}}_{W}^{-}.}
\end{equation}
\end{theorem}

Note that ${\mathrm{grad}}\,\Phi _{n,m}$ is a Laurent polynomial and hence is
regular in the punctured Riemann sphere ${\C}P_{1}\backslash \left(
\left\{ 0\right\} \cup \left\{ \infty \right\} \right) =\C \backslash
\left\{ 0\right\} $.   Hence the factorization problem (\ref{riem}) has the
following geometric meaning. The projective line ${\C}P_{1}$ is covered
by two domains $U_{+}={\C}P_{1}\backslash \left\{ \infty \right\} $ and $%
U_{-}={\C}P_{1}\backslash \left\{ 0\right\} $.   The function $\exp tdh(L)$
is regular in $U_{+}\cap U_{-}={\C}\backslash \left\{ 0\right\} $ and may
be regarded as the transition function of a holomorphic vector bundle over $%
{\C}P_{1}$. Factorization problem \reff{riem} amounts to an analytic
trivialization of this bundle. It is well known (see \cite{SegalPressley})
that not all vector bundles over $\Bbb{C}P_{1}$ are analytically trivial:
each $n$-dimensional bundle breaks up into a sum of line bundles, and their
degrees, $d_{1},...,d_{n}\in \Z$ form a full system of holomorphic
invariants of the given bundle. In the language of transition functions this
means that $\exp tdh(L)$ admits a factorization of the form
\[
\exp tdh(L)=g_{+}\left( z,t\right) d\left( z\right) g_{-}\left( z,t\right) ^{-1},
\]
where $d\left( z\right) =diag\left( z^{d_{1}},...,z^{d_{n}}\right) $.   Thus
formula (\ref{curve}) requires that all partial indices $d_{1},...,d_{n}$
are zero. One can prove that this is true at least for $t$ sufficiently
small \cite{Gohberg}. The exceptional values of $t\in \C$ for which
problem \reff{riem} does not admit a solution form a discrete set in $\C$;
at these points the solution $L\left( t\right) $ has a pole: the
trajectory of the Lax equation goes off to infinity.

%%%%%%%%%%%%%%%%%%%%%%%%%%%%%%%%%%%%%%%%%%%%%%%%%%%%%%%%%%%%%%%%%%%%%%%%%%%

\newsubsection{Riemann Problem and Spectral Curves}

Theorem \rref{Riemann} provides a link between the Hamiltonian scheme of
Section 3 and the algebro-geometric methods of the finite-band integration
theory (see \cite{alg-geom}). Namely, the formula \reff{curve} for the
trajectories immediately implies that the Lax equations linearize on the
Jacobian of the spectral curve associated with the Lax matrix. The proof is
so short and simple that I would like to reproduce it here.

Let $\frak{g}=\frak{gl}\left( n,\C \right) $ and $L\left( z\right) =\sum
X_{i}z^{i},X_{i}\in \frak{g}$,    a matrix-valued Laurent polynomial. Let us
consider the algebraic curve $\Gamma _{0}\subset {\C}\backslash
\left\{ 0\right\} \times C$ defined by the characteristic equation
\begin{equation}
\det (L(z)-\lambda )=0;  \label{curv}
\end{equation}
we may regard $z,\lambda $ as meromorphic functions defined on $\Gamma _{0}$%
. Assume that the spectrum of $L\left( z\right) $ is simple for generic $z$
(this key technical assumption is satisfied in most applications). For each
nonsingular point $P\subset \Gamma _{0}$ which is not a branching point of $%
\lambda $ there is a one-dimensional eigenspace $E\left( P\right) \subset
\C ^{n}$ of $L\left( z\left( P\right) \right) $ with eigenvalue $\lambda
\left( P\right) $.   This gives a holomorphic line bundle on $\Gamma _{0}$
defined everywhere except for singular points and branching points. Let $%
\Gamma $ be the nonsingular, compact model of $\Gamma _{0}$.   One can show that the eigenvector
bundle extends to a holomorphic line bundle $E\longrightarrow \Gamma $ on
the whole smooth curve $\Gamma $.\footnote{Indeed, the mapping $P\mapsto E\left(
P\right) $ determines a meromorphic mapping $\Gamma \rightarrow {\C}%
P_{n-1}$ (this is a corollary of the elementary analytic perturbation
theory). By a standard theorem, any such mapping is actually holomorphic,
and hence the eigenvector bundle extends to $\Gamma $.  }

The spectral curve $\Gamma $ with two distinguished meromorphic functions $z$
and $\lambda $ and the line bundle $E\longrightarrow \Gamma $ constitute the
set of spectral data for $L\left( z\right) $. The evolution determined by a
Lax equation of motion leaves the spectral curve (7.3) invariant, and the
dynamics of the line bundle $E$ is easy to describe. Let $h$ be the
Hamiltonian of our Lax equation; set $M=dh\left( L\right) $;  since $[L,M]$ =
0 pointwise and the spectrum of $L$ is simple, the eigenvectors of $L\left(
z\right) $ are also the eigenvectors of $M\left( z\right) $,
\[
M(z\left( P\right) )v=\mu \left( P\right) v,\;P\in \Gamma ,\;v\in
E(P)\subset \C ^{n},
\]
where $\mu \left( P\right) $ is a meromorphic function on $\Gamma $.   Define
the domains $V_{\pm }\subset \Gamma $ by $V_{\pm }=\left\{ P\in \Gamma
;z\left( P\right) ^{\pm 1}\neq \infty \right\} $.   Clearly, $V_{+}\cup
V_{-}=\Gamma $ and $\mu $ is regular in the intersection $V_{+}\cap
V_{-}=\Gamma _{0}$.   Recall that a line bundle over a curve is specified by
its \emph{transition function} (with values in the multiplicative group $\C^{*}$)
with respect to some covering; tensor product of bundles corresponds
to the ordinary product of transition functions. The equivalence classes of
line bundles form an abelian group ${\mathrm{Pic\,}}\Gamma $ with respect to
the tensor product. Let $F_{t}$ be the line bundle on $\Gamma $ determined
by the transition function $\exp t\mu \left( P\right) $ with respect to the
covering $\left\{ V_{+},V_{-}\right\} $ For all $t\in \C$ the bundles $%
F_{t}$ have degree zero and form a 1-parameter subgroup in the Picard group $%
{\mathrm{Pic}}_{0}\,\Gamma $ of equivalence classes of holomorphic line
bundles of degree zero on $\Gamma $ which, by Abel's theorem, is canonically
isomorphic to \ ${\mathrm{Jac}}\,\Gamma $,    the Jacobian of $\Gamma $.   (See,
for instance, \cite{Griffiths}.)

\begin{theorem}
\label{th7.1} The line bundle $E$ regarded as a point of ${\mathrm{Pic\,}}%
\Gamma $ evolves linearly with time, $E\left( t\right) =E\otimes F_{-t}$
\end{theorem}

\begin{proof} Since the Lax matrix evolves by similarity transformation,
its eigenvectors evolve linearly. Let $g_{\pm }\left( t\right) $ be the
solution of the Riemann problem \reff{riem}. In view of \reff{curve}, the
moving eigenspace $E_{t}\left( P\right) $ regarded as a subspace of $\Gamma
\times \C ^{n}$,    is expressed as
\begin{equation}
E_{t}\left( P\right) =g_{+}\left( t,z\left( P\right) \right) ^{-1}E\left(
P\right) \;  \label{eig}
\end{equation}
over$\;V_{+}$,    and
\begin{equation}
E_{t}\left( P\right) =g_{-}\left( t,z\left( P\right) \right) ^{-1}E\left(
P\right) \;  \label{eigmin}
\end{equation}
over$\;V_{-}$.   In other words, $g_{\pm }\left( t,z\right) ^{-1}$ define
isomorphisms between $E\left( P\right) $ and $E_{t}\left( P\right) $ over $%
V_{\pm }$.   The transition function in $V_{+}\cap V_{-}$ which matches these
two isomorphisms is $g_{+}\left( t,z\left( P\right) \right) ^{-1}g_{-}\left(
t,z\left( P\right) \right) |_{E_{t}\left( P\right) }$.   It is easy to check
that
\[
g_{+}\left( t,z\right) g_{-}\left( t,z\right) ^{-1}=\exp tdh\left(
L_{0}\right)
\]
 implies
\[
g_{-}\left( t,z \right) g_{+}\left(
t,z \right) ^{-1}=\exp tdh\left( L\left( t\right) \right) ;
\]
hence,
\begin{equation}\label{pic}
g_{+}\left( t,z\left( P\right) \right) ^{-1}g_{-}\left( t,z\left( P\right)
\right) |_{E_{t}\left( P\right) }=e^{-tM\left( t,z\left( P\right) \right)
}|_{E_{t}\left( P\right) }=e^{-t\mu \left( P\right) },
\end{equation}
or $E_{t}\left( P\right) =E\left( P\right) \otimes F_{-t}$,    as claimed.
\end{proof}
The eigenvector $\psi \left( t,P\right) =\left( \psi _{1},...,\psi
_{n}\right) \in E\left( t,P\right) \subset \Bbb{C}^{n}$ is called the \emph{%
Baker-Akhiezer function} of $L\left( z,t\right) $.   From (\ref{eig}, \ref
{eigmin}) it follows that the Baker-Akhiezer function $\psi \left(
t,P\right) $ in the domains $V_{\pm }\subset \Gamma $ may be written in the
form
\[
\psi _{\pm }\left( t,P\right) =g_{\pm }\left( t,z\left( P\right) \right)
^{-1}\psi \left( P\right) ,
\]
so that
\[
\psi _{+}\left( t,P\right) =e^{-t\mu \left( P\right) }\psi _{-}\left(
t,P\right) .
\]
Since $\partial _{t}g_{\pm }\cdot g_{\pm }^{-1}=M_{\pm }$ (see the proof of
theorem \ref{fact}), we have
\[
\frac{d}{dt}\psi _{\pm }\left( t,P\right) =-M_{\pm }\left( t,z\left(
P\right) \right) \psi _{\pm }\left( t,P\right) .
\]
Using the machinery of Algebraic Geometry, it is possible to construct the
Baker-Akhiezer function explicitly, in terms of the Riemann theta functions
and Abelian integrals. This, in turn, allows to obtain an explicit solution
of the Riemann problem. Let us explain how the matrices $g_{\pm }\left(
t,z\right) $ may be reconstructed from $\psi _{\pm }\left( t,P\right) $.
Suppose that $z\in \Bbb{C}$ is not a ramification point of $\Gamma $ (i.e.,
all eigenvalues of $L(z)$ are distinct); let $P_{1},...,P_{n}$ be the points
of $\Gamma $ which lie over $z$.   Let us arrange the column vectors $\psi
_{\pm }\left( t,P_{1}\right) ,...,\psi _{\pm }\left( t,P_{n}\right) $ in a $%
n\times n$ matrix $\hat{\psi}_{\pm }\left( t,z\right) $.   Put
\begin{equation}
g_{\pm }\left( t,z\right) =\hat{\psi}_{\pm }\left( t,\lambda \right) \hat{%
\psi}_{\pm }\left( 0,\lambda \right) ^{-1}.  \label{gplus}
\end{equation}
Note that if we change the ordering of branches $P_{1},...,P_{n},\hat{\psi%
}_{\pm }$ is multiplied on the right by a permutation matrix and hence $%
g_{\pm }$ remains invariant.

\begin{proposition}
(i) $g_{\pm }\left( t,z\right) $ satisfies the differential equation
\begin{equation}
\frac{dg_{\pm }\left( t,z\right) }{dt}=-M_{\pm }\left( t,z\right) g_{\pm
}\left( t,z\right) ,\;M_{\pm }\left( t,z\right) =\left(
%TCIMACRO{\TeXButton{grad}{{\rm grad\,}} }
%BeginExpansion
{\mathrm {grad\,}}%
%EndExpansion
h\left[ L\left( t,z\right) \right] \right) _{\pm }.  \label{ev}
\end{equation}
(ii) $g_{\pm }$ are entire functions of $z^{\pm 1}$.   (iii) $g_{\pm }$ solve
the factorization problem
\[
g_{+}\left( t,z\right) g_{-}\left( t,z\right) ^{-1}=\exp t%
%TCIMACRO{\TeXButton{grad}{{\rm grad\,}} }
%BeginExpansion
{\mathrm {grad\,}}%
%EndExpansion
h\left[ L\left( 0,z\right) \right] .
\]
\end{proposition}

The key assertion (ii) is an easy consequence of the differential equation (%
\ref{ev}); indeed, by (\ref{gplus}) $g_{\pm }$ is the fundamental solution
of (\ref{ev}) normalized by $g_{\pm }\left( 0,z\right) =Id$; hence it is
holomorphic in the domain where $M_{\pm }\left( t,z\right) $ is nonsingular,
i.e., in ${\C}P_{1}\backslash \left\{ \infty \right\} $ and ${\C}P_{1}\backslash
\left\{ 0\right\} $,    respectively.

%%%%%%%%%%%%%%%%%%%%%%%%%%%%%%%%%%%%%%%%%%%%%%%%%%%%%%%%%%%%%%%%%%%%%%%%%%%%

\newsection{More Examples}

%%%%%%%%%%%%%%%%%%%%%%%%%%%%%%%%%%%%%%%%%%%%%%%%%%%%%%%%%%%%%%%%%%%%%%%%%%%%

\newsubsection{Twisted Loop Algebras}

The decomposition of the loop algebra we used so far is based on its \emph{%
standard grading }(i.e., grading by the powers of the loop parameter $z)$.
This grading is by no means unique, and it is possible to use other gradings
to produce new examples of Lax representations. Another possibility (which
actually absorbs the former one) is to bring into play \emph{twisted loop
algebras. }Let $\sigma $ be an automorphism of $\frak{g}$ of order $n$.   The
twisted loop algebra $L(\frak{g},\sigma )$ is the subalgebra of $L\frak{g}$
defined by
\begin{equation}
L(\frak{g},\sigma )=\left\{ x\in L\frak{g};\sigma \left( x\left(z
\right) \right) =x\left( \epsilon z \right) \right\} ,  \label{tw}
\end{equation}
where $\epsilon =\exp \frac{2\pi i}{n}$ is the root of unity. Equivalently, $%
L(\frak{g},\sigma )\subset L\frak{g}$ is the stable subalgebra of the
automorphism ${\sigma }:L\frak{g\rightarrow }L\frak{g}$ such that $x^{%
{\sigma }}\left( z \right) =\sigma \left( x\left( \epsilon
z \right) \right) $ for all $x\in L\frak{g.}$ The isomorphism class of
the twisted loop algebra depends on the properties of $\sigma $; one can
show that when $\sigma $ is an \emph{inner} automorphism, the stable
subalgebra $L(\frak{g},\sigma )$ is isomorphic to $L\frak{g}$ as an abstract
Lie algebra; however, the grading which is induced on $L(\frak{g},\sigma
)\subset L\frak{g}$ by the standard grading in $L\frak{g}$ is different.
The classical theorem due to V.Kac asserts that all different gradings on $L%
\frak{g}$ may be obtained in this way. (Among the integrable systems that
may be very naturally constructed along these lines one may quote \emph{%
periodic Toda lattices} which are associated with the decomposition of loop
algebras derived from their\emph{\ principal grading}, see \cite{RS}
for details.)

The most interesting case for applications is when $\sigma $ is an outer
automorphism of order 2 (an involution). In this case we may assume that $%
\frak{g}$ and $L\frak{g}$ are real. Here is the key example: $\frak{g}=\frak{%
gl}(n),\sigma \left( X\right) =-X^{t}$ ( $t$ denotes transposition). The
loop algebra $L\frak{g}^{\sigma }$ consists of Laurent polynomials
\[
X\left( z\right) =\sum X_{k}z^{k},
\]
where $X_{2p}=-X_{2p}^{t},X_{2p+1}=X_{2p+1}^{t}$.   Antisymmetric matrices
belong to the Lie algebra $so(n)$ of the orthogonal group, which describes
kinematics of the rigid body; on the other hand, symmetric matrices are
reminiscent of the quadratic form associated with kinetic energy ( inertia
tensor). Thus, the twisted loop algebra seems to be a good candidate to set
up the stage for applications to the mechanics of the rigid body. Let us
equip $L\frak{g}^{\sigma }$ with the inner product
\begin{equation}
\left\langle X,Y\right\rangle =-%
{\rm Res}\,%
_{z=0}\frac{1}{z}%
{\rm tr}\,
X\left( z\right) Y\left( z\right)  \label{inntw}
\end{equation}
which sets $L\frak{g}^{\sigma }$ into duality with itself (mind the
difference with (\ref{inn}): the factor $z^{-1}$ makes the coupling respect
parity; the minus sign makes the inner product positive on $so(n)$). Let $r$
be the classical r-matrix associated with the standard decomposition $L\frak{%
g}^{\sigma }=L\frak{g}_{+}^{\sigma }{\dot{+}}L\frak{g}_{-}^{\sigma }$,
as in \reff{split}. Obviously,
\[
\left( L\frak{g}_{+}^{\sigma }\right) ^{*}\simeq \left( L\frak{g}%
_{-}^{\sigma }\right) ^{\perp }=\oplus _{k\leq 0}\frak{g}\cdot
z^{k},\;\left( L\frak{g}_{-}^{\sigma }\right) ^{*}\simeq \left( L\frak{g}%
_{+}^{\sigma }\right) ^{\perp }=\oplus _{k>0}\frak{g}\cdot z^{k}.
\]
Coadjoint orbits of $L\frak{g}_{r}^{\sigma }$ are direct products of orbits
of $L\frak{g}_{+}^{\sigma }$ lying in $\left( L\frak{g}_{-}^{\sigma }\right)
^{\perp }$ and orbits of $L\frak{g}_{-}^{\sigma }$ lying in $\left( L\frak{g}%
_{+}^{\sigma }\right) ^{\perp }$.   Lax matrices describing the motion of the
rigid body and related mechanical systems belong to the simplest $ad^{*}$%
-invariant subspace $L\frak{g}_{-1,1}^{\sigma }$ consisting of matrices
\[
L\left( z\right) =az^{-1}+l+bz,l\in so\left( n\right) ,a=a^{t},b=b^{t}.
\]

\begin{exercise}  %{exercise}
(i) All monomials $bz\in L\frak{g}^{\sigma }$ are 1-point orbits of $L\frak{g%
}_{-}^{\sigma }$.   (ii) Coadjoint representation of $L\frak{g}_{+}^{\sigma }$
in the subspace $\left\{ az^{-1}+l\right\} \subset \left( L\frak{g}%
_{+}^{\sigma }\right) ^{*}$ factors through its finite-dimensional quotient $%
L\frak{g}_{+}^{\sigma }/z^{2}L\frak{g}_{+}^{\sigma }$.   \newline
(iii) The quotient algebra $L\frak{g}_{+}^{\sigma }/z^{2}L\frak{g}%
_{+}^{\sigma }$ is isomorphic to the semidirect product $\frak{g}%
_{0}^{\sigma }=so\left( n\right) \ltimes sym(n)$ of the orthogonal algebra $%
so\left( n\right) $ and the space of symmetric matrices (with zero Lie
bracket).
\end{exercise} %{exercise}

The associated Lie group $G_{0}^{\sigma }$ is the semidirect product of $%
SO\left( n\right) $ and the additive group of the linear space $sym(n)$ of
symmetric $n\times n$-matrices. Its coadjoint orbits are easy to describe;
here is a simple example:

\begin{exercise}  %{exercise}
Fix a unit vector $e\in {\R}^{n}$ and let $a=e\otimes e$ be the rank one
orthogonal projection operator onto $\Bbb{R\cdot }e\subset {\R}^{n}$.   Let
$T^{*}S^{n-1}\ $be the cotangent bundle of the sphere $S^{n-1}=SO\left(
n\right) \cdot e\subset {\R}^{n}$ realized as the subbundle of $%
S^{n-1}\times {\R}^{n}$,   %
\[
T^{*}S^{n-1}=\left\{ \left( x,p\right) \in S^{n-1}\times \Bbb{R}%
^{n};\left\langle p,x\right\rangle =0.\right\}
\]
There is natural map $\pi :T^{*}S^{n-1}\rightarrow \mathcal{O}_{a}$ onto the
coadjoint orbit of $G_{0}^{\sigma }$ passing through the monomial $z^{-1}a$,   %
\[
\pi :\left( x,p\right) \longmapsto z^{-1}x\otimes x+p\wedge x.
\]
($\pi $ is actually a double covering.)
\end{exercise} %{exercise}

The Lax matrix associated with this orbit has the form\footnote{We denote by $p\wedge x$ the $n\times n$-matrix with entries
$p_i x_j-p_jx_i$;  in a similar way, the entries of $x\otimes x$ are $x_ix_j$.  }
\[
L\left( z\right) =z^{-1}x\otimes x+p\wedge x+z\,b,\;x\in SO\left( n\right) ,\;p\in
{\R}^{n},\; \left\langle p,x\right\rangle =0;
\]
the associated phase space describes the point moving on a sphere. The
constant matrix $b\in sym\left( n\right) $ does not affect kinematics, but
is quite useful to produce interesting Hamiltonians. The simplest
Hamiltonian is
\[
H=-\frac{1}{4}%
{\mathrm {Res}}\,%
_{z=0}\,%
{\mathrm {tr}}\,
\frac{1}{z}L\left( z\right) ^{2}=\frac{1}{2}\left\langle p,p\right\rangle -%
\frac{1}{2}\left\langle bx,x\right\rangle ;
\]
it describes the so called \emph{Neumann problem} (point moving on a sphere
in a quadratic potential).

\begin{exercise} %{exercise}
Describe the set of commuting integrals of motion for the Neumann problem
which are \emph{quadratic} in momenta.
\end{exercise}   %{exercise}

We may generalize this example in various ways; a useful remark is that we
may avoid a too detailed description of the coadjoint orbits. Instead, we
may produce a map $\pi $ onto such orbit, or a union of orbits, which is
compatible with the Poisson structure but need not be a bijection (and so
possibly introduces some extra variables). This idea is implemented in the
following statement. Set $\;K=SO\left( n\right) ,\frak{k}=so(n)$ and let $%
T^{*}K\simeq K\times \frak{k}$ be the cotangent bundle (equipped with its
standard Poisson bracket). Fix $a\in sym\left( n\right) $ and consider the
mapping
\[
\pi :T^{*}K\longrightarrow \left( \frak{g}_{0}^{\sigma }\right) ^{*}:\left(
k,\rho \right) \longmapsto k\left( \rho +a\right) k^{-1}
\]

\begin{proposition}
$\pi $ is a Poisson mapping (i.e., maps canonical Poisson brackets in $%
T^{*}K $ onto Lie-Poisson brackets in $\left( \frak{g}_{0}^{\sigma }\right)
^{*}$); its image is a union of coadjoint orbits of the semidirect product $%
G_{0}^{\sigma }=K\times sym\left( n\right) $.
\end{proposition}

\noindent[In this statement we identified $\frak{g}_{0}^{\sigma }$ with $%
Mat\left( n\right) $ as a linear space and also used the  inner
product $\left\langle X,Y\right\rangle =-%
{\rm tr}\,
XY$ to identify the dual space $\frak{g}_{0}^{\sigma }$ with $Mat\left(
n\right) $.  ]

The cotangent bundle $T^{*}K$ is naturally interpreted as the phase space of
a rigid body in $\Bbb{R}^{n}$;  we get a family of Lax matrices parametrized
by points of $T^{*}K:$%
\begin{equation}
L\left( z\right) =z^{-1}kak^{-1}+k\rho k^{-1}+zb;  \label{manakov}
\end{equation}
Hamiltonians which may be derived from \reff{manakov} include the so called
\emph{Manakov case} of the motion of a free top in ${\R}^{n}$ (for $n=3$
this is the classical \emph{Euler top}), or more generally, the Manakov top
in a quadratic potential.

\begin{remark}
One may wonder, what is the relation of the low-dimensional Neumann system
to the ``big'' phase space $T^{*}K$\ (we have $\dim T^{*}S^{n-1}=2n-2$\ and $%
\dim T^{*}K=n\left( n-1\right) $). The answer is that, for special choices
of $a\in sym\left( n\right) $,    the Hamiltonians associated with the Lax
matrix \reff{manakov} possess high symmetry (resulting from the redundancy
introduced by $\pi )$. The Neumann system is the result of Hamiltonian
reduction of the ``big system'' with respect to this symmetry group.
\end{remark}

One may notice that the use of the twisted loop algebra was indeed crucial:
the built-in symmetry of the Lax matrix accounts both for the correct
kinematics of the rigid body (antisymmetric matrices) and for the symmetry
of the related quadratic forms (notably, of the kinetic energy). Further
generalization is straightforward: we must scan the list of semisimple Lie
algebras and their involutions and look for nice-looking opportunities. In
this way we get the following list:

\bigskip

\begin{tabular}{|l|l|l|}
\hline
\bfseries{Algebra} & \bfseries{Involution} & \bfseries{Related systems} \\
\hline
$\frak{gl}(n)$ & $X\mapsto -X^{t}$ & Manakov top, Neumann system, etc. \\
\hline
$so(n,1)$ & $X\mapsto -X^{t}$ & Lagrange top, spherical pendulum, etc. \\
\hline
$so\left(p,q\right) ,p>q\geq 2$ & $X\mapsto -X^{t}$ & Kowalevski top and its
generalizations \\ \hline
$so(n,n)$ & $X\mapsto -X^{t}$ & Interacting Manakov tops \\ \hline
$G_2\subset so(4,3)$ & $X\mapsto -X^{t}$ & Exotic integrable top on $SO(4)$
\\ \hline
\end{tabular}

\medskip In all cases, Lax matrices belong to the subspace $L\frak{g}%
_{-1,1}^{\sigma }$;  to get particular examples (for instance, the Kowalevski
top) one sometimes has to perform additional Hamiltonian reduction; we
refer the reader to \cite{RS} for details.

Let us finally discuss the implications of the twisting automorphism for the
geometry of the spectral curve and for the linearization theorem.

\begin{proposition}
(i) Let us assume that the Lax matrix $L\left( z\right) \in L\frak{g}%
^{\sigma }$ and $\sigma $ is an \emph{inner automorphism} of $\frak{g},%
\mathrm{ord\,}\sigma =m$.   In that case the spectral curve $\Gamma =\left\{
\left( z,\lambda \right) \in \Bbb{C}^{2};\det \left( L\left( z\right)
-\lambda \right) =0\right\} $ admits an automorphism $\hat{\sigma}:\left(
z,\lambda \right) \mapsto \left( \epsilon z,\lambda \right) $ (here $\epsilon =
\exp 2\pi i/n$ is the root of unity); this
automorphism lifts to $\mathrm{Pic\,}\Gamma $ and the transition function (%
\ref{pic}) which determines the evolution of the eigenbundle of $L$ is \emph{%
invariant} under $\hat{\sigma}$.   (ii) Suppose that $\sigma $ is an \emph{%
outer involution,} $\sigma \left( X\right) =-X^{t}$.   Then the spectral curve
admits an automorphism $\hat{\sigma}:\left( z,\lambda \right) \mapsto \left(
-z,-\lambda \right) $;  the transition function is \emph{anti-invariant}
under $\hat{\sigma}$,    i.e., $\hat{\sigma}:\exp t\mu (z,\lambda )\mapsto \exp
\left( -t\mu (z,\lambda )\right) $.
\end{proposition}

The check of both assertions is obvious: An inner automorphism preserves the
eigenvalues of a matrix; by contrast, $\sigma \left( X\right) =-X^{t}$
changes the sign of the eigenvalues. The logarithm of the transition
function $\mu \left( z,\lambda \right) $ is the eigenvalue of the gradient
${\mathrm {grad\,}} H\left[ L\right] \left( z\right) $ (more precisely, one of
its branches associated with the eigenvector of $L\left( z\right) $ which
corresponds to $\lambda )$.   Since ${\mathrm {grad\,}}H\left[ L\right] \in L\frak{g}^{\sigma }$,
$\mu \left( z,\lambda \right) $
is invariant when $\sigma $ is inner and changes sign when it's derived from
transposition.

%%%%%%%%%%%%%%%%%%%%%%%%%%%%%%%%%%%%%%%%%%%%%%%%%%%%%%%%%%%%%%%%%%%%%%%%%%%%

\newsubsection{Rational Lax Matrices}

Let us now describe how to deal with Lax matrices which are \emph{rational
functions} of the spectral parameter. As we mentioned, it is possible to
trace down the corresponding coadjoint orbits inside the dual space of an
appropriate completion of the standard loop algebra, but it is more
practical to choose our basic Lie algebra in a different way.

Let $D=\left\{ z_{1},...,z_{N}\right\} \subset {\C}P_{1}$ be a finite
set; we assume that $\infty \in D$.   For $z_{j}\in D$ let $\lambda _{j}$ be
the local parameter on ${\C}P_{1}$ at $z=z_{j}$,    i.e., $\lambda
_{j}=z-z_{j}$ if $z_{j}\neq \infty $ and $\lambda _{\infty }=z^{-1}$ for $%
z_{j}=\infty $.   We define the \emph{local algebra} $\frak{G}_{z_{j}}$ as the
algebra of formal Laurent series in local parameter with coefficients in a
little Lie algebra $\frak{g},\frak{G}_{z_{j}}=\frak{g}\left( \left(
\lambda _{j}\right) \right) $.   (We may assume that $\frak{g}$ is the matrix
algebra with the standard inner product.) If $z_{j}\neq \infty $,    let $\frak{%
G}_{z_{j}}^{+}$ be the algebra of formal Taylor series in local parameter;
for $z_{j}=\infty $ we set $\frak{G}_{\infty }^{+}=\lambda _{\infty }\frak{g}%
\left[ \left[ \lambda _{\infty }\right] \right] $ (in other words, $\frak{G}%
_{\infty }^{+}$ consists of formal Taylor series without constant term). Put
\[
\frak{G}_{D}=\bigoplus_{z_{j}\in D}\frak{G}_{z_{j}},\;\frak{G}%
_{D}^{+}=\bigoplus_{z_{j}\in D}\frak{G}_{z_{j}}^{+}
\]
(direct sum of Lie algebras). Let $\frak{g}(D)$ be the algebra of rational
functions on ${\C}P_{1}$ with coefficients in $\frak{g}$ which are
regular outside $D$;  it is naturally embedded into $\frak{G}_{D}$ (the
embedding assigns to each $X\in \frak{g}(D)$ the collection of its Laurent
series at each point of $D$).

\begin{proposition}
\label{mittag}(i) $\frak{G}_{D}=\frak{g}(D){\dot{+}}\frak{G}_{D}^{+}$
(direct sum of linear spaces). \newline (ii) The $\Bbb{C}$-bilinear inner product on $%
\frak{G}_{D}$%
\begin{equation}
\left\langle X,Y\right\rangle =\sum_{z_{j}\in D}{\mathrm{Res}}_{z_{j}}%
{\mathrm {tr}}\,
X_{j}Y_{j}d\lambda _{j}  \label{res}
\end{equation}
is invariant and nondegenerate. (iii) $\frak{g}(D)$ and $\frak{G}_{D}^{+}$
are isotropic subspaces with respect to (\ref{res}); moreover, $\frak{g}(D)%
\frak{\simeq }\left( \frak{G}_{D}^{+}\right) ^{*}$.   (iv) Coadjoint orbits of
$\frak{G}_{D}^{+}$ in $\frak{g}(D)$ are finite-dimensional.
\end{proposition}

\emph{Sketch of a proof.} An element $X=\left(
X_{j}\right) _{z_{j}\in D}$ is a finite collection of Laurent series;
stripping each of them of its positive part we get a set of \emph{principal
parts} at $z_{j}\in D$;  let $X^{0}$ be the unique rational function with
these principal parts; by construction, $X-X^{0}\in \frak{G}_{D}^{+}$ (mind
the special role of $\infty $ which fixes the normalization condition!). In
brief, we can say that the decomposition
$\frak{G}_{D}=\frak{g}(D){\dot{+}}\frak{G}_{D}^{+}$ is equivalent to the
Mittag-Leffler theorem for rational functions. Isotropy of  $\frak{g}(D)$ and
 $\frak{G}_{D}^{+}$
means that the inner product restricted to these subspaces is identically zero; this
condition assures that the associated classical r-matrix is skew symmetric.
For $\frak{G}_{D}^{+}$ this assertion is immediate, since (for $z\neq\infty$)
the product of two
Taylor
series has zero residue; for $z=\infty$ the residue disappears because of the
normalization condition. The isotropy of $\frak{g}(D)$ is a reformulation of the classical
theorem: the sum of residues of a rational function is zero.

Since $\frak{G}_{D}^{+}$ is a direct sum of local algebras, its coadjoint
orbits are direct products of the coadjoint orbits of each local factor; it
is easy to see that the coadjoint orbits of the local algebra $\frak{G}%
_{z_{j}}^{+}$ are modelled on rational functions with a single pole at $%
z=z_{j}$;  moreover, the subspace of rational functions with prescribed order
of singularity at this point is stable under the coadjoint action of $\frak{G%
}_{z_{j}}^{+}$.   Clearly, this subspace is finite-dimensional, which proves
(iv).

\begin{exercise} %{exercise}
Describe coadjoint orbits in the subspace of functions admitting only simple
poles.
\end{exercise}   %{exercise}

Our main theorem immediately applies in this setting and provides an ample
set of integrals of motion in involution for Lax equations with rational Lax
matrix.

\begin{remark}
(1) Since we are interested only in Lax operators which are global rational
functions on the Riemann sphere, we consider only coadjoint orbits of $\frak{%
G}_{D}^{+}\subset \left( \frak{G}_{D}\right) _{r}$; \ this is legitimate,
since $\left( \frak{G}_{D}\right) _{r}$\ splits into direct sum of two
complementary subalgebras,
\[
\left( \frak{G}_{D}\right) _{r}\simeq \frak{G}_{D}^{+}\oplus \frak{g}(D),
\]
and hence its orbits are direct products of orbits lying in $\frak{g}(D)$\
and in $\frak{G}_{D}^{+}$;  in other words, we take orbits which project into
zero in $\frak{G}_{D}^{+}$.  \newline
\ (2) The global algebra $\frak{G}_{D}^{+}$\ is decomposed into direct sum
of local factors, $\oplus _{z_{j}\in D}\frak{G}_{z_{j}}^{+}$;  coadjoint
orbits of each local algebra are the same that we encountered for the
ordinary loop algebra. What makes things different, is the way these orbits
are embedded into the bigger algebra; this embedding affects the choice of
the invariant Hamiltonians as well as the formulation of the factorization
problem.
\end{remark}

The use of formal series is well adapted for the study of coadjoint orbits
in $\frak{g}(D)$;  in order to be able to define Lie groups associated with
our Lie algebras, we must change the topology by replacing formal series
with convergent ones. Let ${\mathcal{G}}_{z_{j}}^{W}$ be the group of germs of
functions with values in $G$ which are regular in some punctured disc around
$z_{j}\in {\C}P_{1}$ (with topology of uniform absolute convergence), $(%
{\mathcal{G}}_{z_{j}}^{+})^{W}\subset {\mathcal{G}}_{z_{j}}^{W}$ its subgroup
consisting of functions regular in the entire small disc, and $G(D)$ the
group consisting of holomorphic mappings $\Bbb{C}P_{1}\backslash
D\rightarrow G$.   The infinitesimal decomposition described in proposition
\ref{mittag} (i) corresponds to the following multiplicative problem:

\emph{Given a set of local meromorphic functions }$g_{1}$,   \emph{\ }$%
...,g_{N} $\emph{, }$g_{j}\in \mathcal{G}_{z_{j}}^{W}$,   \emph{\ find a global
meromorphic function }$g_{0}$\emph{\ which is regular in the punctured
sphere }${\C}P_{1}\backslash D$\emph{\ such that }$g_{0}g_{j}^{-1}$\emph{%
\ is regular in some small disc around }$z_{j}$.

This is the standard \emph{multiplicative Cousin problem}; its geometrical
meaning is the same as for the matrix Riemann problem discussed above: it
corresponds to the trivialization of a vector bundle over ${\C}P_{1}$
(defined with respect to a different covering of the sphere).

\begin{exercise} %{exercise}
Reformulate the global factorization theorem in this setting.
\end{exercise}   %{exercise}

One is of course tempted to generalize the above construction replacing $%
{\C}P_{1}$ with an arbitrary Riemann surface. There is an obvious
obstruction which comes from the \emph{Mittag-Leffler theorem for curves: }a
global meromorphic function on a curve $\Gamma $ with prescribed principal
parts exists if and only if these principal parts satisfy a set of linear
constraints; roughly, the sum of residues
\[
\sum_{z_{j}\in D}%
{\mathrm {Res}}\,%
_{z_{j}}%
{\mathrm {tr}}\,
X_{j}\omega
\]
must be zero for all holomorphic differentials $\omega \in H^{1}(\Gamma
)\otimes \frak{g.}$ The trouble is that the constrained data do not form a
Lie subalgebra inside the global algebra $\frak{G}_{D}$,    and hence one
cannot find a complement of $\frak{G}_{D}^{+}$ which is a Lie subalgebra.
When $\Gamma $ is elliptic, this obstruction may be overcome by imposing
additional automorphy conditions, see e.g. \cite{RS}.

%%%%%%%%%%%%%%%%%%%%%%%%%%%%%%%%%%%%%%%%%%%%%%%%%%%%%%%%%%%%%%%%%%%%%%%%%%%

\newsection{Zero Curvature Equations}

In applications to integrable PDE's, Lax matrices are replaced by first
order matrix differential operators. The systematic treatment of these
applications is based on the use of \emph{double loop algebras}, or, more
precisely, of their central extensions. Let us start with discussion of the
central extension of the ordinary loop algebra.

%%%%%%%%%%%%%%%%%%%%%%%%%%%%%%%%%%%%%%%%%%%%%%%%%%%%%%%%%%%%%%%%%%%%%%%%%%%%%%

\newsubsection{Central Extensions of Loop Algebras}

Set $\frak{g}=\frak{gl}(n)$ and let $\frak{G}=C^{\infty }\left( S^{1};\frak{g%
}\right) $ be the Lie algebra of smooth functions on the circle with values
in $\frak{g}$ and with pointwise commutator. (Mind that in the present
setting we choose topology in our loop algebra in a different way! This is
because we are willing to treat functions of $x\in S^{1}$ as dynamical
variables for our future evolution equations.) We equip $\frak{G}$ with the
invariant inner product
\begin{equation}
\left\langle X,Y\right\rangle =\int_{0}^{2\pi }%
{\rm tr}\,
XYdx;  \label{int}
\end{equation}
accordingly, we get an embedding $\frak{G}\subset \frak{G}^{*}$ which
defines the \emph{smooth dual} of $\frak{G}$.   Put
\begin{equation}
\omega \left( X,Y\right) =\int_{0}^{2\pi }%
{\rm tr}\,
X\cdot \frac{dY}{dx}dx;  \label{maur}
\end{equation}

\begin{exercise}  %{exercise}
$\omega $ is a skew symmetric bilinear form which satisfies the cocycle
condition
\begin{equation}
\omega \left( \left[ X,Y\right] ,Z\right) +\omega \left( \left[ Y,Z\right]
,X\right) +\omega \left( \left[ Z,X\right] ,Y\right) =0.  \label{cocycl}
\end{equation}
\end{exercise}  %{exercise}

Put $\widehat{\frak{G}}=\frak{G}\oplus {\C}$ and define the Lie bracket
in $\widehat{\frak{G}}$ by
\[
\left[ \left( X,c\right) ,\left( Y,c^{\prime }\right) \right] =\left( \left[
X,Y\right] ,\omega \left( X,Y\right) \right) ,\;X,Y\in \frak{G}%
,\;c,c^{\prime }\in \Bbb{C}.
\]
(The Jacobi identity is equivalent to \reff{cocycl}.) Notice that $\frak{c}%
=\{\left( 0,c\right) ;c\in \Bbb{C}\}\subset \widehat{\frak{G}}$ is central
in $\widehat{\frak{G}}$ and $\frak{G}$ may be identified with the quotient
algebra, $\frak{G}=\widehat{\frak{G}}/\frak{c}$.   The (smooth) dual of $%
\widehat{\frak{G}}$ may be identified with $\frak{G}\oplus {\R}$.

\begin{proposition}
The coadjoint representation of $\widehat{\frak{G}}$ is given by
\begin{equation}
ad^{*}X\cdot \left( L,e\right) =\left( \left[ X,L\right] +e\partial
_{x}X,0\right) .  \label{coad}
\end{equation}
(Note that this representation is trivial on the center $\frak{c}\subset
\widehat{\frak{G}}$ and therefore it may be regarded as a representation of $%
\frak{G}.)$
\end{proposition}

Let $\Bbb{G}=C^{\infty }\left( S^{1};G\right) $ be the Lie group associated
with the Lie algebra $\frak{G}$.   The coadjoint representation of $\frak{G}$
in $\widehat{\frak{G}}^{*}$ may easily be integrated to a representation of $%
\Bbb{G}$.

\begin{proposition}
\label{Ad}The coadjoint representation of $G$ in $\widehat{\frak{G}}^{*}$ is
given by
\begin{equation}
Ad\,
^{*}g\cdot \left( L,e\right) =\left( gLg^{-1}+e\partial _{x}g\cdot
g^{-1},e\right) .  \label{Coad}
\end{equation}
\end{proposition}

Proposition \rref{Ad} admits a remarkable geometric interpretation. Consider
the \emph{auxiliary linear differential equation}
\begin{equation}
e\frac{d\psi }{dx}=L\psi  \label{aux}
\end{equation}
(we regard it as a differential equation on the line with periodic potential
$L$).

\begin{exercise}  %{exercise}
Coadjoint representation (\ref{Coad}) leaves invariant the hyperplanes $%
e=const$ in $\widehat{\frak{G}}^{*}$;  on each hyperplane $e=const\neq 0$ it
is equivalent to the gauge transformation of the potential $L$ in the linear
equation (\ref{aux}) induced by the natural action of $\Bbb{G}=C^{\infty
}\left( S^{1};G\right) $ on its solutions by left multiplication, $g:\psi
\mapsto g\cdot \psi $.
\end{exercise}  %{exercise}

Let $\psi _{0}\in C^{\infty }\left( {\R};G\right) $ be the fundamental
solution of (\ref{aux}) normalized by $\psi _{0}\left( 0\right) =id$;  $%
T_{L}=\psi _{0}\left( 2\pi \right) \in G$ is called the \emph{monodromy matrix} of $%
L$.

\begin{theorem}
\emph{(Floquet)} Two potentials $L,L^{\prime }\in C^{\infty }\left( S^{1};\frak{g}%
\right) $ lie on the same coadjoint orbit in $\widehat{\frak{G}}^{*}$ (with
fixed $e\neq 0)$ if and only $T_{L}$ and $T_{L^{\prime }}$ are conjugate in $%
G$.
\end{theorem}

\emph{Sketch of a proof.} Without restricting the generality we may assume
that $T_{L}=T_{L^{\prime }}$.   Let $\psi _{L},\psi _{L^{\prime }}$ be the
fundamental solutions normalized by $\psi _{L}\left( 0\right) =\psi
_{L^{\prime }}\left( 0\right) =id$;  put $g(x)=\psi _{L}\left( x\right) \psi
_{L^{\prime }}\left( x\right) ^{-1}$.   Clearly, $g$ is $2\pi $-periodic on
the line and $g\cdot \psi _{L^{\prime }}=\psi _{L}$,    hence $%
Ad\,
^{*}g\cdot L^{\prime }=L$.

\begin{corollary}
All coadjoint orbits lying in the hyperplanes $e=const\neq 0$ have finite
codimension; the ring of Casimir functions is generated by spectral
invariants of the monodromy.
\end{corollary}

The Hamiltonian mechanics in $\widehat{\frak{G}}^{*}$ may be defined with
the help of the elementary calculus of variations. Let $\varphi \left[
L\right] $ be a smooth functional of the potential $L,X_{\varphi }=\mathrm{%
grad}\varphi \left[ L\right] \in \frak{G}$ its Frechet derivative defined by
\[
\frac{d}{ds}\varphi \left[ L+s\eta \right] =\int_{0}^{2\pi }%
{\mathrm {tr}}\,
X_{\varphi }\left( x\right) \eta \left( x\right) dx.
\]
The Lie-Poisson bracket of two functionals $\varphi _{1},\varphi _{2}$ is
given by
\[
\left\{ \varphi _{1},\varphi _{2}\right\} \left[ L\right] =\int_{0}^{2\pi }%
{\mathrm {tr}}\,
\left( L\left( x\right) \left[ X_{\varphi _{1}}\left( x\right) ,X_{\varphi
_{2}}\left( x\right) \right] +eX_{\varphi _{1}}\partial _{x}X_{\varphi
_{2}}\right) dx.
\]

\begin{proposition}
The Hamiltonian equation of motion on $\widehat{\frak{G}}^{*}$ with
Hamiltonian $\varphi $ is equivalent to the following differential equation
for the potential $L:$%
\begin{equation}
\frac{\partial L}{\partial t}=-\left[ X_{\varphi },L\right] -e\frac{\partial
X_{\varphi }}{\partial x}.  \label{zero}
\end{equation}
\end{proposition}

Equation  \reff{zero} has a nice geometrical meaning. Let us consider
 the $\frak{g}$-valued differential form
\[
Ldx+X_{\varphi }dt;
\]
 it may be regarded as a connection form of a  connection
on ${\R}^{2}$ (with values in $\frak g$); equation \reff{zero} then means that this connection
 has zero curvature (hence the term ``zero curvature
equation''.) We would like to use the central extension $\widehat{\frak{G}}$
as a building block to construct integrable equation; there are already two
reassuring points:

\begin{enumerate}
\item  The description of coadjoint orbits in $\widehat{\frak{G}}^{*}$
automatically leads to the auxiliary linear problem \reff{aux}.

\item  Equations of motion associated with $\widehat{\frak{G}}$ are zero curvature
equations, as desired.
\end{enumerate}

\noindent However, there are also two major drawbacks:

\begin{enumerate}
\item  There is only a finite number of independent Casimirs (one can take
e.g. the coefficients of the characteristic polynomial $\det \left(
T_{L}-\lambda \right) )$.

\item  The Casimirs are highly nonlocal functionals of the potential.
\end{enumerate}

\noindent By contrast, in order to get integrable PDE's we need an \emph{infinite} number of
conservation laws; these conservation laws are usually expressed as
integrals of\emph{\ local densities} which are polynomial in the matrix
coefficients of the potential $L$ and its derivatives in $x$.   The way to
resolve these difficulties is suggested by the auxiliary linear equation (%
\ref{aux}): in order to characterize the potential, we need to know the
monodromy \emph{for all energies}; in other words, we must introduce into
\reff{aux} spectral parameter. Algebraically, this means that we have to
modify the choice of our basic Lie algebra.

%%%%%%%%%%%%%%%%%%%%%%%%%%%%%%%%%%%%%%%%%%%%%%%%%%%%%%%%%%%%%%%%%%%%%%%%%%%%

\newsubsection{Double Loop Algebras}

Let us put $\frak{G}=C^{\infty }\left( S^{1};\frak{g}\right) $ as before;
let ${\mathbf{g=}} \frak{G}\otimes {\C}\left( \left( z\right) \right) $ be
the algebra of formal Laurent series with values in $\frak{G}$.   (In other
words we can say that $\mathbf{g}$ is the \emph{double loop algebra} of $\frak{g}$;
the different roles of the variables $x,z$ are imposed by our choice of its
central extension.) We equip $\mathbf{g}$ with the inner product
\begin{equation}
\left\langle X,Y\right\rangle =%
{\mathrm {Res}}\,%
_{z=0}\int
{\mathrm {tr}}\,
X(x,z)Y\left( x,z\right) dx=\frac{1}{2\pi i}\int
{\mathrm {tr}}\,
X(x,z)Y\left( x,z\right) dxdz.  \label{restr}
\end{equation}

The 2-cocycle $\omega $ on $\mathbf{g}$ is defined by
\begin{equation}
\omega \left( X,Y\right) =\left\langle X,\frac{dY}{dx}\right\rangle .
\label{omega}
\end{equation}

Let $\widehat{\mathbf{g}}$ be the central extension of $\mathbf{g}$ defined
by this cocycle.\footnote{\label{ref}In the case of the loop algebra the
definition of the 2-cocycle $\omega $ is essentially unique; for the double
loop algebra there is a possibility to introduce into \reff{omega} a weight
factor $\phi (z)$ which does not depend on $x$;  this weight factor will
modify the auxiliary linear problem. This freedom is useful in applications;
our choice in \reff{omega} is the simplest one possible.} As before, we may
identify the dual of $\widehat{\mathbf{g}}$ with ${\mathbf{g}}\oplus \Bbb{C}$.
Formula \reff{coad} for the coadjoint representation remains valid. We
conclude that the coadjoint representation for $\widehat{\mathbf{g}}$
coincides with (infinitesimal) gauge transformation associated with the
auxiliary linear problem \reff{aux}, where this time $L\in \mathbf{g,}$
i.e., it is a formal series in $z$ with coefficients in $\frak{G}$.   In other
words, $z$ plays the role of spectral parameter in the auxiliary linear
problem, as desired. There are some troubles with the definition of the
associated Lie group, but let us ignore them for the moment. Notice that if $%
L$ is a polynomial in $z,z^{-1}$,    the monodromy matrix $T_{L}$ is a
well-defined analytic function of $z$ (with values in $G=GL\left( n\right) $%
) which is holomorphic in ${\C}\backslash \left\{ 0\right\} $.

Our choice of the basic Lie algebra makes it easy to define the other key
element of our scheme, the classical r-matrix. Set ${\mathbf{g}}_{+}=\frak{G}%
\otimes {\C}\left[ \left[ z\right] \right] ,{\mathbf{g}}_{-}=\frak{G}%
\otimes z^{-1}{\C}\left[ z^{-1}\right] $.   Clearly, ${\mathbf{g}}={\mathbf{g}}%
_{+}{\dot{+}}{\mathbf{g}}_{-}$ as a linear space. Both subalgebras $%
{\mathbf{g}}_{+}$ and ${\mathbf{g}}_{-}$ are isotropic with respect to the inner
product \reff{restr} which sets them into duality. As before, we put
\begin{equation}
r=P_{+}-P_{-},  \label{skew}
\end{equation}
where $P_{+},P_{-}$ are the projection operators onto ${\mathbf{g}}_{+}$ and $%
{\mathbf{g}}_{-}$,    respectively, and define the r-bracket on $\mathbf{g}$ by $%
\left[ X,Y\right] _{r}=\frac{1}{2}\left( \left[ rX,Y\right] +\left[
X,rY\right] \right) $.   In this way we get the algebra ${\mathbf{g}}_{r}$,    but
this is still not quite what we need to apply our main theorem, since our
basic algebra is $\widehat{\mathbf{g}}$,    not $\ {\mathbf{g}!}$ As it happens,
the theorem survives central extension.

\begin{lemma}
\label{or}Let $\omega $ be a 2-cocycle on $\mathbf{g}\ $ and $r\in End\,%
\mathbf{g}$ a linear operator which satisfies the modified Yang-Baxter
identity. Set
\[
\omega _{r}\left( X,Y\right) =\frac{1}{2}\left( \omega \left( rX,Y\right)
+\omega \left( X,rY\right) \right) .
\]
Then $\omega _{r}$ is a 2-cocycle on $\mathbf{g}_{r}$.
\end{lemma}

\begin{exercise}  %{exercise}
Prove lemma \ref{or} (the proof is abstract and uses only manipulation with
the Jacobi and the Yang-Baxter identities).
\end{exercise}  %{exercise}

Let $\widehat{\mathbf{g}_{r}}$ be the central extension of ${\mathbf{g}}_{r}$
associated with $\omega _{r}$.   It is easy to see that $\left( \widehat{%
\mathbf{g}},\widehat{{\mathbf{g}}_{r}}\right) $ is a double Lie algebra and we
may apply our main idea: \emph{Casimirs of }$\widehat{\mathbf{g}}$\emph{\
regarded as Hamiltonians with respect to the Lie-Poisson bracket of }$%
\widehat{{\mathbf{g}}_{r}}$\emph{\ give rise to generalized Lax equations. }%
Actually, there is one more simplification, which is due to our choice of $%
\omega $ (see \reff{omega}):

\begin{exercise}  %{exercise}
Suppose that $r\in End\,\mathbf{g}$ is skew symmetric with respect to the
inner product on $\mathbf{g}$;  then $\omega _{r}=0$.
\end{exercise}  %{exercise}

The r-matrix \reff{skew} clearly satisfies this condition; hence the
algebra $\widehat{{\mathbf{g}}_{r}}={\mathbf{g}}_{r}\oplus {\C}$ splits and
the Lie-Poisson brackets for $\widehat{{\mathbf{g}}_{r}}$ and ${\mathbf{g}}_{r}$
coincide. Explicitly, this means that the Poisson bracket of two smooth
functionals $\varphi _{1},\varphi _{2}$ defined on ${\mathbf{g}}_{r}^{*}\simeq
\mathbf{g}$ is given by
\begin{equation}\label{local}
\left\{ \varphi _{1},\varphi _{2}\right\} _{r}\left[ L\right] =\frac{1}{2\pi
i}\int \mathrm{tr\,}\left( \left[ \mathrm{grad\,}\varphi _{1},\mathrm{grad\,}%
\varphi _{1}\right] _{r}\left( x,z\right) \cdot L\left( x,z\right) \right)
dxdz,  \label{varbr}
\end{equation}
where ${\mathrm{grad\,}}\varphi _{i}\left[ L\right] \left( x,z\right) \in
\mathbf{g}$, $i=1,2$,    is the Frechet derivative.

\begin{remark}
The skew symmetry of $r$\ makes the above discussion of the cocycle $\omega
_{r}$\ void; however, as already noticed (see footnote \rref{ref}), we may
modify the cocycle $\omega $\ by a weight factor $\phi \left( z\right) $,   \
and in that case our Poisson bracket will contain derivatives $\partial _{x}$%
\ of the gradients.\footnote{%
Poisson bracket of functionals which does not contain derivatives $\partial
_{x}$ of the gradients is sometimes called \emph{ultralocal}; in more
complicated cases, Poisson brackets may contain derivatives \emph{%
(non-ultralocal case) }or even be non-local, i.e., contain integral
operators.}
\end{remark}

The antisymmetry of $r$ makes the description of coadjoint orbits very
simple. In the absence of cocycle we must deal with the orbits of $\mathbf{g}%
_{r}$;  note that the r-matrix \reff{skew} is acting only on the variable $z$
in the double loop algebra, and hence the other variable $x$ becomes a
parameter. Let us consider the `little' loop algebra $L\frak{g}=\frak{g}%
\otimes {\C}\left( \left( z\right) \right) $ and the associated algebra $L%
\frak{g}_{r}$ which we have already discussed in Section 6. The
`big' algebra ${\mathbf{g}}_{r}$ consists of smooth periodic functions on the
line with values in $L\frak{g}_{r}$ and with pointwise commutator.

\begin{proposition}
Fix an orbit ${\mathcal{O}}\subset \left( L\frak{g}_{r}\right) ^{*}$;  then $%
{\mathbf{O}}=Map(S^{1},{\mathcal{O}})$ is an orbit of ${\mathbf{g}}_{r}$.
\end{proposition}

More generally, we may vary orbits $\mathcal{O}$ lying over different points
of $S^{1}$ (i.e., consider smooth families of orbits in $\left( L\frak{g}%
_{r}\right) ^{*}$ parametrized by $S^{1}$).

\begin{example}
\label{Heis} Let $\frak{g}=sl\left( 2\right) $;  then the matrices $s\in
\frak{g}$,   %
\begin{equation}
s=\left(
\begin{array}{ll}
s_{3} & s_{1}+is_{2} \\
s_{1}-is_{2} & -s_{3}
\end{array}
\right) ,s_{j}\in \Bbb{C},s_{1}^{2}+s_{2}^{2}+s_{3}^{2}=1,  \label{spin}
\end{equation}
form a coadjoint orbit ${\mathcal{S}}_{1}\subset sl(2)$.   Check that ${\mathcal{O%
}}_{H}=\left\{ z^{-1}\ s,s\in {\mathcal{S}}_{1}\right\} \subset L\frak{g}%
\simeq L\frak{g}^{*}$ is a coadjoint orbit of $L\frak{g}_{+}\subset L\frak{g}%
_{r}$.   The corresponding orbit $\mathbf{O}_{H}\subset \mathbf{g}$ is
parametrized by a triple of $2\pi $-periodic functions $s_{j},j=1,2,3$,
satisfying the constraint \reff{spin}. The associated linear differential
operator is
\begin{equation}\label{wh}
\frac{d}{dx}-\frac{1}{z}s\left( x\right) .
\end{equation}
One can show that the simplest local Hamiltonian associated with \reff{wh} is
\[
H=-\frac{1}{2}\mathrm{tr}\int s_x^2\, dx;
\]
the corresponding nonlinear equation describes the Heisenberg ferromagnet:
\[
s_t=\left[s,s_{xx}\right].
\]
\end{example}

\begin{example}
\label{Schroed}Set
\[
\sigma =\left(
\begin{array}{ll}
1 & 0 \\
0 & -1
\end{array}
\right) .
\]
The matrices
\[
U=\left(
\begin{array}{ll}
0 & u \\
v & 0
\end{array}
\right) +z\sigma ,u,v\in \Bbb{C},
\]
form a coadjoint orbit of the Lie algebra $L\frak{g}_{-}\subset L\frak{g}%
_{r} $.   The corresponding orbit ${\mathbf{O}}_{S}\subset {\mathbf{g}}$ is
parametrized by a pair of functions $u,v\in C^{\infty }(S^{1}, \C )$;  the
associated linear operator is
\[
\frac{d}{dx}-U;
\]
this is (essentially) the Lax operator for the so called split nonlinear Schroedinger
equation\footnote{%
For simplicity, in both examples we deal with complex Lie algebras; in
order to pass to a real form we must choose an {\em anti-involution} of
our basic loop algebra; this is of course a necessary step in order to
make the auxiliary linear operators genuine skew selfadjoint operators.}
\[
i\partial_t u=-u_{xx}+u^2v,\\
-i\partial_t v=-v_{xx}+v^2u.
\]
\end{example}

In a more general way, for $\frak{g}=\frak{gl}(n)$ we can easily construct
coadjoint orbits associated with linear differential operators of the form
\begin{equation}
\frac{d}{dx}-U\left( x,z\right) ,  \label{lin}
\end{equation}
where the potential $U$ is a Laurent polynomial in $z$.

%%%%%%%%%%%%%%%%%%%%%%%%%%%%%%%%%%%%%%%%%%%%%%%%%%%%%%%%%%%%%%%%%%%%%%%%%%%%%

\newsubsection{Monodromy Map}

Examples \rref{Heis}, \rref{Schroed} show that our approach is indeed
reasonable: starting with double loop algebras, we arrived at the natural
auxiliary linear problems with spectral parameter of the form which is
familiar in spectral theory. The next question is to construct an
appropriate class of Hamiltonians. Formally, the Hamiltonians are spectral
invariants of the auxiliary linear operator, i.e., the spectral invariants
of its monodromy matrix. The monodromy matrix $M_{U}$ of \reff{lin} is a
holomorphic function in ${\C}P_{1}\backslash \left( \left\{ 0\right\}
\cup \left\{ \infty \right\} \right) $;  any functional of the monodromy
which is invariant under conjugation gives rise to a zero curvature equation
on the coadjoint orbits of ${\mathbf{g}}_{r}$.   The mapping ${\mathbf{M}}%
:U\rightsquigarrow M_{U}$ is the \emph{direct spectral transform} for
\reff{lin}. We may regard $\mathbf{M}$ as a mapping from $\mathbf{g}$ into the
group of matrix-valued functions which are regular in the punctured Riemann
sphere. When the Poisson structure is \emph{ultralocal} (i.e., the r-matrix
is skew with respect to the inner product in $\mathbf{g}$), the spectral
transform $\mathbf{M}$ has an important property: the target space carries a
natural Poisson bracket and $\mathbf{M}$ is a Poisson mapping (i.e., it
preserves Poisson brackets). We shall briefly outline this computation,
since it plays an important role in the theory and brings into play an
important class of Poisson brackets (the so called \emph{Sklyanin brackets}%
). To simplify the notation, let us start with ordinary loop algebras.

Let $\frak{g}=\frak{gl}(n)$,    and $r\in {\mathrm {End}}\,%
\frak{g}$ a linear operator which satisfies the modified Yang-Baxter
identity \reff{mcybe} and is skew symmetric with respect to the inner
product on $\frak{g}$.   Let $\frak{G}=C^{\infty }(S^{1},\frak{g})$;  we equip $%
\frak{G}$ with the inner product \reff{int}; the Poisson bracket of
functionals on $\frak{G}^{*}\simeq \frak{G}$ is given by
\begin{equation}
\left\{ \varphi _{1},\varphi _{2}\right\} _{r}\left[ L\right]
=\int_{0}^{2\pi }%
{\rm tr}\,
\left( \left[
{\mathrm {grad}}\,%
\varphi _{1}\left[ L\right] \left( x\right) ,%
{\mathrm {grad}}\,%
\varphi _{2}\left[ L\right] \left( x\right) \right] _{r}L\left( x\right)
\right) dx.  \label{rbrack}
\end{equation}

For $L\in \frak{G}$ let $\psi _{L}$ be the fundamental solution of \reff{aux}
normalized by $\psi _{L}(0) =id$ and $M_{L}\in GL(n)$ the monodromy matrix.
Fix a smooth function $\varphi \in
C^{\infty }\left( GL(n) \right) $ and put $h_{\varphi }\left[
L\right] =\varphi \left( M_{L}\right) $.   Let $\nabla \varphi ,\nabla
^{\prime }\varphi \in \frak{g}$ be the \emph{left} and \emph{right gradients} of $\varphi $
on $G=GL\left( n\right) $ defined by
\[
\frac{d}{ds}_{s=0}\varphi (e^{sX}x)={\mathrm{tr\,}}\left( X\cdot \nabla
\varphi \left( x\right) \right) ,\;\frac{d}{ds}_{s=0}\varphi (xe^{sX})=%
{\mathrm {tr}}\,
\left( X\cdot \nabla ^{\prime }\varphi \left( x\right) \right) ,X\in \frak{g}%
.
\]

\begin{lemma}
The Frechet derivative of the functional $h_{\varphi}$ is given by
\begin{equation}
{\mathrm {grad}}\,%
h_{\varphi }\left[ L\right] (x)=\psi \left( x\right) \nabla
^{\prime }\varphi \left( M_{L}\right) \psi \left( x\right) ^{-1}.
\label{varder}
\end{equation}
\end{lemma}

\begin{corollary}
The Frechet derivative satisfies the differential equation
\begin{equation}
\frac{dX}{dx}=\left[ L,X\right]  \label{grad}
\end{equation}
and the boundary conditions
\begin{equation}
X\left( 0\right) =\nabla ^{\prime }\varphi \left( M_{L}\right) ,X\left( 2\pi
\right) =\nabla \varphi \left( M_{L}\right) .  \label{bound}
\end{equation}
\end{corollary}

\emph{Sketch of a proof.} Taking variation of the both sides of \reff{aux},
we get
\begin{equation}
\partial _{x}\delta \psi _{L}=L\delta \psi _{L}+\left( \delta L\right)
\psi _{L}.  \label{v}
\end{equation}
Set $\delta \psi _{L}=\psi _{L}Y$,    where $Y$ is an unknown function, $Y\in
C^{\infty }\left( {\R},\frak{g}\right) ,Y\left( 0\right) =0$;  \reff{v}
yields $\partial _{x}Y=\psi ^{-1}\delta L\psi $,    whence
\[
Y\left( x\right) =\int_{0}^{x}\psi _{L}^{-1}\left( y\right) \delta L\left(
y\right) \psi _{L}\left( y\right) \,dy.
\]
Since $M_{L}=\psi _{L}\left( 2\pi \right) $,    we get
\[
M_{L}^{-1}\delta M_{L}=\int_{0}^{2\pi }\psi _{L}^{-1}\cdot \delta L\cdot
\psi _{L}\,dy.
\]
Now,
\[
\delta \varphi \left( M_{L}\right) =%
{\mathrm {tr}}\,
\left( \nabla ^{\prime }\varphi \left( M_{L}\right) \cdot M_{L}^{-1}\delta
M_{L}\right) = \int_{0}^{2\pi } %
{\mathrm {tr}} \,
\psi _{L}(y) \nabla ^{\prime }\varphi \left( M_{L}\right) \psi
_{L}^{-1}(y) \cdot \delta L(y) \,dy,
\]
which implies \reff{varder}. Taking derivatives of the both sides of
\reff{varder} yields \reff{grad}.

\begin{proposition}
The Poisson bracket of two functionals $h_{\varphi _{1}},h_{\varphi _{2}}$
on $\frak{G}_{r}^{*}$ is given by
\begin{equation}
\left\{ h_{\varphi _{1}},h_{\varphi _{2}}\right\} \left[ L\right]
=h_{\left\{ \varphi _{1},\varphi _{2}\right\} }\left[ L\right] ,  \label{b}
\end{equation}
where the Poisson bracket of $\varphi _{1},\varphi _{2}\in C^{\infty }\left(
G\right) $ is defined by
\begin{equation}
\left\{ \varphi _{1},\varphi _{2}\right\} _{G}=\frac{1}{2}%
{\mathrm {tr}}\,
\left( r\left( \nabla \varphi _{1}\right) \nabla \varphi _{2}-r\left( \nabla
^{\prime }\varphi _{1}\right) \nabla ^{\prime }\varphi _{2}\right) .
\label{skl}
\end{equation}
\end{proposition}

\begin{corollary}
Let us equip the group $G=GL\left( n\right) $ with the Poisson bracket
\reff{skl}. Then the monodromy map $\mathbf{M}:\frak{G}_{r}^{*}\rightarrow
G:L\rightsquigarrow M_{L}$ preserves Poisson brackets.
\end{corollary}

\begin{proof} Set $X_{i}=%
{\mathrm {grad}}\,%
h_{\varphi _{i}}$,    $i=1,2$.   We have
\[\arr{2.0}{ll}{\ds
\left\{ h_{\varphi _{1}},h_{\varphi _{2}}\right\} \left[ L\right]
=\int_{0}^{2\pi }%
{\mathrm {tr}}\,
\left( \left[ X_{1},X_{2}\right] _{r}L\right) dx \\
\qquad \ds =\frac{1}{2}\int_{0}^{2\pi }%
{\mathrm {tr}}\,
\left( \left[ rX_{1},X_{2}\right] +\left[ X_{1},rX_{2}\right] \right) L\,dx
\\
\qquad \ds =\frac{1}{2}\int_{0}^{2\pi }%
{\mathrm {tr}}\,
\left( \left[ L,X_{2}\right] rX_{1}+\left[ L,X_{1}\right] rX_{2}\right) \,dx
\\
\qquad \ds =\frac{1}{2}\int_{0}^{2\pi }\frac{d}{dx}%
{\mathrm {tr}}
\left( rX_{1}\cdot X_{2}\right) \,dx,}
\]
where we used the definition of the r-bracket, the cyclic invariance of
trace, the differential equation \reff{grad} satisfied by $X_{i}$ and,
finally, the skew symmetry of $r$.   Evaluating the last integral and taking
into account the boundary conditions \reff{bound} for $X_{i}$,    we get
\reff{b}.
\end{proof}

Formula \reff{skl} defines a remarkable Poisson bracket (\emph{the Sklyanin
bracket}) on the group manifold itself. The Jacobi identity for this bracket
is not obvious (though it follows from our computation). Its  properties
will be discussed in some detail in Section 9.

\begin{exercise}  %{exercise}
Show that central functions on $G$ (i.e., functions which satisfy $\varphi
\left( xy\right) =\varphi \left( yx\right) $ for all $x,y\in G)$ commute
with respect to the Sklyanin bracket.
\end{exercise}  %{exercise}

The above discussion applies to loop algebra $\frak{G}=C^{\infty }(S^{1},%
\frak{g})$;  the generalization to the double loop algebra is
straightforward: in our computation, we must replace the finite dimensional
algebra $\frak{g}$ with its loop algebra $L\frak{g}$;  accordingly, smooth
functions $\varphi _{1},\varphi _{2}\in C^{\infty }\left( G\right) $ are
replaced by \emph{smooth functionals} on the corresponding loop group, their
left and right gradients are replaced by the left and right variational
derivatives, etc. Spectral invariants of the auxiliary linear problem
\reff{lin} correspond to \emph{central functionals} on the loop group.
An example of such a functional is given by \emph{evaluation functionals}
$H_{n,w}\left[ U\right] =%
{\mathrm {tr}}\,
M_{U}^{n}(w), w\in \Bbb{C}P_{1}\backslash \left( \left\{ 0\right\} \cup
\left\{ \infty \right\} \right) $.

\begin{exercise}  %{exercise}
Compute the variational derivative of $H_{n,w}$ with respect to \ $U$.
\end{exercise}  %{exercise}

The fundamental drawback of these functionals is, however, their
\emph{nonlocality. }The description of \emph{local} functionals is outlined in the
next paragraph.

%%%%%%%%%%%%%%%%%%%%%%%%%%%%%%%%%%%%%%%%%%%%%%%%%%%%%%%%%%%%%%%%%%%%%%%%%%%%%

\newsubsection{Formal Diagonalization and Local Conservation Laws}

In contrast to evaluation functionals, local conservation laws are related
to the asymptotic expansions of the monodromy matrix at its essential
singularities, i.e., for $z=0,\infty $.   This implies some additional
difficulties:

\begin{enumerate}
\item  These functionals are not defined everywhere on the double loop
algebra.

\item  The associated formal series are in most cases divergent.
\end{enumerate}

Let us assume that the potential $U\left( x,z\right) $ in the auxiliary
linear problem \reff{lin} is a Laurent polynomial,
\[
U=\sum_{-N}^{M}U_{k}z^{k},U_{k}\in C^{\infty }\left( S^{1},\frak{gl}%
(n)\right) ;
\]
let $J_{0}=U_{-N},J_{\infty }=U_{M}$ be its lowest and highest coefficients..

\begin{definition}
\label{reg}$U$ is called \emph{regular} if
\begin{enumerate}
 \renewcommand{\theenumi}{\roman{enumi}} \renewcommand{\labelenumi}{(%
\theenumi)}\item the matrices $J_{0}\left( x\right) ,J_{\infty }\left( x\right) $
are semisimple,

\item  the centralizers of $J_{0}(x),J_{\infty }(x)$ in $\frak{g}=\frak{gl}%
(n)$ are conjugate for all $x\in S^{1}$.
\end{enumerate}
\end{definition}

We have seen that the set of Laurent polynomials of fixed degree is a
Poisson subspace for the r-bracket. It is easy to check that the regularity
condition holds for entire coadjoint orbits in this subspace, and hence it
is a characteristic of our phase space. Our next theorem allows to construct
for regular Lax operators two series of local Hamiltonians which are
associated with the poles of $U$ on ${\C}P_{1}$.   For concreteness, we
shall describe the construction of the series associated with the pole at
infinity. Performing a suitable gauge transformation we may assume that the
leading coefficient at infinity $J_{\infty }(x)$ satisfies a stronger
condition:

\emph{(ii}$^{\prime }$\emph{) the centralizer of }$J_{\infty }$\emph{\ in }$%
\ \frak{g}=\frak{gl}(n)$\emph{\ is a fixed subalgebra }$\frak{g}^{J_{\infty
}}\subset \frak{g}$\emph{\ which does not depend on }$x$.

(Note that since, by construction, local Hamiltonians are gauge invariant,
this stronger condition does not restrict generality.) Let $\frak{g}%
_{J_{\infty }}\subset \frak{g}^{J_{\infty }}$ be the commutant of $\frak{g}%
^{J_{\infty }}$,
\[
\frak{g}_{J_{\infty }}=\left\{ X\in \frak{g};\left[ X,Y\right] =0\hbox{ for
all }Y\in \frak{g}^{J_{\infty }}\right\} .
\]

\begin{theorem}
\label{NF}(On normal form at infinity) There exists a formal gauge
transformation
\[
\Phi ^{\infty }=Id+\sum_{m=1}^{\infty }\Phi _{m}z^{-m},\Phi _{m}\in
C^{\infty }\left( S^{1},Mat\left( n\right) \right) ,
\]
which transforms the differential operator $\partial _{x}-U$ into normal
form,
\begin{equation}
\left( \Phi ^{\infty }\right) ^{-1}\circ \left( \frac{d}{dx}-U\right) \circ
\Phi ^{\infty }=\frac{d}{dx}-D^{\infty },  \label{inter}
\end{equation}
where
\[
D^{\infty }=\sum_{m=-M}^{\infty }D_{m}^{\infty }z^{-m},\,D_{m}^{\infty }\in
C^{\infty }\left( S^{1},\,\frak{g}^{J_{\infty }}\right) ,\,D_{-M}^{\infty
}=J_{\infty };
\]
matrix coefficients of $\Phi _{m}^{\infty },\,D_{m}^{\infty }$ are expressed
as polynomials of the coefficients of $U$ and its derivatives in $x$.  %
\footnote{%
In various applications $J_{0},J_{\infty }$ are regular matrices with
distinct eigenvalues. In that case $\frak{g}^{J_{0}}=\frak{g}_{J_{0}}$ and $%
\frak{g}^{J_{\infty }}=\frak{g}_{J_{\infty }}$ are abelian subalgebras;
hence theorem \ref{NF} means that the potential $U$ may be transformed to
\emph{diagonal form} by a formal gauge transformation. When the eigenvalues
of $J_{0},J_{\infty }$ have multiplicities, the potential may be transformed
only to \emph{block diagonal form.}}
\end{theorem}

\emph{Sketch of a proof.} The intertwining relation $\left( \ref{inter}%
\right) $ is equivalent to the differential equation
\begin{equation}
\left( \frac{d}{dx}-U\right) \Phi ^{\infty }=-\Phi ^{\infty }D^{\infty };
\label{entre}
\end{equation}
which may be solved recurrently in powers of the local parameter $z^{-1}$.
The first nontrivial coefficients $\Phi _{1},D_{-M+1}$ satisfy
\begin{equation}
J_{\infty }\Phi _{1}^{\infty }-\Phi _{1}^{\infty }J_{\infty
}=D_{-M+1}^{\infty }-U_{M-1}.  \label{recur}
\end{equation}
This equation for $\Phi _{1}$ admits a solution if and only if the r.h.s. is
in the image of $ad\,J_{\infty }\in
End\,%
\frak{g}$.   Assumption (i) above implies that
\[
\frak{g}=\mathrm{Im}\,ad\,J_{\infty }{\dot{+}}\ker ad\,J_{\infty },
\]
moreover, by (ii$^{\prime }$) $\mathrm{Im}\,ad\,J_{\infty }$ and $\ker
ad\,J_{\infty }=\frak{g}^{J_{\infty }}$ do not depend on $x$.   Hence, $%
D_{-M+1}^{\infty }\in \frak{g}^{J_{\infty }}$ is uniquely determined from
the solvability condition of $\left( \ref{recur}\right) $ and
\[
\Phi _{1}^{\infty }=\left( ad\,J_{\infty }\right) ^{-1}\left(
D_{-M+1}^{\infty }-U_{M-1}\right) .
\]
If the coefficients $\Phi _{1}^{\infty },...,\Phi _{m}^{\infty
},D_{-M+1}^{\infty },...,D_{-M+m}^{\infty }$ are already determined, we get
for $\Phi _{m+1}^{\infty }$ the relation of the form
\begin{equation}
ad\,J_{\infty }\cdot \Phi _{m+1}^{\infty }=-F_{m}\left( U,\Phi _{1}^{\infty
},...,\Phi _{m}^{\infty },D_{-M+1}^{\infty },...,D_{-M+m}^{\infty }\right) ,
\label{rec}
\end{equation}
where $F_{m}$ depends on $U$ and on the already determined coefficients and
their derivatives. By the same argument, \reff{rec} allows
to determine $D_{-M+m+1}^{\infty },\Phi _{m+1}^{\infty }$.

\begin{remark}
The coefficients $\Phi _{m}^{\infty },D_{m}^{\infty }$ are determined from
\reff{rec} not completely canonically, since we must fix in some way the
operator $\left( ad\,J_{\infty }\right) ^{-1}$.   One can show that this
freedom corresponds to the possibility to perform gauge transformations
\begin{eqnarray}
\frac{d}{dx}-D^{\infty } &\rightsquigarrow &\exp \left( -\phi \right) \circ
\left( \frac{d}{dx}-D^{\infty }\right) \circ \exp \phi ,  \label{jau} \\
\phi  &=&Id+\sum_{m=1}^{\infty }\phi _{m}z^{-m},\phi _{m}\in \frak{g}%
^{J_{\infty }}.  \nonumber
\end{eqnarray}
\end{remark}

The formal series $\Phi ^{\infty }$ is sometimes called the \emph{formal
Baker function at infinity} of the operator $L=\partial _{x}-U$.   For $\alpha
\in \frak{g}_{J_{\infty }}\otimes \Bbb{C}\left[ z,z^{-1}\right] $ and put
\begin{equation}
H_{\alpha }^{\infty }\left[ U\right] =%
{\mathrm {Res}}\,%
_{z=0}\int_{0}^{2\pi }%
{\mathrm {tr}}\,
\alpha \left( z\right) D^{\infty }(x,z)\,dx.  \label{halpha}
\end{equation}

\begin{theorem}
\label{zer}
\begin{enumerate}
\renewcommand{\theenumi}{\roman{enumi}} \renewcommand{\labelenumi}{(%
\theenumi)}

\item  Functionals $H_{\alpha }^{\infty }$ do not depend on the freedom in
the definition of the normal form.

\item  All functionals $H_{\alpha }^{\infty }$ are in involution with
respect to the Poisson bracket \reff{rbrack} on ${\mathbf{g}}_{r}^{*}$.

\item  Hamiltonian equation of motion defined by $H_{\alpha }^{\infty }$ on $%
{\mathbf{g}}_{r}^{*}$ have the form of zero curvature equations.
\end{enumerate}
\end{theorem}

\begin{lemma}
Gauge transformations \reff{jau} leave the density
${\mathrm {tr}}\,\alpha \left( z\right) D^{\infty }(x,z)$
 invariant up to a total
derivative.
\end{lemma}

\emph{Sketch of a proof.} Gauge transformations \reff{jau} map
$D^{\infty }$ into
$e^{-\phi} D^{\infty }e^{\phi} - e^{-\phi}\partial _{x} e^{\phi}   $.   By a standard formula,
\[
\partial _{x}\left( \exp \phi \right)=
\frac{e^{-ad\phi}-Id}{-ad\phi}\cdot \partial_x\phi=\\
\left(Id-\frac{1}{2}ad\phi+\frac{1}{3!}(ad\phi)^2+...\right)\cdot\partial_x\phi.
\]
Hence
\begin{eqnarray*}
{\mathrm {tr}}\,
\alpha \exp \left( -\phi \right) \partial _{x}\left( \exp \phi \right) &=&%
{\mathrm {tr}}\,
\alpha \cdot \left( \frac{e^{-ad\phi }-Id}{-ad\phi }\cdot \partial _{x}\phi
\right) = \\
{\mathrm {tr}}\,
\left( \frac{e^{-ad\phi }-Id}{-ad\phi }\cdot \alpha \right) \cdot \partial
_{x}\phi &=&%
{\mathrm {tr}}\,
\alpha \cdot \partial _{x}\phi =\partial _{x}\left(
{\mathrm {tr}}\,
\alpha \phi \right) ;
\end{eqnarray*}
where we also used
the invariance of trace, the condition $\alpha \in \frak{g}_{J_{\infty }}$
which assures that it commutes with $\phi $ and, finally, the condition $%
\partial _{x}\alpha =0$.

\begin{lemma}
The Frechet derivative of $H_{\alpha }^{\infty }$ is given by
\begin{equation}
{\mathrm {grad}}\,%
H_{\alpha }^{\infty }=\Phi ^{\infty }\alpha \left( \Phi ^{\infty }\right)
^{-1},  \label{freche}
\end{equation}
where $\Phi ^{\infty }$ is the formal Baker function.
\end{lemma}

\emph{Sketch of a proof.} Taking variations of both sides of (\ref{entre}),
we get
\[
\delta D^{\infty }=\left( \Phi ^{\infty }\right) ^{-1}\delta U\Phi ^{\infty
}+\left[ D^{\infty },\left( \Phi ^{\infty }\right) ^{-1}\delta \Phi ^{\infty
}\right] -\partial _{x}\left( \left( \Phi ^{\infty }\right) ^{-1}\delta \Phi
^{\infty }\right) .
\]
Hence
\begin{eqnarray*}
\lefteqn{\delta H_{\alpha }^{\infty }= } \\
&&{\mathrm {Res}}\,%
_{z=0}\int_{0}^{2\pi }\left\{
{\mathrm {tr}}\,%
\Phi ^{\infty }\alpha \left( \Phi ^{\infty }\right) ^{-1}\delta U+%
{\mathrm {tr}}\,%
\left( \partial _{x}\alpha -\left[ D^{\infty },\alpha \right] \right) \left(
\Phi ^{\infty }\right) ^{-1}\delta \Phi ^{\infty }\right\} dx,
\end{eqnarray*}
where we used the invariance of trace and integrated by parts; the
contribution of the second term vanishes, since $\partial _{x}\alpha =\left[
D^{\infty },\alpha \right] =0$.   $\Box $

\begin{corollary}
The Frechet derivative $X={\mathrm {grad}}\, H_{\alpha }^{\infty }$ satisfies
the differential equation
\begin{equation}
\partial _{x}X=\left[ U,X\right] .  \label{di}
\end{equation}
\end{corollary}

Indeed, \reff{freche} and \reff{entre} imply
\[
\partial _{x}X=\left( U\Phi ^{\infty }-\Phi ^{\infty }D^{\infty }\right)
\alpha \left( \Phi ^{\infty }\right) ^{-1}
-\Phi ^{\infty }\alpha \left( \Phi ^{\infty }\right) ^{-1}\left( U\Phi
^{\infty }-\Phi ^{\infty }D^{\infty }\right) \left( \Phi ^{\infty }\right)
^{-1}
\]
\[
=\left[ U,X\right] -\Phi ^{\infty }\left[ D^{\infty },\alpha \right] \left(
\Phi ^{\infty }\right) ^{-1}=\left[ U,X\right] .
\]
Note that geometrically \reff{di} is equivalent to
\begin{equation}
ad\,_{\widehat{\mathbf{g}}}^{*}\,
{\rm grad}\,%
H_{\alpha }^{\infty }\left[ U\right] \cdot U=0,  \label{casim}
\end{equation}
where $ad\,_{\widehat{\mathbf{g}}}^{*}$ is the coadjoint representation of
the Lie algebra $\widehat{\mathbf{g}}$ (the central extension of $\mathbf{g}$%
); this is precisely the property which characterizes the Casimirs of a Lie
algebra (cf. proposition \rref{casimirs}). In the present setting $H_{\alpha
}^{\infty }$ is not a true Casimir function: it is defined only for regular
elements $U\in \mathbf{g}$ with fixed highest coefficient. However, a short
inspection of the proof of theorem \rref{AKS} shows that it uses only
\reff{casim}; the last assertion of theorem \rref{zer} now follows.

In a similar way, we may define the second series of local Hamiltonians
which is associated with the pole at $z=0$;  one can show that the
Hamiltonians from these two families mutually commute (this does not follow
immediately from the arguments above, but may be proved in a similar way).
In a more general way, if the potential $U$ is a rational function on
${\C}P_{1}$(cf. Section 7.2), we may associate a series of local
Hamiltonians to each of its poles; the corresponding Frechet derivatives are
formal Laurent series in local parameter at the pole.

%%%%%%%%%%%%%%%%%%%%%%%%%%%%%%%%%%%%%%%%%%%%%%%%%%%%%%%%%%%%%%%%%%%%%%%%%%%%%

\newsubsection{Higher order differential operators}

In applications, it is quite common to deal with Lax representations which
contain higher order differential operators; the most famous example is the
KdV equation associated with the Schroedinger operator on the line
\[
D_{2}=-\frac{d^{2}}{dx^{2}}+u\left( x\right) .
\]
In order to put these operators into our framework, one needs extra work. We
shall outline the procedure without going into details. First of all, an $n$%
-th order differential equation
\begin{equation}
D_{n}=\frac{d^{n}\psi }{dx^{n}}+u_{n-2}\frac{d^{n-2}\psi }{dx^{n-2}}%
+...+u_{0}\psi +z\psi =0  \label{sc}
\end{equation}
may be written as a first order matrix equation,
\begin{equation}
\frac{d}{dx}\varphi +L\varphi =0,  \label{first}
\end{equation}
where
\begin{equation}
L=\left(
\begin{array}{llll}
0 & 1 & ... & 0 \\
\vdots & \ddots & \ddots & \vdots \\
0 & \ddots & 0 & 1 \\
u_{0}+z & \ldots & u_{n-2} & 0
\end{array}
\right) ,\ \varphi =\left(
\begin{array}{l}
\psi \\
\psi ^{\prime } \\
\vdots \\
\psi ^{\left( n-1\right) }
\end{array}
\right) .  \label{n}
\end{equation}
However, the companion matrix in (\ref{n}) contains ``too much zeros'' and
cannot be directly associated with a coadjoint orbit of the loop algebra.
Let us observe first of all that choosing the column vector $\varphi $ in (%
\ref{n}) in this particular form is not quite canonical:   we may
add to $\psi ^{\left( k\right) }$, $k=1,2,...,n-1$,      an
arbitrary linear combination (possibly, with variable coefficients) of $\psi
,\psi ^{\prime },...,\psi ^{\left( k-1\right) }$; this freedom amounts to
a gauge transformation $\varphi \left( x\right) \mapsto n\left( x\right)
\cdot \varphi \left( x\right) $,    where $n$ is a lower triangular (unipotent)
matrix. The potential $L$ in (\ref{n}) becomes an arbitrary matrix of the
form
\begin{equation}
L=\left(
\begin{array}{lllll}
\ast & 1 & 0 & \cdots & 0 \\
\ast & * & 1 & \ddots & \vdots \\
\vdots & \ddots & * & \ddots & 0 \\
&  &  & \ddots & 1 \\
\ast & \cdots &  & * & *
\end{array}
\right)  \label{tri}
\end{equation}
The companion matrix in (\ref{n}) is the result of \emph{gauge fixing; }%
indeed, we have

\begin{proposition}
For each potential $L$ of the form (\ref{tri}) there exists a unique lower
triangular gauge transformation which transforms it into the canonical form (%
\ref{n}).
\end{proposition}

Conclusion: The space of n-th order differential operators is the \emph{%
quotient space of the set of all potentials $ L $ of the form \reff{tri} modulo the gauge action of the lower triangular group}. This
quotient space is modelled on the set of companion matrices. With a little
skill in symplectic geometry one may describe this quotient space in terms
of Hamiltonian reduction (the key point is to observe that potentials of the
form (\ref{tri}) form a level surface for the moment map associated with our
gauge action).

There is one more difficulty: the term of highest degree in $z$ in the
potential $L\left( z\right) $ is a \emph{nilpotent matrix}, and so the
expansion procedure which yields local integrals of motion does not work. To
tackle with this trouble let us recall that the loop parameter $z$ is in
fact associated with a special grading \emph{(the standard grading)} of the
loop algebra; in this grading, the constant matrix in (\ref{first}) is
\begin{equation}
\left(
\begin{array}{lllll}
0 & 1 & 0 & \cdots & 0 \\
& 0 & 1 & \ddots & \vdots \\
\vdots &  & \ddots & \ddots & 0 \\
0 &  &  & \ddots & 1 \\
z & \cdots &  & 0 & 0
\end{array}
\right) .  \label{circ}
\end{equation}
If we use the so called \emph{principal grading }of the loop algebra
instead, (\ref{circ}) is replaced, after rescaling, with
\[
\zeta \cdot \left(
\begin{array}{lllll}
0 & 1 & 0 & \cdots & 0 \\
& 0 & 1 & \ddots & \vdots \\
\vdots &  & \ddots & \ddots & 0 \\
0 &  &  & \ddots & 1 \\
1 & \cdots &  & 0 & 0
\end{array}
\right) ,
\]
which is already a semisimple matrix with different eigenvalues; now local
conservation laws may constructed by expansion in $\zeta ^{-1}$ in the usual
way.

\newsubsection{Dressing and Solutions of Zero Curvature Equations}

Formula (\ref{freche}) for the Frechet derivative of a local Hamiltonian
makes it clear that the direct use of the global factorization theorem
(theorem \ref{fact}) to solve zero curvature equations is impossible.
Indeed, ${\mathrm {grad}}\, H_{\alpha }^{\infty }\left[ U\right] $
is a \emph{formal} series in local parameter $z^{-1}$;  it is therefore
impossible even to define the
1-parameter subgroup $\exp t\,{\mathrm {grad}}\,\/H_{\alpha }^{\infty }\left[ U\right] $;
this reflects real analytic difficulties which exist in the study of the initial
value problem with arbitrary initial data for integrable PDE's. Our way around
this difficulty is to introduce the following definition.

\begin{definition}
A differential operator $\partial _{x}-U$ is called \emph{strongly regular
at zero (at infinity)} if $U$ satisfies the conditions imposed in definition
\ref{reg} and, moreover, its formal Baker function at zero (at infinity) is
convergent.
\end{definition}

One may of course strongly doubt the merits of this definition. The point
is, however, that strongly regular operators form a\emph{\ homogeneous space:%
} there is a natural action of the loop group (called \emph{dressing
transformations}) on the set of potential which preserves strong regularity.
Moreover, commuting Hamiltonian flows generated by local Hamiltonians are
naturally included into the dressing group as a maximal commutative
subgroup. Examples of strongly regular operators include solitonic and
finite-band solutions; on the other hand, it is very difficult to give their
complete characteristics in local terms (i.e., to tell which initial data in
the space $C^{\infty }(S^{1},\frak{G})$ correspond to strongly regular
operators). The group action in question was discovered by Zakharov and
Shabat \cite{ZSh} (though at the time they did not notice the composition
law), and later by Sato and his school \cite{Sato}. A subtle question in the
theory of dressing transformations is the treatment of boundary conditions.
In our discussion above we used \emph{periodic} boundary conditions; the
main motivation was the Floquet theorem which gives an accurate description
of coadjoint orbits and Casimir functions for the central extension of the
loop algebra. Dressing transformations do \emph{not} preserve periodicity.%
\footnote{%
There is a different version of dressing which uses the \emph{Riemann
problem in the half-plane; }under some additional restrictions, this version
is adapted to the study of rapidly decreasing solutions.} One more
interesting point is the relation between dressing and the Poisson brackets:
since the dressing action is defined in terms of the Riemann problem and, on
the other hand, the Poisson structure is derived from the classical r-matrix
related to the \emph{same} Riemann problem, one anticipates some relation
between the two matters. However, the simplest guess appears to be wrong:
dressing transformations do \emph{not} preserve Poisson brackets on the
phase space of integrable PDE's. The exact relation is more subtle: the loop
group itself carries a natural Poisson bracket, again derived from the
classical r-matrix, and dressing is an example of a \emph{Poisson group
action }\cite{rims}. The same Poisson structure on the loop group plays an
important role in the study of difference Lax equations, and also yields a
semiclassical approximation in the theory of Quantum groups.

In this paragraph we shall discuss the simplest facts about dressing
transformations. We start with the motivation of the main definitions.

Let $\Bbb{G}$ be the double loop group consisting of maps
\[
g:\Bbb{R}\times \Bbb{C}P_{1}\backslash \left( \left\{ 0\right\} \cup \left\{
\infty \right\} \right) \rightarrow GL\left( n\right)
\]
which are holomorphic with respect to the 2nd argument. The Lie algebra $%
\mathbf{g}$ of $\Bbb{G}$ consists of maps
\[
U:\Bbb{R}\times \Bbb{C}P_{1}\backslash \left( \left\{ 0\right\} \cup \left\{
\infty \right\} \right) \rightarrow \frak{gl}\left( n\right)
\]
which are holomorphic with respect to the 2nd argument. Informally, we may
call elements of $\Bbb{G}$ \emph{wave functions}. Define the mapping
\[
p:\Bbb{G}\rightarrow \mathbf{g:}\psi \longmapsto U_{\psi }=\partial _{x}\psi
\cdot \psi ^{-1};
\]
$\Bbb{G}$ acts on itself by left multiplications; we have
\begin{equation}
g\cdot U_{\psi }\stackrel{\rm def}{=}U_{g\psi }=gU_{\psi }g^{-1}+\partial
_{x}g\cdot g^{-1},  \label{jauge}
\end{equation}
in other words, left multiplication on $\Bbb{G}$ induces gauge transformations on
the set of ``potentials'' $U$.   Conversely, if $U\in \mathbf{g}$,    let $\psi
_{U}$ be the fundamental solution of the differential equation on the line
\begin{equation}
\frac{d\psi }{dx}=U(x,z)\psi ,  \label{d}
\end{equation}
normalized by $\psi _{U}\left( 0\right) =Id$.   The mapping
\[
\mathbf{\psi }:{\mathbf{g}}\rightarrow \Bbb{G}:U\mapsto \psi _{U}
\]
is a right inverse of $p$.   Let $\mathcal{G}\subset \Bbb{G}$ be the subgroup
consisting of maps which do not depend on $x\in {\R}$.

\begin{lemma}
\label{loopgr}(i) ${\mathcal{G}}\subset \Bbb{G}$ is the stationary subgroup of
the zero potential on the line. (ii) ${\mathcal{G}}^{U}=\psi _{U}{\mathcal{G}}%
\psi _{U}^{-1}\subset \Bbb{G}$ is the stationary subgroup of
$U\in \mathbf{g.}$
\end{lemma}

Let $\Bbb{G}_{+}\subset \Bbb{G}$ be the subgroup consisting of functions
which are regular in ${\C}P_{1}\backslash \left\{ \infty \right\} $ with
respect to the second argument, and $\Bbb{G}\_\subset \Bbb{G}$ the subgroup
of functions which are regular in ${\C}P_{1}\backslash \left\{ 0\right\} $
and satisfy the normalization condition $g_{-}\left( \infty \right) =Id$.
The \emph{factorization problem} in $\Bbb{G}$ consists in representing $g\in
\Bbb{G}$ as $g=g_{+}g_{-}^{-1},g_{\pm }\in \Bbb{G}_{\pm }$;  the first
argument of $g(x,z)$ is regarded as a parameter.

\begin{theorem}
Formula
\begin{equation}
dr_{\left( x,y\right) }\psi =\left( \psi xy^{-1}\psi ^{-1}\right)
_{+}^{-1}\psi x=\left( \psi xy^{-1}\psi ^{-1}\right) _{-}^{-1}\psi y
\label{dr}
\end{equation}
defines a \emph{right} group action $dr:\left( {\mathcal{G}}\times {\mathcal{G}}%
\right) \times \Bbb{G}\longrightarrow \Bbb{G}$.
\end{theorem}

The definition looks rather exotic; in particular, the composition law for
dressing transformations is not at all obvious. Note that the equality in
\reff{dr} is closely related to the fact that
$\psi xy^{-1}\psi ^{-1}\in {\mathcal{G}}^{U_{\psi }}$
and hence the two factors $\left( \psi xy^{-1}\psi ^{-1}\right) _{\pm }$
define the same gauge transformation of the potential $U_{\psi }$.   To check
the composition law we shall give a geometric interpretation of \reff{dr}.
To avoid lengthy notation we shall use a model example.

Let $K$ be a group admitting a factorization into product of its subgroups $%
K_{\pm }$;  set $D(K)=K\times K$.   Let $K^{\delta }\subset K\times K$ be the
diagonal subgroup, $K^{\delta }=\left\{ \left( x,x\right) ;x\in K\right\} $,
and $K_{r}=K_{+}\times K_{-}$.

\begin{lemma}
$D(K)=K_{r}\cdot K^{\delta }$;  in other words, the factorization problem in $%
D(K)$,
\begin{equation}
\left( x,y\right) =\left( \eta _{+},\eta _{-}\right) \cdot \left( \xi ,\xi
\right) ,\,\eta _{\pm }\in K_{\pm },\,\xi \in K,  \label{dd}
\end{equation}
is uniquely solvable.\footnote{$D(K)$ is called \emph{the double} of $K$;
we have already used a similar construction for Lie algebras in Section 4.
One may notice that the construction below uses only factorization
in $D(K)\ $and the subgroups $K_{\pm }$ need not be complementary in $K$. In
Section 9 we shall once again encounter this construction in
the study of Poisson Lie groups.}
\end{lemma}

Indeed, we have
\[
\eta _{\pm }=\left( xy^{-1}\right) _{\pm },\;\xi =\left( xy^{-1}\right)
_{+}^{-1}x=\left( xy^{-1}\right) _{-}^{-1}y.
\]

\begin{corollary}
The quotient space $K_{r}\backslash D(K)$ of left coset classes $\mathrm{mod}%
\,K_{r}$ is modelled on the diagonal subgroup $K^{\delta }$;  the projection $%
\pi :D\left( K\right) \rightarrow K^{\delta }$ is given by
\[
\pi (x,y)=\left( xy^{-1}\right) _{+}^{-1}x=\left( xy^{-1}\right) _{-}^{-1}y.
\]
\end{corollary}

The group $D(K)$ acts on itself by right translations. Consider the
commutative diagram
\smallskip
\begin{center}
\begin{picture}(150,100)(10,10)
\put(152,110){\makebox(0,0)[lb]{$D(K)$}}
\put(150,65){\makebox(0,0)[lb]{$K_r\backslash D(K)$}}
\put(157,10){\makebox(0,0)[lb]{$K$}}
\put(22,110){\makebox(0,0)[lb]{$D(K)\times D(K)$}}
\put(10,65){\makebox(0,0)[lb]{${K_r\backslash D(K)}\times D(K)$}}
\put(28,10){\makebox(0,0)[lb]{$K\times D(K)$}}
\put(120,115){\makebox(0,0)[lb]{$m$}}
\put(165,90){\makebox(0,0)[lb]{${\pi}$}}
\put(58,90){\makebox(0,0)[lb]{${\pi}\times id$}}
\put(120,16){\makebox(0,0)[lb]{${\mathrm dr}$}}
\put(100,112){\vector(1,0){45}}
\put(100,68){\vector(1,0){45}}
\put(100,13){\vector(1,0){45}}
\put(161,105){\vector(0,-1){30}}
\put(53,105){\vector(0,-1){30}}
\put(162,55){\line(0,-1){30}}
\put(160,55){\line(0,-1){30}}
\put(53,55){\line(0,-1){30}}
\put(51,55){\line(0,-1){30}}
\end{picture}
\end{center}
\begin{proposition}
The right action $K\times D(K)\stackrel{dr}{\longrightarrow }K$ induced by
the identification of $K^{\delta }\subset D\left( K\right) $ with the coset
space $K_{r}\backslash D(K)$ is given by
\begin{equation}
dr(x,y):k\longmapsto \left( kxy^{-1}k^{-1}\right) _{+}kx=\left(
kxy^{-1}k^{-1}\right) _{-}ky.  \label{dres}
\end{equation}
\end{proposition}

A comparison of \reff{dr} and \reff{dres} explains the mystery around
the definition. In our model setting we assumed for simplicity that the
factorization problem is globally solvable. In general this is of course
not true; however, under reasonable conditions it is solvable on an open
dense subset of the big group, and hence the diagonal subgroup may be
identified with a ``big cell'' in the quotient space. Thus the situation
is not much different from the treatment of e.g. fractional linear
transformations on the line.

Returning back to \reff{dr}, note that the diagonal subgroup
$\mathcal{G} ^{\delta }\subset D\left( {\mathcal{G}}\right) $ acts by
$dr\left( g,g\right):\psi \mapsto \psi g$;  this action amounts to a simple change of the
normalization of the wave function and does nor affect the potential
$U=\partial _{x}\psi \psi ^{-1}$.   On the other hand, the subgroup
${\mathcal{G}} _{r}={\mathcal{G}}_{+}\times {\mathcal{G}}_{-}$
preserves the normalization condition $\psi \left( 0\right) =Id$.
Hence we may define an action ${\mathcal{G}}_{r}\times
{\mathbf{g}}\rightarrow {\mathbf{g}}$ on the space of
potentials with the help of commutative diagram
\begin{equation} \label{DR}
\begin{picture}(100,50)(10,20)
\put(110,70){\makebox(0,0)[lb]{$\Bbb{G}$}}
\put(110,20){\makebox(0,0)[lb]{$\mathbf{g}$}}
\put(10,70){\makebox(0,0)[lb]{${\mathcal {G}}_r\times \Bbb{G}$}}
\put(10,20){\makebox(0,0)[lb]{${\mathcal {G}}_r\times \mathbf{g}$}}
\put(70,75){\makebox(0,0)[lb]{$\mathrm {dr}$}}
\put(118,46){\makebox(0,0)[lb]{$\mathbf{\psi }$}}
\put(32,46){\makebox(0,0)[lb]{${id\times \mathbf{\psi }}$}}
\put(70,28){\makebox(0,0)[lb]{$\mathrm {dr}$}}
\put(45,72){\vector(1,0){55}}
\put(45,23){\vector(1,0){55}}
\put(112,30){\vector(0,1){35}}
\put(25,30){\vector(0,1){35}}
\end{picture}
\end{equation}
Let \ ${\mathbf{g}}_{M,N}\subset {\mathbf{g}}$ be the subspace of Laurent
polynomials with the degrees of pole at zero (at infinity) not exceeding $M$
(resp., $N)$

\begin{proposition}
Dressing action on $\mathbf{g}$ preserves ${\mathbf{g}}_{M,N}$.
\end{proposition}

\emph{Sketch of a proof.} Compare two equivalent formulae for dressing which
follow from \reff{dr}; the first one shows that dressing does not increase
the degree of pole at $0$,    the second one, that it does not affect infinity.

This argument explains the key idea of the dressing method: indeed, the
most striking property of dressing is the fact that it preserves the structure
of poles of the Lax operator. A slight
refinement of the same argument shows that \emph{dressing preserves symplectic leaves of the
r-bracket} in ${\mathbf{g}}_{M,N}\subset {\mathbf{g}}\simeq {\mathbf{g}}_{r}^{*}$
(here $r$ is the standard r-matrix associated with the factorization problem
in $\Bbb{G}$).

%\begin{exercise}
%Check this assertion for coadjoint orbits in the subspace $\mathbf{g}%
%_{1,1}\subset \mathbf{g.}$
%\end{exercise}

\begin{proposition}
Dressing preserves strong regularity.
\end{proposition}

\emph{Sketch of a proof. }Formal Baker functions at zero and at infinity of
the dressed operator are given by
\begin{eqnarray*}
\Phi _{0}^{g} &=& \left( \psi g_{+}g_{-}^{-1}\psi ^{-1}\right) _{+}^{-1}\Phi
_{0}, \\
\Phi _{\infty }^{g} &=& \left( \psi g_{+}g_{-}^{-1}\psi ^{-1}\right)
_{-}^{-1}\Phi _{\infty }.
\end{eqnarray*}
Clearly the gauge factors $\left( \psi g_{+}g_{-}^{-1}\psi ^{-1}\right)
_{+}^{-1},\left( \psi g_{+}g_{-}^{-1}\psi ^{-1}\right) _{-}^{-1}$ expand
into convergent series in local parameter around zero and infinity,
respectively; hence the same is true for the dressed wave functions.

In applications, dressing is usually applied to \emph{trivial,} or free, Lax
matrices. Let us assume that the leading coefficient at infinity is a
diagonal matrix with distinct eigenvalues. By definition, free Lax operator
has the form
\[
L_{free}=\frac{d}{dx}-D(z),
\]
where $D(z)$ is a constant diagonal matrix (with coefficients which are
polynomial in $z)$.   Our next assertion shows that the factorization theorem
survives for regular potentials; moreover, the dynamical flows associated with all
Lax equations (derived from a given Lax operator) correspond to
the  action of an abelian subgroup of the ``big'' dressing group
(essentially, the group of diagonal loops which are regular at infinity).

\begin{proposition}
Assume that $L$ is obtained from $L_{free}$ by dressing, $L=L_{free}^{g}$.
The integral curve of the Hamiltonian equation of motion with the
Hamiltonian $H_{\alpha }$ defined (\ref{halpha}) which starts at $L$ is
given by
\[
L(t) =g_{\pm }(t) ^{-1}\circ L\circ g_{\pm }(t) ,
\]
where $g_{\pm }(t) $ are regarded as multiplication operators on the line and
$g_{+}(t,x)$,    $g_{-}(t,x)$ are the solutions of the factorization
problem\footnote{This problem is nontrivial, since in order to get a nonzero Hamiltonian $%
H_{\alpha }$,    the constant diagonal matrix $\alpha \left( z^{-1}\right) $
must have pole at zero.}
\begin{eqnarray*}
g_{+}\left( t,x\right) g_{-}\left( t,x\right) ^{-1} &=& \psi _{free}\left(
x\right) \exp t\alpha \left( z^{-1}\right) \cdot g\cdot \psi _{free}\left(
x\right) ^{-1}, \\
\psi _{free}\left( x\right) &=& \exp xD(z).
\end{eqnarray*}
\end{proposition}

%%%%%%%%%%%%%%%%%%%%%%%%%%%%%%%%%%%%%%%%%%%%%%%%%%%%%%%%%%%%%%%%%%%%%%%%%%%%%

\newsection{Difference Equations and Poisson Lie Groups}

%%%%%%%%%%%%%%%%%%%%%%%%%%%%%%%%%%%%%%%%%%%%%%%%%%%%%%%%%%%%%%%%%%%%%%%%%%%%%

\newsubsection{Motivation: Zero curvature equations on the lattice}

We have already mentioned that integrable systems which are associated with
\emph{difference operators} require a special treatment; in this case the
underlying Poisson structures are \emph{nonlinear}, and hence the geometric
setting we considered so far, based on the use of the Lie-Poisson brackets,
must be generalized. Nonlinear equations associated with a finite difference
operator may be regarded as lattice analogues of zero curvature equations.
They are usually written in the form

\begin{equation}
\frac{dL_{m}}{dt}=L_{m}M_{m+1}-M_{m}L_{m},m\in \Bbb{Z}.  \label{dlax}
\end{equation}

Equation \reff{dlax} is the compatibility condition for the linear system
\begin{eqnarray}
\psi _{m+1} &=&L_{m}\psi _{m},  \label{daux} \\
\frac{d\psi _{m}}{dt} &=&M_{m}\psi _{m},\; m\in {\Z}.  \nonumber
\end{eqnarray}
This system of equations is covariant under the gauge transformations of the form
\begin{equation}
\psi_n\mapsto g_m\psi_m,\; L_m\mapsto g_{m+1}L_mg_m^{-1},\;M_m\mapsto g_mM_mg_m^{-1}
+\partial_tg_m\cdot g_m^{-1}.
\end{equation}

The use of difference operators associated with a 1-dimensional lattice is
particularly well-suited for the study of ``multi-particle'' problems. Let
us assume that ``potentials'' $L_{m}$ are periodic, $L_{m+N}=L_{m}$; the
period $N$ may be interpreted as the number of copies of an elementary
system. In this way we get families of Hamiltonians which remain integrable
for all $N$.   The phase spaces for such systems are direct products of
``one-particle'' phase spaces.

It is natural to suppose that the dynamics associated with difference Lax
equations develops on submanifolds of a matrix Lie group $G$ (or of a loop
group, if there is a spectral parameter), rather than on Lie algebras or
their duals.\footnote{%
Typical dynamical systems of this kind are the classical analogues of
lattice models in quantum statistics, although some of the systems which we
mentioned earlier (e.g., Toda lattices, certain tops, etc.) also admit
difference Lax representations.} As before, we are looking for a geometric
theory which should simultaneously account for the Poisson structure of the
phase space, the origin of conservation laws, and the reduction of dynamics
to factorization problems. An extension of the geometric scheme described in
Section 3 to lattice systems is based on the theory of \emph{%
Poisson Lie groups} introduced by Drinfeld \cite{dr} following the
pioneering work of Sklyanin \cite{skl}.\footnote{This latter paper was in turn
motivated by the Quantum Inverse Scattering Method developed by Faddeev,
Takhtajan and Sklyanin (cf. \cite{Fad}) and by the work of Baxter on Quantum
Statistical Mechanics (cf. \cite{bax}).} Unlike the Lie-Poisson brackets
discussed before, this new class of Poisson brackets was virtually unknown
in geometry.

Very briefly, the motivation for the formal definitions we are going to
discuss is as follows. As in the continuous case, the natural Hamiltonians
associated with the zero curvature equations should be \emph{gauge invariant.}
Let us assume that $L_{m+N}=L_{m}$.   Consider the monodromy mapping which assigns to the set of
local Lax matrices their ordered product,
\[
T:G^{N}\rightarrow G:\left( L_{0},...,L_{N-1}\right) \longmapsto T_{L}=%
\stackrel{\curvearrowleft }{\prod_{k}}L_{k}.
\]
A version of the Floquet
 theorem asserts that two difference operators with
periodic coefficients are gauge equivalent if and only if their monodromy
matrices are conjugate. Thus one expects the integrals of motion of equation
\reff{dlax} to be of the form
\[
h_{k}=%
{\rm tr}\,
T_{L}^{k}.
\]
This will hold if the monodromy itself satisfies a Lax equation,
\begin{equation}
\frac{dT_{L}}{dt}=\left[ T_{L},A_{L}\right]  \label{dnov}
\end{equation}
In more formal terms, let $F_{t}:G^N\longrightarrow G^N $ be the dynamical
flow associated with \reff{dlax} and $\bar{F}_{t}:G\mathbf{\longrightarrow }G$
the corresponding flow associated with \reff{dnov}; then the following
diagram should be commutative:
\begin{equation} \label{fl}
\begin{picture}(100,50)(10,20)
\put(110,70){\makebox(0,0)[lb]{$G^N$}}
\put(110,20){\makebox(0,0)[lb]{$G$}}
\put(10,70){\makebox(0,0)[lb]{$G^N$}}
\put(10,20){\makebox(0,0)[lb]{$G$}}
\put(60,28){\makebox(0,0)[lb]{${\bar{F_t}}$}}
\put(100,46){\makebox(0,0)[lb]{$T_L$}}
\put(20,46){\makebox(0,0)[lb]{$T_L$}}
\put(60,61){\makebox(0,0)[lb]{${F_t}$}}
\put(23,72){\vector(1,0){80}}
\put(23,23){\vector(1,0){80}}
\put(115,65){\vector(0,-1){35}}
\put(15,65){\vector(0,-1){35}}
\end{picture}
\end{equation}
We want to equip our phase spaces with Poisson structures which are
compatible with all mappings in this diagram. Moreover, we would like to
keep to our geometric picture suggested by theorem \ref{AKS}; this means
that we must find \emph{two} Poisson brackets in each space, so that

\begin{enumerate}
\renewcommand{\theenumi}{\roman{enumi}} \renewcommand{\labelenumi}{(%
\theenumi)}
\item  \emph{Spectral invariants of the monodromy are Casimir functions for
the first structure.}

\item  \emph{They are in involution with respect to the second one and
generate difference Lax equations (respectively, ordinary Lax equations for
the monodromy).}

\item  \emph{The flows }$F_{t},\bar{F}_{t}$\emph{\ preserve }intersections%
\emph{\ of symplectic leaves for the two brackets.}

\item  \emph{Vertical arrows in the diagram (\ref{fl}) are Poisson mappings
with respect to both structures.}

\item  \emph{Finally, the equations of motion (both upstairs and downstairs)
should reduce to a factorization problem.}
\end{enumerate}

It is remarkable that all these conditions may be satisfied with the help of
a single ingredient, the classical r-matrix, the same one which is
responsible for the factorization problem. As compared to the previous case
(that of Lie algebras), we need only one extra property (which actually is
satisfied in most of the examples we considered beforehand): \emph{the Lie
algebra }$\frak{g}$\emph{\ of our Lie group }$G$\emph{\ must be equipped
with an invariant inner product and the r-matrix }$r\in End\,\frak{g}$\emph{\
must be a skew-symmetric operator. }The construction which provides Poisson
brackets satisfying all these conditions is rather nontrivial (in fact, an
important message is that this is possible at all!); it may be naturally
divided into two separate problems:

\begin{enumerate}
\item  \emph{Given an r-matrix,} find the brackets on ${\mathbf{G}}=G^{N}$ and
on $G$ which have spectral invariants of the monodromy as their Casimir
functions.

\item  Find a Poisson bracket on ${\mathbf{G}}=G^{N}$ which yields zero
curvature equations \reff{dlax} as the equations of motion and assures that
the monodromy map is compatible with the Poisson brackets.
\end{enumerate}

The key point in both questions is that the r-matrix
is fixed \emph{in advance} and we must arrange the brackets with its help
(otherwise, there are too many options and the problem is not well posed!).

The second question is better known than the first one; in fact, it is this
question that has led to the theory of \emph{Poisson groups.} The key step
is the following simplifying assumption:

\begin{itemize}
\item  \emph{Dynamical variables associated with different factors in
${\mathbf{G}}=\underbrace{G\times ...\times G}_{N}$
commute with each other.}
\end{itemize}

By induction, the monodromy $T:G^{N}\longrightarrow G$ is a Poisson mapping
if \emph{the product map }$m:G\times G\rightarrow G$\emph{\ preserves
Poisson brackets.}

\begin{definition}
Poisson bracket on a Lie group $G$ satisfying the property above is called
\emph{multiplicative;} a Lie group equipped with multiplicative bracket is
called a \emph{Poisson Lie group.}
\end{definition}

{\footnotesize Let us explain this condition in a more explicit way. Let
$\varphi,\psi \in C^{\infty}(G)$;  put $\Phi(x,y)=\varphi(xy), \Psi(x,y)=\psi(xy)$,
$\Phi,\Psi \in  C^{\infty}(G\times G)$.   In order to compute the  Poisson bracket
$\{\Phi, \Psi\}$  we regard them as functions of  \emph{two} variables, that is,
we compute derivatives of $\Phi, \Psi$ with respect to  $x$ for fixed $y$ and with
respect to  $y$ for fixed $x$ and take the sum of these two terms; on the other
hand we may compute the bracket $\{\varphi,\psi \}$ for functions
of  \emph{one} variable  $z\in G$ and then insert $z=xy$.   Multiplicativity means
that the two results coincide.}

%%%%%%%%%%%%%%%%%%%%%%%%%%%%%%%%%%%%%%%%%%%%%%%%%%%%%%%%%%%%%%%%%%%%%%%%%%%%%%%%%%

\newsubsection{Key example: Sklyanin bracket}

Let us assume that the Lie algebra $\frak{g}$ of $G$ carries an invariant
inner product and $r\in {\mathrm{End}}\,\frak{g}$ is skew and satisfies the
modified Yang-Baxter identity. For $\varphi \in C^{\infty }\left( G\right) $
let $\nabla \varphi ,\nabla ^{\prime }\varphi \in \frak{g}$ be its
\emph{left and right gradients} defined by
\begin{eqnarray*}
\left\langle \nabla \varphi \left( x\right) ,X\right\rangle &=& \left( \frac{d%
}{ds}\right) _{s=0}\varphi \left( e^{sX}\cdot x\right) , \\
\left\langle \nabla ^{\prime }\varphi \left( x\right) ,X\right\rangle &=&
\left( \frac{d}{ds}\right) _{s=0}\varphi \left( x\cdot e^{sX}\right) ,X\in
\frak{g.}
\end{eqnarray*}

\begin{proposition}
The bracket on $G\ $%
\begin{equation}
\left\{ \varphi ,\psi \right\} =\frac{1}{2}\left( \left\langle r\left(
\nabla \varphi \right) ,\nabla \psi \right\rangle -\left\langle r\left(
\nabla ^{\prime }\varphi \right) ,\nabla ^{\prime }\psi \right\rangle
\right)   \label{skl1}
\end{equation}
is multiplicative and satisfies the Jacobi identity.\footnote{\label{Ob}%
Formula \reff{skl1} seems to be not the simplest bracket which can be
arranged using an antisymmetric operator: why not take $\left\{ \varphi
,\psi \right\} ^{g}\left( x\right) =\left\langle r\left( \nabla \varphi
\right) ,\nabla \psi \right\rangle $,    or $\left\{ \varphi ,\psi \right\}
^{d}\left( x\right) =\left\langle r\left( \nabla ^{\prime }\varphi \right)
,\nabla ^{\prime }\psi \right\rangle $ ? The reason is this: when $r$
satisfies the modified Yang-Baxter identity, neither of these brackets
satisfies Jacobi. However, the obstructions cancel each other when we take
the \emph{difference}, or the \emph{sum} of the two (and precisely in these
two cases)! We shall return to this question in Section 9.4 below.}%

\end{proposition}

\begin{remark}
Formula \reff{skl1} coincides with \reff{skl}, which we deduced from the
study of the monodromy map in the continuous case.

{\footnotesize This is of course not a coincidence. To explain why the Poisson bracket
for the monodromy on the circle should be multiplicative let us consider the auxiliary
problems \reff{aux}  with potentials $L$ consisting of two separate patches, so that
${\mathrm supp}\,L$ is the union of two disjoint  intervals,
${\mathrm supp}\,L=I'\cup I''$. Let us denote by
${\frak G}_{I'},\;{\frak G}_{I''}$ the set of all potentials supported on
$I',\,I''$, respectively; then
${\frak G}_{I'},\;{\frak G}_{I''}\subset{\frak G}$ are Poisson submanifolds,
and moreover,
${\frak G}_{I'\cup I''}={\frak G}_{I'}\times {\frak G}_{I''}$, again as Poisson
manifolds (which means that $L'\in{\frak G}_{I'},L'\in{\frak G}_{I'}$ may be treated
as independent variables with respect to our Poisson structure).\footnote{At this point
we use the crucial property of \reff{local}: the Poisson operator is a multiplication operator.}  Clearly,
  for  $L=L'+L'',\; L'\in {\frak G}_{I'}, L''\in {\frak G}_{I''}$ we have
$M_L=M_{L''}M_{L'}$ and the Poisson bracket for the monodromy matrix may be computed
in two different ways: either by computing the Poisson brackets for the monodromy
matrices $M_{L''}, M_{L'}$ regarded as independent variables, or alternatively by computing
 directly the monodromy $M_L$ for the potential $L=L'+L''$ decomposed into two separate patches.
 The two results should of course coincide, and this means precisely that the Poisson bracket
 for the monodromies should be multiplicative.
}
\end{remark}

Note that $\nabla \varphi \left( x\right) =Ad\,x\cdot \nabla ^{\prime
}\varphi \left( x\right) $,    or, in the matrix case, $\nabla \varphi \left(
x\right) =x\cdot \nabla ^{\prime }\varphi \left( x\right) \cdot x^{-1}$,    so
we may rewrite (\ref{skl1}) using only left gradients:
\[
\left\{ \varphi ,\psi \right\} \left( x\right) =%
{\mathrm {tr}}\,
\left( \eta _{r}\left( x\right) \cdot \left( \nabla \varphi \wedge \nabla
\psi \right) \right) ,
\]
where we set
\[
\eta _{r}\left( x\right) =\frac{1}{2}\left( r-Ad\,x^{-1}\circ r\circ
Ad\,x\right)
\]
and identify $\nabla \varphi \wedge \nabla \psi \in \frak{g}\wedge \frak{g}$
with an antisymmetric linear operator on $\frak{g}$,    using the inner
product. The function $\eta _{r}:\frak{g}\rightarrow \mathrm{End}\,\frak{g}$
satisfies the following remarkable functional equation:\footnote{This condition is usually expressed by
saying that $\eta _{r}$ is  a \emph{1-cocycle} (in fact, a coboundary) on $G$.
In these lectures we shall not use this language.}
\begin{equation}
\eta _{r}\left( xy\right) =\eta _{r}\left( x\right) +Ad\,x^{-1}\circ \eta
_{r}\left( y\right) \circ Ad\,x,  \label{cocy}
\end{equation}
  One may check
that this functional
equation   is exactly equivalent to the multiplicativity of \reff{skl1}.

Assume that $G=GL\left( n\right) $ is a matrix group. Let us consider
``tautological'' functions $\phi _{ij}$ on \ $G$ which assign to a matrix $%
L\in G$ its matrix coefficients, $\phi _{ij}\left( L\right) =L_{ij}$;
clearly, the ring of polynomials ${\C}\left[ \phi _{ij}\right]$
 is   dense in $C^{\infty }\left( G\right) $,    and the Poisson
bracket on $G$ is completely specified by its values on the ``generators'' $%
\phi _{ij}$.   Let us identify $r\in {\mathrm {End}}\,%
\frak{gl}(n)$ with an element of $\frak{gl}(n)\otimes \frak{gl}(n)\simeq
Mat\left( n^{2}\right) $.

\begin{proposition}
The Poisson bracket \reff{skl1} of matrix coefficients is given by
\begin{equation}
\left\{ \phi _{ij},\phi _{km}\right\} \left( L\right) =\left[ r,L\otimes
L\right] _{ikjm}.  \label{matr}
\end{equation}
\end{proposition}

The commutator in the r.h.s is computed in $Mat\left( n^{2}\right) $.
Usually people do not distinguish $\phi _{ij}$ and its values and write this
formula (with  suppressed matrix indices!) as

\begin{equation}\label{rll}
\left\{ L \, \stackrel{\otimes }{,} \, L \right\} =\left[ r, L \otimes L \right] .
\end{equation}
Formula \reff{rll} has served as the original definition of the \emph{Sklyanin bracket.}
Note that the r.h.s. in \reff{rll} is a \emph{quadratic} expression in
matrix coefficients (this is to be compared with the Lie-Poisson bracket
which is \emph{linear}).

Let us note some important properties of the Sklyanin bracket.
\begin{enumerate}
\renewcommand{\theenumi}{\roman{enumi}} \renewcommand{\labelenumi}{(%
\theenumi)}

\item \emph{\ The bracket is identically zero at the unit element  of the group.} (Indeed
at $x=e$ right and left gradients coincide).

\item \emph{\ Linearizing the bracket at the origin of group, we get the
structure of a Lie algebra in the cotangent space} $T_{e}^{*}G=\frak{g}^{*}:$
if $\xi ,\zeta \in \frak{g}^{*},X\in \frak{g}$,    choose $\varphi ,\psi \in
C^{\infty }\left( G\right) $ in such a way that $\nabla \varphi \left(
e\right) =\xi ,\nabla \psi \left( e\right) =\zeta $ and set
\[
\left\langle \left[ \xi ,\zeta \right] _{*},X\right\rangle =\left( \frac{d}{%
ds}\right) _{s=0}\{\varphi ,\psi \}\left( e^{sX}\right) =\left( \frac{d}{ds}%
\right) _{s=0}%
{\mathrm {tr}}\,
\left( \eta _{r}\left( e^{sX}\right) \cdot \left( \xi \wedge \zeta \right)
\right)
\]
(the second formula checks that the bracket $\left[ \xi ,\zeta \right] _{*}$
does not really depend on the choice of $\varphi ,\psi $ and so is well
defined).
\end{enumerate}
\begin{proposition}
\label{bial}The bracket $\left[ \xi ,\zeta \right] _{*}$ coincides with the
\emph{r-bracket }(up to the identification of\emph{\ }$\frak{g}$ and $\frak{g%
}^{*}$ induced by the inner product\footnote{%
Formula \reff{rr} explains the choice of normalization in \reff{skl1}: we
wanted to get the same thing as in \reff{rbr}.}:
\begin{equation}
\left[ \xi ,\zeta \right] _{*}=\frac{1}{2}\left( \left[ r\xi ,\zeta \right]
+\left[ \xi ,r\zeta \right] \right) .  \label{rr}
\end{equation}
\end{proposition}
Up to dualization, \reff{rr} coincides with \reff{rbr} which was our
starting point in Section 3. In the present setting we get some
extra properties which follow from multiplicativity of the bracket. Set
\[ \delta _{r}\left( X\right) =\left( \frac{d}{ds}\right) _{s=0}
\eta _{r}\left( e^{sX}\right) . \]
Explicitly, we get
\[ \delta _{r}\left( X\right) =adX\circ r-r\circ adX. \]
\begin{proposition}
\label{coc}(i) We have
\begin{equation}
{\mathrm {tr}}\,
\left( \delta _{r}\left( X\right) \circ \left( \xi \wedge \zeta \right)
\right) =\left\langle \left[ \xi ,\zeta \right] _{*},X\right\rangle .
\label{du}
\end{equation}
(ii) The mapping $\delta _{r}:\frak{g}\rightarrow {\mathrm {End}}\,
\frak{g}$ satisfies the functional equation
\begin{equation}
\delta _{r}\left( \left[ X,Y\right] \right) =\left[ adX,\delta _{r}\left(
Y\right) \right] -\left[ adY,\delta _{r}\left( X\right) \right] .
\label{cocycle}
\end{equation}
\end{proposition}

Equation \reff{du} shows that $\delta _{r}$ is the dual of the commutator
map $\frak{g}^{*}\wedge \frak{g}^{*}\rightarrow \frak{g}^{*}$.\footnote{In formal
terms, equation
\reff{cocycle} means that $\delta _{r}$ is a \emph{1-cocycle} on $\frak{ g}$
(with values in ${\mathrm {End}}\, \frak{g}$).}

\begin{definition}
\label{big}A pair $\left( \frak{g},\frak{g}^{*}\right) $ is called a \emph{%
Lie bialgebra} if (i) $\frak{g}$ and $\frak{g}^{*}$ are set in duality as
linear spaces, (ii) both $\frak{g}$ and $\frak{g}^{*}$ are Lie algebras,
(iii) the dual of the commutator map $\left[ ,\right] _{*}:\frak{g}%
^{*}\wedge \frak{g}^{*}\rightarrow \frak{g}^{*}$ satisfies the functional equation
\reff{du}.
\end{definition}

\begin{remark}
One can show that (iii) implies that in, the dual way, the mapping $\delta
_{*}:\frak{g}^{*}\rightarrow \frak{g}^{*}\wedge \frak{g}^{*}$ which dualizes
the commutator $\left[ ,\right] :\frak{g}\wedge \frak{g}\rightarrow \frak{g}$
is a 1-cocycle on $\frak{g}^{*}$,    and so this definition is symmetric.
\end{remark}

It is instructive to compare the definitions of Lie bialgebras and of the
double Lie algebras introduced in Section 3. These definitions are
\emph{different} and use different notions of the classical r-matrix. In the
case of double Lie algebras there are two Lie brackets on the \emph{same
underlying linear space}; the classical r-matrix is a linear operator $r\in
{\mathrm {End}}\, \frak{g}$;
in the case of Lie bialgebras there are two Lie brackets which
are defined on \emph{dual spaces} $\frak{g}$ and $\frak{g}^{*}$.   The
motivation for these definitions are very much different as well: as we saw,
double Lie algebras provide a natural setting for the Involutivity theorem
(theorem \rref{AKS}); Lie bialgebras naturally arise in the study of
multiplicative Poisson brackets on Lie groups. Proposition \rref{bial}
specifies the setting in which these two notions merge together: we must
assume that $\frak{g}$ carries an\emph{\ invariant inner product} which
allows to identify $\frak{g}$ and $\frak{g}^{*}$ and that
$r\in {\mathrm {End}}\, \frak{g}$ is \emph{skew}. One more natural condition
is the \emph{modified Yang-Baxter equation} (which assures that there is an underlying
factorization problem). When all three conditions are satisfied, we say that
$\left( \frak{g},\frak{g}^{*}\right) $ is a \emph{factorizable Lie
bialgebra. }Factorizable Lie bialgebras and the associated Poisson Lie
groups provide a natural environment for all applications to lattice
integrable systems.

Before turning to lattice zero curvature equations let us discuss ordinary Lax
equations on Lie groups. Here is a version of the factorization theorem
(theorem \rref{fact}) which applies in this setting. Let $G$ be a matrix
Lie group; we assume that the
Poisson bracket on $G$ is given by \reff{skl1} and that its tangent Lie
bialgebra $\left( \frak{g},\frak{g}^{*}\right) $ is factorizable. Let $%
I\left( G\right) \in C^{\infty }\left( G\right) $ be the algebra of central
functions on $G$ ($\varphi \in C^{\infty }\left( G\right) $ is central if $%
\varphi \left( xy\right) =\varphi \left( yx\right) $ for all $x,y\in G$).

\begin{theorem}
\label{Gr}(i) All central functions are in involution with respect to the
Sklyanin bracket \reff{skl1}. (ii) Hamiltonian equation on $G$ with
Hamiltonian $h\in I\left( G\right) $ may be written in Lax form$\footnote{%
In this formula the velocity vector $dL/dt$ belongs to the tangent space $%
T_{L}G$; in order to be more accurate, we may rewrite \reff{la} as an
equality in the Lie algebra:
\[ L^{-1}dL/dt=M_{\pm }-Ad\,L^{-1}\cdot M_{\pm } \]
}$
\begin{equation}
\frac{dL}{dt}=LM_{\pm }-M_{\pm }L,  \label{la}
\end{equation}
where $M_{\pm }=r_{\pm }\left( \nabla h\left( L\right) \right) $.   (iii) The
integral curve $L\left( t\right) $ of (\ref{la}) with $L\left( 0\right)
=L_{0}$ is given by
\begin{equation}
L\left( t\right) =g_{\pm }\left( t\right) ^{-1}L_{0}g_{\pm }\left( t\right) ,
\label{sol}
\end{equation}
where $g_{+}\left( t\right) ,g_{-}\left( t\right) $ are the solutions of the
factorization problem in $G$
\begin{equation}
g_{+}\left( t\right) g_{-}\left( t\right) ^{-1}=\exp t\nabla h\left(
L_{0}\right)   \label{pr}
\end{equation}
associated with $r$.
\end{theorem}

As before, the direct proof of theorem \rref{Gr} is easy: to check that
\reff{sol} is an integral curve of \reff{la} just compute the derivative of the
r.h.s in \reff{sol}. As in Section 3.3, there exists also a geometric
proof which explains the background machinery. Below, we shall briefly
outline the corresponding construction.

%%%%%%%%%%%%%%%%%%%%%%%%%%%%%%%%%%%%%%%%%%%%%%%%%%%%%%%%%%%%%%%%%%%%%%%%%%%%%%%%%%%%%%

\newsubsection{Duality for Poisson Lie groups}

As already noted, Lie bialgebras possess a remarkable symmetry:  if $\left( \frak{g},%
\frak{g}^{*}\right) $ is a Lie bialgebra, the same is true for $\left(
\frak{g}^{*},\frak{g}\right)$.   Hence the \emph{dual group} $G^{*}$
(which corresponds to $G^*$) also carries a
multiplicative Poisson bracket. In the case of factorizable Poisson groups
this dual bracket may be pushed forward to $G$ by means of the factorization map.
 Thus we get \emph{two}
brackets on $G$ which fit into our geometric treatment of Lax equations. The
best way to understand this duality is to notice that both $G$ and $G^{*}$
are Poisson subgroups of a bigger Poisson group, the \emph{double of $G$}.

Let $\left( \frak{g},\frak{g}^{*}\right) $be a Lie bialgebra; the linear
space $\frak{d}=\frak{g}\oplus \frak{g}^{*}$ carries a natural inner product
\begin{equation}
\left\langle \left\langle \left( X,F\right) ,\left( X^{\prime },F^{\prime
}\right) \right\rangle \right\rangle =\left\langle F,X^{\prime
}\right\rangle +\left\langle F^{\prime },X\right\rangle .  \label{dub}
\end{equation}
The following key theorem was discovered by Drinfeld.
\begin{theorem}
There exists a unique structure of the Lie algebra on $\frak{d}$ such that:
(i) $\frak{g},\frak{g}^{*}\subset \frak{d}$ are Lie subalgebras. (ii) The
inner product \reff{dub} is invariant.
\end{theorem}

\begin{corollary}
Let $P_{\frak{g}},P_{\frak{g}^{*}}$ be the projection operators onto $\frak{g%
},\frak{g}^{*}$ in the decomposition $\frak{d}=\frak{g}\oplus \frak{g}^{*}$.
Set $r_{\frak{d}}=P_{\frak{g}}-P_{\frak{g}^{*}}$; then $r_{\frak{d}}$
defines on $\frak{d}$ the structure of a factorizable Lie bialgebra.
\end{corollary}

The pair $\left( \frak{d},\frak{d}^{*}\right) $ is called the \emph{Drinfeld
double} of $\left( \frak{g},\frak{g}^{*}\right)$.   When the initial Lie
bialgebra $\left( \frak{g},\frak{g}^{*}\right) $ is itself factorizable, the
description of the double is very simple. Consider the Lie algebra $\frak{d}=%
\frak{g}\oplus \frak{g}$ (direct sum of two copies of $\frak{g}$) and equip
it with the inner product
\begin{equation}
\left\langle \left\langle \left( X,Y\right) , \left( X^{\prime%
}, Y^{\prime }\right) \right\rangle  \right\rangle = \left\langle X,X^{\prime }\right\rangle
-\left\langle Y,Y^{\prime }\right\rangle ,  \label{mi}
\end{equation}
where $\left\langle ,\right\rangle $ is the invariant inner product on $%
\frak{g}$.

\begin{proposition}
The double of a factorizable Lie algebra is canonically isomorphic to
$\frak{d}=\frak{g}\oplus \frak{g.}$
\end{proposition}

\emph{Sketch of a proof.} We have already seen in Section 4 that
there are two natural homomorphisms $r_{\pm }:\frak{g}^{*}\rightarrow \frak{g}$
given by (\ref{pm}); their combination yields an embedding $\frak{g}%
^{*}\subset \frak{g}\oplus \frak{g}$.   Let $\frak{g}^{\delta }\subset \frak{g}%
\oplus \frak{g}$ be the diagonal subalgebra,
$\frak{g}^{\delta }=\{(X,X); \, X\in \frak{g}\}$.   As discussed in Section 4,
${\frak{d}}={\frak{g}}^{\delta }{\dot{+}}{\frak{g}}^{*}$; it is easy to
check that the skew symmetry of $r$ and the choice of the inner product in
$\frak{d}$ imply that $\frak{g}^{*}$ and $\frak{g}^{\delta }$ are isotropic
with respect to the inner product (\ref{mi}); this is equivalent to the skew
symmetry of $r_{\frak{d}}=P_{\frak{g}}-P_{\frak{g}^{*}}$.   In matrix
notation, $r_{\frak{d}}\in {\mathrm {End}}\,\left( \frak{g}\oplus \frak{g}\right)$
is given by a $2\times 2$ block
matrix:
\begin{equation}
r_{\frak{d}}=\left(
\begin{array}{ll}
r & 2r_{+} \\
2r_{-} & -r
\end{array}
\right) .  \label{block}
\end{equation}

The Lie group which corresponds to $\frak{d}$ is $D\left( G\right) =G\times
G$.   Let $G^{\delta },G^{*}$ be its subgroups which correspond to $\frak{g}%
^{\delta },\frak{g}^{*}$.   Clearly, $G^{\delta }\subset D\left( G\right) $ is
the diagonal subgroup. As in (\ref{dd}), we may associate with the r-matrix $%
r_{\frak{d}}$ a factorization problem in $D\left( G\right) $.   Let us assume
for simplicity that it is globally solvable, i.e. $D\left( G\right) \simeq
G\cdot G^{*}$.

\begin{proposition}
Let us equip $D\left( G\right) $ with the Sklyanin bracket associated with $%
r_{\frak{d}}$.   Then $G^{\delta },G^{*}\subset D\left( G\right) $ are
Poisson subgroups (i.e. they are Poisson submanifolds and the induced
Poisson structure is multiplicative).\footnote{%
More precisely, the bracket induced on $G^{*}\subset D\left( G\right) $ has
opposite sign, due to the minus sign in $r_{\frak{d}}=P_{\frak{g}}-P_{\frak{g%
}^{*}}$.  }
\end{proposition}

The bracket induced on $G^{\delta }$ coincides with the original Sklyanin
bracket associated with $r$; the Poisson bracket on the dual group $G^{*}$
is described in the following way. Consider the mapping $m:D\left( G\right)
\rightarrow G:\left( x,y\right) \mapsto xy^{-1}$; its restriction to
$G^{*}\subset D\left( G\right) $ is a diffeomorphism.

\begin{proposition}
The Poisson bracket on $G$ induced by $m:G^{*}\rightarrow G$ is given by
\begin{eqnarray}
\left\{ \varphi ,\psi \right\} _{*}&=&1/2\left\langle r\nabla \varphi ,\nabla
\psi \right\rangle +1/2\left\langle r\nabla ^{\prime }\varphi ,\nabla
^{\prime }\psi \right\rangle   \nonumber \\
&&  -\left\langle r_{+}\nabla \varphi ,\nabla ^{\prime }\psi \right\rangle
-\left\langle r_{-}\nabla ^{\prime }\varphi ,\nabla \psi \right\rangle ,
\label{dualbr}
\end{eqnarray}
where $\nabla \varphi ,\nabla \psi $ and $\nabla ^{\prime }\varphi ,\nabla
^{\prime }\psi $ are left and right gradients of $\varphi ,\psi $.
\end{proposition}

Formula \reff{dualbr} looks rather complicated; however, the bracket $%
\left\{ ,\right\} _{*}$ is very remarkable.

\begin{proposition}
\label{dgroup}(i) Symplectic leaves of $\left\{ ,\right\} _{*}$ coincide
with conjugacy classes in $G$.   (ii) Casimir functions of $\left\{
,\right\} _{*}$ are precisely the central functions on $G$.   (iii) The
bracket (\ref{dualbr}) vanishes at the unit element $e\in G$; the induced
Lie bracket on the tangent space coincides with the original Lie bracket on $%
\frak{g}$.
\end{proposition}

Thus the bracket \reff{dualbr} provides the missing ingredient of our
geometric picture: we have got now \emph{two Poisson structures} on the same
underlying manifold $G$ and Lax equations preserve intersections of two
systems of symplectic leaves.

The symplectic leaves of the Sklyanin bracket also admit a description in
terms of the factorization problem. Let us identify $G$ with the quotient
space $D\left( G\right) /G^{*}$ using the factorization $D\left( G\right)
=G^{\delta }\cdot G^{*}$ Let us denote by $\pi $ the canonical projection $%
D\left( G\right) \rightarrow D\left( G\right) /G^{*}$; define the action $%
G^{*}\times G\rightarrow G$ using the commutative diagram
\smallskip
\begin{center}
\begin{picture}(150,100)(10,10)
\put(152,110){\makebox(0,0)[lb]{$D(G)$}}
\put(150,65){\makebox(0,0)[lb]{$D(G)/G^*$}}
\put(157,10){\makebox(0,0)[lb]{$G$}}
\put(30,110){\makebox(0,0)[lb]{$G^*\times D(G)$}}
\put(24,65){\makebox(0,0)[lb]{$G^*\times D(G)/G^*$}}
\put(33,10){\makebox(0,0)[lb]{$G^*\times G$}}
\put(120,115){\makebox(0,0)[lb]{$m$}}
\put(148,90){\makebox(0,0)[lb]{${\pi}$}}
\put(23,90){\makebox(0,0)[lb]{${id\times \pi}$}}
\put(108,16){\makebox(0,0)[lb]{$Dress$}}
\put(100,112){\vector(1,0){45}}
\put(100,68){\vector(1,0){45}}
\put(100,13){\vector(1,0){45}}
\put(161,105){\vector(0,-1){30}}
\put(53,105){\vector(0,-1){30}}
\put(162,55){\line(0,-1){30}}
\put(160,55){\line(0,-1){30}}
\put(53,55){\line(0,-1){30}}
\put(51,55){\line(0,-1){30}}
\end{picture}
\end{center}
%\[
%\begin{CD} G^*\times D(G) @>m>>D( G) \\ @V{id\times\pi}VV @V{\pi}V V\\
%G^*\times D(G)/G^* @>>> D(G)/G^* \\ @| @| \\ G^*\times G @>Dress>> G \end{CD}%
%.
%\]

(Here $m$ is the group multiplication in $D\left( G\right) $ restricted to
the subgroup $G^{*}\subset D\left( G\right) .)$ By analogy with the
definition of dressing transformations, this action is called \emph{dressing
action.}

\begin{proposition}
Symplectic leaves of the Sklyanin bracket in $G$ coincide with the orbits of
$G^{*}$ in $G$ with respect to the dressing action.
\end{proposition}

{\footnotesize More explicitly, the dressing prescription is as follows:
given $x\in G$,    $\left( h_{+},h_{-}\right) \in G^{*}$ solve the factorization
problem in \ $D\left( G\right) $;%
\[
\left( h_{+}x,h_{-}x\right) =\left( x^{\prime }g_{+},x^{\prime }g_{-}\right)
,x^{\prime }\in G,\left( g_{+},g_{-}\right) \in G^{*};
\]
then $Dress\left( h_{+},h_{-}\right) \cdot x=x^{\prime }$.   This immediately
yields the following formula in terms of the factorization problem in $G:$%
\begin{equation}
Dress\left( h_{+},h_{-}\right) \cdot x=h_{+}x\left(
x^{-1}h_{+}^{-1}h_{-}x\right) _{+}=h_{-}x\left(
x^{-1}h_{+}^{-1}h_{-}x\right) _{-},  \label{dract}
\end{equation}
where $x^{-1}h_{+}^{-1}h_{-}x=\left( x^{-1}h_{+}^{-1}h_{-}x\right) _{+}\cdot
\left( x^{-1}h_{+}^{-1}h_{-}x\right) _{-}^{-1}$ is the factorization in $G$
associated with the original r-matrix.}

\begin{exercise}     %{exercise}
Check this formula using the definition of dressing action.
\end{exercise}   %{exercise}

Dressing action may be regarded as a nonlinear analog of the coadjoint
representation, as it is clear from the following simple assertion.

\begin{proposition}
Dressing action leaves the unit $e\in G$ invariant; the linearization of the
dressing action in the tangent space $T_{e}G\simeq \frak{g}$ coincides with
the coadjoint representation of $G^{*}$ in $\frak{g}^{**}\simeq \frak{g}$.
\end{proposition}

In applications it is natural to assume that $G$ is an \emph{algebraic loop
group} consisting of matrices whose coefficients are rational functions of $%
z$.   Orbits of the dressing action of $G^{*}$ in this loop group are
finite-dimensional, and we get a description of phase spaces for Lax
equations which is largely parallel to the case of coadjoint orbits of $G_{r}
$ discussed in Section 6.

One more application of the dual Poisson structure described by \reff{dualbr}
is the accurate description of the Poisson properties of the dressing
transformations from Section 8.6. We keep to the notation introduced
in lemma \ref{loopgr}. Let $\mathcal{G}$ be the loop group; its Lie algebra
$L\frak{g} =\frak{g}[z,z^{-1}]$ is equipped with the inner product
\[
\left\langle X,Y\right\rangle ={\mathrm{Res}}_{z=0}{\mathrm{tr}}
\left( X\left(z\right) Y\left( z\right) \right)
\]
and with the standard r-matrix $r=P_{+}-P_{-}$ associated with the
Riemann problem. The loop group $\mathcal{G}$ equipped with the
corresponding Sklyanin bracket becomes a Lie-Poisson group. Let
${\mathcal{G}}_{r}\simeq {\mathcal{G}}^{*} = {\mathcal{G}}_{+}\times {\mathcal{G}}_{-}$
be the dual group equipped with its natural Poisson structure \reff{dualbr}.

\begin{theorem}
The dressing action described by the commutative diagram \reff{DR} is a
Poisson group action.
\end{theorem}

(We shall not reproduce the proof here; see \cite{rims}, \cite{Uhl}.)

%%%%%%%%%%%%%%%%%%%%%%%%%%%%%%%%%%%%%%%%%%%%%%%%%%%%%%%%%%%%%%%%%%%%%%%%%%%%

\newsubsection{Symplectic double and the free dynamics}

One more important ingredient of the geometric picture outlined in
Section 3.3 is the ``big phase space'' with ``free'' dynamical flow. Its
counterpart in the present setting is provided by the so called \emph{%
symplectic double} of $G$.   Let again $D\left( G\right) =G\times G$ be the
double of $G$; the Sklyanin bracket on $D\left( G\right) $ is given by
\begin{equation}
\left\{ \varphi ,\psi \right\} =\left\langle \left\langle r_{\frak{d}}\nabla
\varphi ,\nabla \psi \right\rangle \right\rangle -\left\langle \left\langle
r_{\frak{d}}\nabla ^{\prime }\varphi ,\nabla ^{\prime }\psi \right\rangle
\right\rangle \stackrel{def}{=}\left\{ \varphi ,\psi \right\} ^{g}-\left\{
\varphi ,\psi \right\} ^{d}  \label{skl-d}
\end{equation}
As noticed in footnote \ref{Ob}, the terms with left and right gradients
separately do not satisfy the Jacobi identity; the obstructions cancel when
the two are combined together. Explicitly,
\begin{equation}
\begin{array}{l}
\{ \{ \varphi _{1},\varphi _{2}\} ^{g},\varphi _{3}\}
^{g}+c.p. = \left\langle \left\langle \left[ \nabla \varphi _{1},\nabla
\varphi _{2}\right] ,\nabla \varphi _{3}\right\rangle \right\rangle ,
\label{Ja} \\
\\
\{ \{ \varphi _{1},\varphi _{2}\} ^{d},\varphi _{3}\}
^{d}+c.p. = -\left\langle \left\langle \left[ \nabla ^{\prime }\varphi
_{1},\nabla ^{\prime }\varphi _{2}\right] ,\nabla ^{\prime }\varphi
_{3}\right\rangle \right\rangle .
%\left\{ \left\{ \varphi _{1},\varphi _{2}\right\} ^{d},\varphi _{3}\right\}
%^{d}+c.p. &=&-\left\langle \left\langle \left[ \nabla ^{\prime }\varphi
%_{1},\nabla ^{\prime }\varphi _{2}\right] ,\nabla ^{\prime }\varphi
%_{3}\right\rangle \right\rangle .
\end{array}
\end{equation}
The two terms cancel, since $\nabla \varphi \left( x\right) =x\cdot \nabla
^{\prime }\varphi \left( x\right) \cdot x^{-1}$ and the inner product is $Ad$%
-invariant. It is important to notice that the crucial minus sign in \reff{Ja}
is due not to the minus in \reff{skl-d}, but rather to the fact that
the action of a group by right translations is its \emph{anti-}%
representation. (There are no terms of ``mixed chirality'' in the
obstruction, since left and right translations commute with each other.)
Thus we get the following assertion:

\begin{proposition}
\label{twor}Let $r,r^{\prime }\in {\mathrm {End}}\,
\frak{d}$ be two arbitrary classical r-matrices satisfying the modified
Yang-Baxter equation; the bracket
\begin{equation}
\left\{ \varphi ,\psi \right\} _{r,r^{\prime }}=\left\langle \left\langle
r\nabla \varphi ,\nabla \psi \right\rangle \right\rangle +\left\langle
\left\langle r^{\prime }\nabla ^{\prime }\varphi ,\nabla ^{\prime }\psi
\right\rangle \right\rangle   \label{two r}
\end{equation}
satisfies the Jacobi identity.
\end{proposition}

Specifically, let us take $r=r^{\prime }=r_{\frak{d}}$; the resulting
Poisson structure is \emph{nondegenerate} (at least if we assume -- as we
always do in this Section -- that the factorization problem in $D$ is
globally solvable), and hence defines a \emph{symplectic structure} on
$D\left( G\right) $.

\begin{definition}
The manifold $D\left( G\right)$ with the Poisson bracket
\begin{equation}
\left\{ \varphi ,\psi \right\} _{+}=\left\langle \left\langle
r_{\frak{d}}\nabla \varphi ,\nabla \psi \right\rangle \right\rangle +\left\langle
\left\langle r_{\frak{d}}\nabla ^{\prime }\varphi ,\nabla ^{\prime }\psi
\right\rangle \right\rangle   \label{plus}
\end{equation}
is called the \emph{symplectic double} of $G$.   We shall denote it $D\left(
G\right) _{+}$ in order to distinguish it from the \emph{Drinfeld double}
equipped with the Sklyanin bracket.
\end{definition}

The bracket \reff{plus} is \emph{not} multiplicative, so from the point of
view of the theory of Poisson groups $D\left( G\right) _{+}$ is not a group!
Instead, we have the following property which shows that $D\left( G\right)
_{+}$ is a \emph{principal homogeneous space} for $D\left( G\right) :$

\begin{proposition}
\label{poiss}Left and right multiplication in $D\left( G\right) $ induce
Poisson mappings $D\left( G\right) \times D\left( G\right)
_{+}\longrightarrow D\left( G\right) _{+},D\left( G\right) _{+}\times %
D\left( G\right) \longrightarrow D\left( G\right) _{+}$.
\end{proposition}

Proposition \rref{poiss} paves the way to use the reduction technique in our
present setting. Let us recall the point of view on reduction adopted in
Section 3.3: if $M$ is symplectic and $K\times M\longrightarrow M$ is
a group action, the reduction is the natural projection map $\pi
:M\longrightarrow M/K$ onto the space of $K$-orbits in $M$.   The key property
which we need to get a Poisson bracket on $M/K$ is this: \emph{Poisson
bracket of two }$G$\emph{-invariant functions on \ }$M$\emph{\ is again }$G$%
\emph{-invariant}. Let us discuss briefly how can one control this property.
For $X\in \frak{k}$ let us denote by $\hat{X}\in Vect\,M$ the vector field on $%
M$ generated by the 1-parameter transformation group $\exp tX$.   We have:
\[
\varphi \in C^{\infty }\left( M\right) \emph{\ is\ }G\emph{-invariant\ }%
\Longleftrightarrow \hat{X}\varphi =0\emph{\ for\ all\ }X\in \frak{k.}
\]
When vector fields $\hat{X}\in Vect\,M$ are Hamiltonian we have simply
\[
\hat{X}\{ \varphi ,\psi \} =\{ \hat{X}\varphi ,\psi \}
+\{ \varphi ,\hat{X}\psi \} =0.
\]

In the case of Poisson group actions vector fields $\hat{X}$ are no longer
Hamiltonian; however, the rate of nonconservation of Poisson brackets by
these vector fields may be characterized very sharply. For $\varphi \in
C^{\infty }\left( M\right) ,x\in M$,    let us denote by $\xi _{\varphi
}\left( x\right) \in \frak{k}^{*}$ the linear functional defined by
\[
\left\langle \xi _{\varphi }\left( x\right) ,X\right\rangle =\frac{d}{dt}%
_{t=0}\varphi \left( \exp tX\cdot x\right) .
\]

\begin{proposition}
Let us assume that $K$ is a Poisson Lie group with Lie bialgebra $\left(
\frak{k},\frak{k}^{*}\right) $.   The mapping $K\times M\longrightarrow M$ is
a Poisson mapping if and only if
\[
\hat{X}\{ \varphi ,\psi \} - \{ \hat{X}\varphi ,\psi \}
-\{ \varphi ,\hat{X}\psi \} =\left\langle \left[ \xi _{\varphi
},\xi _{\psi }\right] _{*},X\right\rangle .
\]
\end{proposition}

When $\hat{X}\varphi =\hat{X}\psi =0$ for all $X\in \frak{k},\xi
_{\varphi }=\xi _{\psi }\equiv 0$,    and hence $\hat{X}\left\{ \varphi ,\psi
\right\} =0$,    which assures the possibility of reduction. In a more general
way, let us say that a subgroup $H\subset K$ is \emph{admissible} if $\hat{X}%
\varphi =\hat{X}\psi =0$ for all $X\in \frak{h}\Longrightarrow \hat{X}%
\left\{ \varphi ,\psi \right\} =0$.

\begin{proposition}
$H\subset K$ is admissible $\Leftrightarrow \frak{h}^{\bot }\subset
\frak{k}^{*}$ is a Lie subalgebra.
\end{proposition}

\emph{Sketch of a proof.} $\hat{X}\varphi =\hat{X}\psi =0$ for all $X\in
\frak{h}$ implies that $\xi _{\varphi },\xi _{\psi }\in \frak{h}^{\bot }$.
When $\frak{h}^{\bot }\subset \frak{k}^{*}$ is a Lie subalgebra, $%
\left\langle \left[ \xi _{\varphi },\xi _{\psi }\right] _{*},X\right\rangle
=0$ for all $X\in \frak{h}$ and hence $\hat{X}\left\{ \varphi ,\psi \right\}
=0$.

As a first example of reduction, let us derive from \reff{plus} the dual
Poisson bracket \reff{dualbr} on $G$.

\begin{proposition}
\label{quot}Consider the action $G\times D\left( G\right)
_{+}\longrightarrow D\left( G\right) _{+}:h:\left( x,y\right) \longmapsto
\left( hx,hy\right) $.   This action is admissible, and the projection map $p:$
$D\left( G\right) _{+}\longrightarrow G:\left( x,y\right) \longmapsto
x^{-1}y $ is constant on its orbits. The quotient Poisson bracket on $%
D\left( G\right) _{+}/G\simeq G$ coincides with \reff{dualbr}.
\end{proposition}

{\footnotesize
%\scriptsize
Here is a simple check of this assertion: Let $\varphi \in
C^{\infty }\left( G\right) $; set $\Phi \left( x,y\right) =\varphi \left(
x^{-1}y\right) $.   Clearly, $\Phi $ is left $G^{\delta }$-invariant; hence
its left gradient $\nabla \Phi $ is in $\left( \frak{g}^{\delta }\right)
^{\bot }\subset \frak{d}$; but the inner product in $\frak{d}$
is so chosen that $\frak{g}^{\delta }\subset \frak{d}$ is a
\emph{maximal isotropic subspace}, i.e., $\left( \frak{g}^{\delta }\right)
^{\bot }=\frak{g}^{\delta }$.   Let us compute the Poisson bracket $\left\{
\Phi _{1},\Phi _{2}\right\} _{+}$ for two functions of such type. Since $%
\nabla \Phi _{1},\nabla \Phi _{2}\in \frak{g}^{\delta }$ and
$r_{\frak{d}}|_{\frak{g}^{\delta}}=id$,    we have
\[
\left\langle \left\langle r_{\frak{d}}\nabla \Phi _{1},\nabla \Phi
_{2}\right\rangle \right\rangle =\left\langle \left\langle \nabla \Phi
_{1},\nabla \Phi _{2}\right\rangle \right\rangle =0.
\]
On the other hand\footnote{Gradients in the l.h.s are computed with respect
to two variables $x,y$;  in the r.h.s they are computed
with respect to a single variable.},
\[
\ \nabla ^{\prime }\Phi \left( x,y\right) \ =\left(
\begin{array}{l}
-\nabla\phantom{\prime } \varphi \left( x^{-1}y\right) \\
\phantom{-}\nabla ^{\prime }\varphi \left( x^{-1}y\right)
\end{array}
\right) \in \frak{g} \oplus \frak{g};
\]
substituting this expression into $\left\langle \left\langle r_{\frak{d}%
}\nabla ^{\prime }\Phi _{1},\nabla ^{\prime }\Phi _{2}\right\rangle
\right\rangle $ and using \reff{block}, we get \reff{dualbr}.}

We can now state the nonlinear version of reduction theorem described in
Section 3.3. Let $\varphi \in I\left( G\right) $ be a central
function. Define the Hamiltonian $h_{\varphi }$ on $D\left( G\right) _{+}$
by $h_{\varphi }\left( x,y\right) =\varphi \left( x^{-1}y\right) $.

\begin{lemma}
(On free dynamics) The Hamiltonian flow on $D\left( G\right) _{+}$ defined
by $h_{\varphi }$ is given by
\begin{equation}
F_{t}:\left( x,y\right) \longmapsto \left( xe^{tX},ye^{tX}\right)
,\;X=\nabla \varphi \left( x^{-1}y\right) .  \label{Free}
\end{equation}
\end{lemma}

\begin{theorem}
(i) The Hamiltonian $h_{\varphi }$ is invariant with respect to the group $%
G^{*}$ which is acting on $D\left( G\right) _{+}$ via
\[
\left( h_{+},h_{-}\right) :\left( x,y\right) \longmapsto \left(
h_{+}xh_{-}^{-1},h_{+}yh_{-}^{-1}\right)
\]
(ii) $G^{*}$ is an admissible subgroup in $D\left( G\right) \times D\left(
G\right) $ (which acts on $D\left( G\right) _{+}$ by left and right
translations). (iii) The mapping
\[
\pi :D\left( G\right) _{+}\longrightarrow G:\left( x,y\right) \longmapsto
y_{+}^{-1}xy_{-}
\]
is constant on $G^{*}$-orbits in $D\left( G\right) _{+}$ and allows to
identify the quotient space $D\left( G\right) _{+}/G^{*}$ with the subgroup $%
G=\left\{ \left( x,e\right) ; x\in G\right\} \subset D\left( G\right) _{+}$.%
\footnote{As usual, for $g\in G$ we denote by $g_{+},g_{-}$ the solutions of the
factorization problem $g=g_{+}g_{-}^{-1}$.  } (iv) The quotient Poisson
structure on $D\left( G\right) _{+}/G^{*}\simeq G$ coincides with the
Sklyanin bracket. (v) The quotient flow $\bar{F}_{t}\ $on $G$ is given by $%
\bar{F}_{t}:x\longmapsto g_{\pm }\left( t\right) ^{-1}xg_{\pm }\left(
t\right) $,    where $g_{+}\left( t\right) ,g_{-}\left( t\right) $ solve the
factorization problem $\exp t\nabla \varphi \left( x\right) =g_{+}\left(
t\right) g_{-}\left( t\right) ^{-1}$,    and satisfies the Lax equation (\ref
{la}).
\end{theorem}

%%%%%%%%%%%%%%%%%%%%%%%%%%%%%%%%%%%%%%%%%%%%%%%%%%%%%%%%%%%%%%%%%%%%%%%%%%

\newsubsection{Lattice zero curvature equations and the twisted double}

So far our geometric construction is restricted to ordinary Lax equations on
a \emph{single copy} of $G$.   In order to put lattice zero curvature
equations into our framework we need one more effort. First of all, let us
state the factorization theorem which applies in this case (cf. Section 9.1).
Let again
\[
T:G^{N}\rightarrow G:\left( L_{0},...,L_{N-1}\right) \longmapsto T_{L}=%
\stackrel{\curvearrowleft }{\prod_{k}}L_{k}
\]
be the monodromy map; choose $\varphi \in I\left( G\right) $ and set $%
H_{\varphi }=\varphi \circ T$.   We define the ``wave function'' $\psi _{m}$
associated with the auxiliary linear problem (\ref{daux}) by
\[
\psi _{m}=\stackrel{\curvearrowleft }{\prod_{0\leq k\leq m-1}}L_{k}
\]
The Poisson structure on $G^{N}$ is defined as the direct product of
Sklyanin brackets on each factor. As usual, we assume that our basic
r-matrix is skew and satisfies the modified Yang-Baxter equation, so that
$\left( \frak{g},\frak{g}^{*}\right) $ is a factorizable Lie bialgebra.

\begin{theorem}
\label{fdcase}(i) The Hamiltonian equation of motion on $G^{N}$ with
Hamiltonian $H_{\varphi }$ may be written as
\begin{equation}
\frac{dL_{m}}{dt}=L_{m}M_{m+1}^{\pm }-M_{m}^{\pm }L_{m},  \label{lzcurv}
\end{equation}
where
\begin{equation}
M_{m}^{\pm }=r_{\pm }\left( \psi _{m}\nabla \varphi \left( T_{L}\right) \psi
_{m}^{-1}\right) .  \label{mop}
\end{equation}
(ii) Its integral curve with origin $L^{0}=\left(
L_{0}^{0},...,L_{N-1}^{0}\right) \ $ is given by
\begin{equation}
L_{m}\left( t\right) =g_{m}^{\pm }\left( t\right) ^{-1}L_{m}^{0}g_{m+1}^{\pm
}\left( t\right) ,  \label{evo}
\end{equation}
where $g_{m}^{+}\left( t\right) ,g_{m}^{-}\left( t\right) $ are the
solutions of the factorization problem
\begin{equation}
g_{m}^{+}\left( t\right) ,g_{m}^{-}\left( t\right) ^{-1}=\psi _{m}^{0}\cdot
\exp t\nabla \varphi \left( T_{L^{0}}\right) \cdot \left( \psi
_{m}^{0}\right) ^{-1}.  \label{evol}
\end{equation}
\end{theorem}

As usual, a direct proof of theorem \rref{fdcase} is easy. One point worth to
be mentioned is the computation of the gradients of the Hamiltonian: for that
end we may use the ``variation of constants'' in the auxiliary linear
problem \reff{daux}, similarly to the case of differential operators on
the circle discussed in Section 8.3. A geometric derivation is not
so straightforward. Let us introduce the following notation in order to
simplify the bulky formulae. We set $\mathbf{G}=G^{N},\frak{G}=\oplus ^{N}%
\frak{g},{\mathbf{L}}=\left( L_{0},...,L_{N-1}\right) \in \mathbf{G}$.   Let $\tau $
be the automorphism of $\mathbf{G}$ induced by cyclic permutation of indices,
$\left( L_{0},...,L_{N-1}\right) ^{\tau }=\left(
L_{1},L_{2},...,L_{N-1},L_{0}\right) $; we denote the corresponding
automorphism of $\frak{G}$ by the same letter.\footnote{%
The twisting automorphism $\tau $ plays the role which is similar to that of
the derivation $\partial _{x}$ for ordinary zero curvature equations; we can
say that its use allows to reproduce for lattice systems the effects of
central extension of loop algebras discussed in Section 8.1.}
Equations \reff{lzcurv}, \reff{evo}, \reff{evol} may be rewritten as
\begin{equation}\label{Quot}
\begin{array}{l}
\quad \frac{d\mathbf{L}}{dt} =\mathbf{LM}_{\pm }^{\tau }-\mathbf{M}_{\pm }
\mathbf{L},   \\
\mathbf{L}\left( t\right) =\mathbf{g}_{\pm }\left( t\right) ^{-1}
\mathbf{L}^{0}\mathbf{g}_{\pm }\left( t\right) ^{\tau },
\end{array}
\end{equation}
where
\[
\mathbf{g}_{+}\left( t\right) \mathbf{g}_{-}\left( t\right) ^{-1}  =\mathbf{%
\psi }^{0}\exp t\nabla \varphi \left( T_{L^{0}}\right) \left( \mathbf{\psi }%
^{0}\right) ^{-1}.
\]
(in the last expression $\exp t\nabla \varphi $ is embedded into $\mathbf{G}=%
G^{N}$ diagonally). The important feature of these formulae is the presence
of the \emph{twisting automorphism} $\tau $.   It plays the key role in the
definition of two more objects: (1) \emph{The second Poisson bracket} on $%
\mathbf{G}$ which has got the spectral invariants of the monodromy as its
Casimirs and (2) \emph{The twisted Poisson structure} on the ``big double'' $%
\mathbf{D}\left( \mathbf{G}\right) =\mathbf{G}\times \mathbf{G}$ which is
responsible for the free dynamics. We shall start with the latter question.

Let $r\in {\mathrm {End}}\, \frak{g}$
be the classical r-matrix associated with the factorization
problem in a single copy of $G$.   The corresponding r-matrix in the ``big''
algebra $\mathbf{g}$ is the direct sum of $r$'s acting in each copy of
$\frak{g}$,
\[
\mathbf{r}=\oplus _{m=0}^{N-1}r.
\]
Note that $\mathbf{r}\circ \tau =\tau \circ \mathbf{r.}$ In a similar way,
the r-matrix of the ``big double'' $\frak{D}=\frak{G}\oplus \frak{G}$ is
\[
\mathbf{r}_{\frak{D}}=\oplus _{m=0}^{N-1}r_{\frak{d}};
\]
in block notation we have
\[
\mathbf{r}_{\frak{D}}=\left(
\begin{array}{ll}
\mathbf{r} & 2\mathbf{r}_{+} \\
2\mathbf{r}_{-} & -\mathbf{r}
\end{array}
\right)
\]
(cf. \reff{block}). Let us define the automorphism $\mathbf{\tau }\in
Aut\left( \mathbf{G}\times \mathbf{G}\right) $ by $\left( \mathbf{x},\mathbf{%
y}\right) ^{\mathbf{\tau }}=\left( \mathbf{x},\mathbf{y}^{\tau }\right) $;
the corresponding automorphism of $\frak{D}=\frak{G}\oplus \frak{G}$ is
again denoted by the same letter. Put $\mathbf{r}_{\frak{D}}^{\mathbf{\tau }%
}=\mathbf{\tau }\circ \mathbf{r}_{\frak{D}}\circ \mathbf{\tau }^{-1}$; in
block notation we have
\[
\mathbf{r}_{\frak{D}}^{\mathbf{\tau }}=\left(
\begin{array}{ll}
\mathbf{r} & 2\mathbf{r}_{+}\circ \mathbf{\tau }^{-1} \\
2\mathbf{\tau \circ r}_{-} & -\mathbf{r}
\end{array}
\right) .
\]
Let us define the \emph{twisted Poisson structure} on $\mathbf{D}\left(
\mathbf{G}\right) $ by
\begin{equation}
\left\{ \varphi ,\psi \right\} _{\mathbf{\tau }}=\left\langle \left\langle
\mathbf{r}_{\frak{D}}\nabla \varphi ,\nabla \psi \right\rangle \right\rangle
+\left\langle \left\langle \mathbf{r}_{\frak{D}}^{\mathbf{\tau }}\nabla
^{\prime }\varphi ,\nabla ^{\prime }\psi \right\rangle \right\rangle .
\label{twist}
\end{equation}
The Jacobi identity for \reff{twist} follows from proposition \rref{twor}.
Our next assertion is the twisted version of proposition \rref{quot}.

\begin{proposition}
The action $\mathbf{G}\times \mathbf{D}\left( \mathbf{G}\right) _{\mathbf{%
\tau }}\longrightarrow \mathbf{D}\left( \mathbf{G}\right) _{\mathbf{\tau }%
}:h:\left( x,y\right) \longmapsto \left( hx,hy\right) $ is admissible; the
projection map $p:\mathbf{D}\left( \mathbf{G}\right) _{\mathbf{\tau }%
}\longrightarrow \mathbf{G}:\left( \mathbf{x},\mathbf{y}\right) \longmapsto
\mathbf{x}^{-1}\mathbf{y}$ is constant on its orbits. The quotient Poisson
bracket on $\mathbf{D}\left( \mathbf{G}\right) _{+}\mathbf{/G}\simeq \mathbf{%
G}$ is given by
\begin{equation}
\left\{ \varphi ,\psi \right\} =\left\langle \left\langle \mathbf{r}_{\frak{D%
}}^{\mathbf{\tau }}\left(
\begin{array}{l}
\nabla \varphi \\
-\nabla ^{\prime }\varphi
\end{array}
\right) ,\left(
\begin{array}{l}
\nabla \psi \\
-\nabla ^{\prime }\psi
\end{array}
\right) \right\rangle \right\rangle .  \label{twbr}
\end{equation}
\end{proposition}

The proof is parallel to that of proposition \rref{quot}. The properties of
the quotient bracket are also quite remarkable. (Note that unlike the
Sklyanin bracket on $\mathbf{G}=G\times ...\times G$ the bracket \reff{twbr}
is \emph{non-local}, due to presence of the twist $\tau)$.   Let us
consider the monodromy map $T:\mathbf{G}\longrightarrow G$.   Assume that
$\mathbf{G}$ carries the Poisson bracket \reff{twbr} and $G$ is equipped
with the bracket \reff{dualbr}.

\begin{proposition}
\label{twstr}(i) The monodromy $T:\mathbf{G}\longrightarrow G$ is a Poisson
mapping. (ii) The spectral invariants of the monodromy are the Casimirs of
the quotient bracket. (iii) Define the gauge action $\mathbf{G}\times
\mathbf{G}\longrightarrow \mathbf{G}:\mathbf{x}:\mathbf{g}\longmapsto
\mathbf{xg}\left( \mathbf{x}^{\tau }\right) ^{-1}$; when $N$ is odd, its
orbits coincide with symplectic leaves of the quotient bracket.
\end{proposition}

In brief, we see that the bracket \reff{twbr} provides us with the missing
ingredient for our geometric picture: it is the sought for \emph{second
bracket} which we need for a geometric treatment of lattice zero curvature
equations.

\begin{lemma}
(on free dynamics)
 Let $\varphi \in I\left( G\right) $; set $h_{\varphi
}=\varphi \, \circ T \, \in C^{\infty }\left( \mathbf{G}\right)$,    $H_{\varphi
}=h_{\varphi }\circ p\in C^{\infty }\left( \mathbf{D}\left( \mathbf{G}%
\right) \right) $.   The integral curves of the Hamiltonian $H_{\varphi }$ in $%
\mathbf{D}\left( \mathbf{G}\right) _{\mathbf{\tau }}$ are given by
\begin{equation}
\left( \mathbf{x}\left( t\right) ,\mathbf{y}\left( t\right) \right) =\left(
x_{0}e^{tX},y_{0}e^{tX}\right) ,X=\nabla \varphi \left( T(\mathbf{x}%
^{-1}y)\right) .  \label{freefl}
\end{equation}
\end{lemma}

\begin{theorem}
\label{last}Set $\mathbf{G}^{*}=\left( G^{*}\right) ^{N}$ and consider the
action $\mathbf{G}^{*}\times \mathbf{D}\left( \mathbf{G}\right) _{\mathbf{%
\tau }}\longrightarrow \mathbf{D}\left( \mathbf{G}\right) _{\mathbf{\tau }}$
given by
\[
\left( \mathbf{h}_{+},\mathbf{h}_{-}\right) :\left( \mathbf{x},\mathbf{y}%
\right) \longmapsto \left( \mathbf{h}_{+}\mathbf{xh}_{-}^{-1},\mathbf{h}_{+}%
\mathbf{y}\left( \mathbf{h}_{-}^{\tau }\right) ^{-1}\right) .
\]
This action is admissible; the quotient space $\mathbf{D}\left( \mathbf{G}%
\right) _{\mathbf{\tau }}/\mathbf{G}^{*}$ may be identified with $\mathbf{G\
}$by means of the map
\[
\pi :\mathbf{D}\left( \mathbf{G}\right) \longrightarrow \mathbf{G:}\left(
\mathbf{x},\mathbf{y}\right) \longmapsto \mathbf{y}_{+}^{-1}\mathbf{xy}%
_{-}^{\tau ^{-1}}
\]
which is constant on the orbits of ${\mathbf{G}}^{*}$,    and the quotient
Poisson structure coincides with the Sklyanin bracket. The flow \reff{freefl}
admits reduction with respect to this action; the quotient flow is
coincides with \reff{Quot}.
\end{theorem}

Theorem \rref{last} fills the last gap in our geometric picture and shows that
the qualitative behaviour of difference zero curvature equations is the same
as in the case of linear phase spaces.

%%%%%%%%%%%%%%%%%%%%%%%%%%%%%%%%%%%%%%%%%%%%%%%%%%%%%%%%%%%%%%%%%%%%%%%%%%%

\begin{acknowledgements}
It is my personal pleasure to thank the Organizing Committee and in particular
Professor A. Ferreira dos Santos and Dr. N. Manojlovic for their kind invitation
and for the inspiring atmosphere they have created during the School.

The
present work  was partially supported by the INTAS Open 00-00055 grant.
\end{acknowledgements}

%\newpage
%%%%%%%%%%%%%%%%%%%%%%%%%%%%%%%%%%%%%%%%%%%%%%%%%%%%%%%%%%%%%%%%%%%%%%%%%%

\address{Math\'ematique Physique\\
Universit\'e de Bourgogne\\
Dijon \\ France}


\received{}
\end{document}